\documentclass[iop,numberedappendix,appendixfloats,twocolappendix]{emulateapj}

\usepackage{amsmath, latexsym, color, verbatim}
\usepackage{graphicx,epsfig,natbib, epstopdf}

\newcommand{\mangadrp}{{\sc mangadrp}}
\newcommand{\idlutils}{{\sc idlutils}}
\newcommand{\idlspec}{{\sc idlspec2d}}

\newcommand{\mangacore}{{\sc mangacore}}
\newcommand{\otwo}{\textrm{[O\,{\sc ii}]}}
\newcommand{\othree}{\textrm{[O\,{\sc iii}]}}
\newcommand{\ntwo}{\textrm{[N\,{\sc ii}]}}
\newcommand{\stwo}{\textrm{[S\,{\sc ii}]}}
\newcommand{\Ha}{\ensuremath{\rm H\alpha}}
\newcommand{\Hb}{\ensuremath{\rm H\beta}}
\newcommand{\kms}{\textrm{km s}$^{-1}$}
\newcommand{\svn}{{\sc svn}}

\slugcomment{AJ, in press}

\shorttitle{The MaNGA Data Reduction Pipeline} 
\shortauthors{Law et al.}

\begin{document}

\title{The Data Reduction Pipeline for the SDSS-IV MaNGA IFU Galaxy Survey}

\author{David R.~Law\altaffilmark{1},
Brian Cherinka\altaffilmark{2},
Renbin Yan\altaffilmark{3},
Brett H.~Andrews\altaffilmark{4},
Matthew A.~Bershady\altaffilmark{5},
Dmitry Bizyaev\altaffilmark{6},
Guillermo A.~Blanc\altaffilmark{7,8,9},
Michael R.~Blanton\altaffilmark{10}, 
Adam S.~Bolton\altaffilmark{11},
Joel R.~Brownstein\altaffilmark{11},
Kevin Bundy\altaffilmark{12},
Yanmei Chen\altaffilmark{13,14},
Niv Drory\altaffilmark{15},
Richard D'Souza\altaffilmark{16},
Hai Fu\altaffilmark{17},
Amy Jones\altaffilmark{16},
Guinevere Kauffmann\altaffilmark{16},
Nicholas MacDonald\altaffilmark{18},
Karen L.~Masters\altaffilmark{19},
Jeffrey A.~Newman\altaffilmark{4},
John K.~Parejko\altaffilmark{18},
Jos\'e R.~S\'anchez-Gallego\altaffilmark{3},
Sebastian F.~S{\'a}nchez\altaffilmark{20}, 
David J.~Schlegel\altaffilmark{21},
Daniel Thomas\altaffilmark{19},
David A.~Wake\altaffilmark{5,22},
Anne-Marie Weijmans\altaffilmark{23},
Kyle B.~Westfall\altaffilmark{19},
Kai~Zhang\altaffilmark{3}
}

\altaffiltext{1}{Space Telescope Science Institute, 3700 San Martin Drive, Baltimore, MD 21218, USA (dlaw@stsci.edu)}
\altaffiltext{2}{Center for Astrophysical Sciences, Department of Physics and Astronomy, Johns Hopkins University, Baltimore, MD 21218, USA}
\altaffiltext{3}{Department of Physics and Astronomy, University of Kentucky, 505 Rose Street, Lexington, KY 40506-0055, USA}
\altaffiltext{4}{Department of Physics and Astronomy and PITT PACC, University of Pittsburgh, 3941 O'Hara St., Pittsburgh, PA 15260, USA}
\altaffiltext{5}{Department of Astronomy, University of Wisconsin-Madison, 475 N. Charter Street, Madison, WI, 53706, USA}
\altaffiltext{6}{Apache Point Observatory, P.O. Box 59, Sunspot, NM 88349, USA}
\altaffiltext{7}{Departamento de Astronom\'ia, Universidad de Chile, Camino del Observatorio 1515, Las Condes, Santiago, Chile}
\altaffiltext{8}{Centro de Astrof\'isica y Tecnolog\'ias Afines (CATA), Camino del Observatorio 1515, Las Condes, Santiago, Chile}
\altaffiltext{9}{Visiting Astronomer, Observatories of the Carnegie Institution for Science, 813 Santa Barbara St, Pasadena, CA, 91101, USA}
\altaffiltext{10}{Center for Cosmology and Particle Physics, Department of Physics, New York University, 4 Washington Place, New York, NY 10003}
\altaffiltext{11}{Department of Physics and Astronomy, University of Utah, 115 S 1400 E, Salt Lake City, UT 84112, USA}
\altaffiltext{12}{Kavli Institute for the Physics and Mathematics of the Universe, Todai Institutes for Advanced Study, the University of Tokyo, Kashiwa, Japan 277-8583 (Kavli IPMU, WPI)}
\altaffiltext{13}{School of Astronomy and Space Science, Nanjing University, Nanjing 210093, China}
\altaffiltext{14}{Key Laboratory of Modern Astronomy and Astrophysics (Nanjing University), Ministry of Education, Nanjing 210093, China}
\altaffiltext{15}{McDonald Observatory, Department of Astronomy, University of Texas at Austin, 1 University Station, Austin, TX 78712-0259, USA}
\altaffiltext{16}{Max Planck Institute for Astrophysics, Karl-Schwarzschild-Str. 1, D-85748 Garching, Germany}
\altaffiltext{17}{Department of Physics \& Astronomy, University of Iowa, Iowa City, IA 52242, USA}
\altaffiltext{18}{Department of Astronomy, Box 351580, University of Washington, Seattle, WA 98195, USA}
\altaffiltext{19}{Institute of Cosmology and Gravitation, University of Portsmouth, Portsmouth, UK}
\altaffiltext{20}{Instituto de Astronomia, Universidad Nacional Autonoma de Mexico, A.P. 70-264, 04510 Mexico D.F., Mexico}
\altaffiltext{21}{Physics Division, Lawrence Berkeley National Laboratory, Berkeley, CA 94720-8160}
\altaffiltext{22}{Department of Physical Sciences, The Open University, Milton Keynes, UK}
\altaffiltext{23}{School of Physics and Astronomy, University of St Andrews, North Haugh, St Andrews KY16 9SS, UK}

\begin{abstract}

Mapping Nearby Galaxies at Apache Point Observatory (MaNGA) is an
optical fiber-bundle integral-field unit (IFU) spectroscopic survey that is one of three core programs in the fourth-generation Sloan Digital Sky Survey (SDSS-IV). 
With a spectral coverage of 3622 -- 10,354 \AA\ and an average footprint of $\sim 500$ arcsec$^2$ per IFU
the scientific data products derived from MaNGA will permit exploration of the internal structure of a statistically large sample of 10,000 
low redshift galaxies in unprecedented detail.
Comprising 174 individually pluggable science and calibration IFUs with a near-constant data stream, MaNGA is expected to obtain 
$\sim$ 100 million raw-frame spectra and $\sim 10$ million 
reduced galaxy spectra over the six-year lifetime of the survey.
In this contribution, we describe the MaNGA Data Reduction Pipeline (DRP) algorithms and centralized metadata framework that produces 
sky-subtracted, spectrophotometrically
calibrated spectra and rectified 3-D data cubes that combine individual dithered observations.
For the 1390 galaxy data cubes released in Summer 2016 as part of SDSS-IV Data Release 13 (DR13), we demonstrate that the MaNGA data 
have nearly Poisson-limited sky subtraction shortward of $\sim$ 8500 \AA\ and reach a typical 
$10\sigma$ limiting continuum surface brightness $\mu = 23.5$ AB arcsec$^{-2}$ in a five arcsec diameter aperture in the $g$ band.
The wavelength calibration of the MaNGA data is accurate to $5$ \kms\  rms, with a median spatial 
resolution of  2.54 arcsec FWHM (1.8 kpc at the median redshift of 0.037) and a median spectral resolution of $\sigma = 72$ \kms.

\end{abstract}






\section{INTRODUCTION}

Over the last twenty years, multiplexed spectroscopic surveys have been valuable tools for 
bringing the power of statistics to bear on the study of galaxy formation.  Using large samples
of tens to hundreds of thousands of galaxies with optical spectroscopy from the Sloan
Digital Sky Survey \citep{york00,dr1.ref} for instance, studies have outlined fundamental relations
between stellar mass, metallicity, element abundance ratios, and star formation history \citep[e.g.,][]{kauffmann03,tremonti04,thomas10}.
However, this statistical power
has historically come at the cost of treating galaxies as point sources, with only a small and biased region
subtended by a given optical fiber contributing to the recorded spectrum.

As technology has advanced, techniques have been developed for {\it imaging spectroscopy}
that allow simultaneous spatial and spectral coverage, with correspondingly greater information density
for each individual galaxy.
Building on early work by (e.g.) \citet{colina99} and \citet{dezeeuw02}, such integral-field spectroscopy has provided a wealth of information.
In the nearby universe for instance, observations from the DiskMass survey \citep{bershady10} have indicated that late-type galaxies tend to have sub-maximal disks
\citep{bershady11}, while Atlas-3D observations \citep{cappellari11a}
showed that early-type galaxies frequently have rapidly-rotating components \citep[especially in low density environments;][]{cappellari11b}.
In the more distant universe, integral-field spectroscopic observations have been crucial in establishing the prevalence of high gas-phase velocity 
dispersions \citep[e.g.,][]{fs09,law09,law12,wisnioski15}, giant kiloparsec-sized ÒclumpsÓ of young stars \citep[e.g.,][]{fs11}, and powerful nuclear outflows \citep{fs14} that may indicate
fundamental differences in gas accretion mechanisms in the young universe \citep[e.g.,][]{dekel09}.

More recently, surveys such as CALIFA \citep[Calar Alto Legacy Integral Field Area Survey,][]{sanchez12,califadr2}, SAMI \citep[Sydney-AAO Multi-object IFS,][]{croom12,allen15}, 
and MaNGA \citep[Mapping Nearby Galaxies at APO,][]{bundy15} have begun to 
combine the information density of integral field spectroscopy with the statistical power of large multiplexed samples.
As a part of the 4th generation of the Sloan Digital Sky Survey
(SDSS-IV), the MaNGA project bundles single fibers from 
the Baryon Oscillation Spectroscopic Survey (BOSS) spectrograph \citep{smee13} into integral-field units (IFUs); over the six-year lifetime of the survey
(2014-2020) MaNGA will obtain spatially resolved optical+NIR spectroscopy of 10,000 galaxies at redshifts $z \sim 0.02 - 0.1$.
In addition to providing insight into the resolved structure of stellar populations, galactic winds, and dynamical evolution in the local universe
\citep[e.g.,][]{belfiore15,li15,wilkinson15}, the MaNGA data set will be an invaluable legacy product with which to help understand
galaxies in the distant universe.  As next-generation facilities come online in the final years of the MaNGA survey, IFU spectrographs such as TMT/IRIS \citep[][]{moore14,wright14},
JWST/NIRSPEC \citep[][]{closs08,birkmann14}, and 
JWST/MIRI-MRS \citep[][]{wells15} will trace the crucial rest-optical bandpass in galaxies out to redshift $z \sim 10$ and beyond.

Imaging spectroscopic surveys such as MaNGA face substantial calibration challenges
in order to meet the science requirements of the survey \citep{yan16b}.
In addition to requiring accurate absolute spectrophotometry from each fiber,
MaNGA must correct for gravitationally-induced flexure variability in the Cassegrain-mounted BOSS spectrographs,
determine accurate micron-precision astrometry for each IFU bundle, and
combine spectra from the individual fibers with accurate astrometric information in order to construct
3-D data cubes that rectify the wavelength-dependent differential atmospheric refraction
and (despite large interstitial gaps in the fiber bundles) consistently deliver high-quality {\it imaging} products.  
These combined requirements have driven a substantial software pipeline development effort throughout the early years of SDSS-IV.

\begin{deluxetable*}{lcccc}
\tablecolumns{5}
\tabletypesize{\scriptsize}
\tablecaption{IFU Data Reduction Software}
\tablehead{
\colhead{Telescope} & \colhead{Spectrograph} & \colhead{IFU}  & \colhead{Pipeline} & \colhead{Reference}}
\startdata
 \multicolumn{5}{c}{Fiber-Fed IFUs}\\
\hline
AAT & AAOMEGA & SAMI & {\sc 2dfdr} & \citet{sharp15}\\
Calar Alto 3.5m & PMAS & PPAK & {\sc p3d} & \citet{sandin10} \\ 
& & & {\sc r3d} & \citet{sanchez06}\tablenotemark{a}\\
& & & IRAF & \citet{martin13}\tablenotemark{c}\\
HET & VIRUS & VIRUS & {\sc cure} & \citet{snigula14}\\
McDonald 2.7m & VIRUS-P & VIRUS-P &{\sc vaccine} & \citet{adams11}\\
& & & {\sc venga} & \citet{blanc13}\\
SDSS 2.5m & BOSS & MaNGA & {\sc mangadrp} & This paper\\
WHT & WYFFOS & INTEGRAL & {\sc iraf} & \\
WIYN & WIYN Bench Spec. & DensePak & {\sc iraf} & \citet{andersen06}\\
& & SparsePak & {\sc iraf} & \\
\hline
 \multicolumn{5}{c}{Fiber $+$ Lenslet-Based IFUs}\\
\hline
AAT & AAOMEGA & SPIRAL & {\sc 2dfdr} & \citet{hopkins13}\\
Calar Alto 3.5m & PMAS & LARR & As PPAK above &  \\ 
Gemini & GMOS & GMOS & {\sc iraf} & \\
Magellan & IMACS & IMACS & {\sc kungifu} & \citet{bolton07}\\
VLT & GIRAFFE & ARGUS & {\sc girbldrs} & \citet{blecha00} \\
& & & {\sc eso cpl}\tablenotemark{b} & \\
 & VIMOS & VIMOS & {\sc vipgi} & \citet{zanichelli05}\\
 & & & {\sc eso cpl}\tablenotemark{b} & \\
\hline
 \multicolumn{5}{c}{Lenslet-Based IFUs}\\
\hline
Keck & OSIRIS & OSIRIS & {\sc osirisdrp} & \citet{krabbe04} \\
UH 2.2m & SNIFS & SNIFS & {\sc snurp} & \\
WHT & OASIS & OASIS & {\sc xoasis} & \\
 & SAURON & SAURON & {\sc xsauron} & \citet{bacon01}\\
\hline
 \multicolumn{5}{c}{Slicer-Based IFUs}\\
\hline
ANU & WiFeS & WiFeS & {\sc iraf} & \citet{dopita10}\\
Gemini & GNIRS & GNIRS & {\sc iraf} & \\
 & NIFS & NIFS & {\sc iraf} & \\
 VLT & KMOS & KMOS & {\sc eso cpl}\tablenotemark{b}, {\sc spark} & \citet{davies13}\\
   & MUSE & MUSE & {\sc eso cpl}\tablenotemark{b} & \citet{weil12}\\
      & SINFONI & SINFONI & {\sc eso cpl}\tablenotemark{b} & \citet{modig07}\\
\enddata
\tablenotetext{a}{See \citet{sanchez12} for details of the implementation for the CALIFA survey.}
\tablenotetext{b}{See http://www.eso.org/sci/software/cpl/}
\tablenotetext{c}{Reference corresponds to the DiskMass survey.}
\label{software.table}
\end{deluxetable*}

Historically, IFU data have been processed with a mixture of software tools ranging from custom built pipelines
\citep[e.g.,][]{zanichelli05} to general
purpose tools capable of performing all or part of the basic data reduction tasks for multiple IFUs.
For fiber-fed IFUs (with or without coupled lenslet arrays) that deliver a pseudo-slit of discrete apertures the raw data
is similar in format to traditional multiobject spectroscopy and has hence been able to build upon an existing code base.
In contrast, slicer-based IFUs produce data in a format more akin to long-slit spectroscopy, while pure-lenslet IFUs
are different altogether with individual spectra staggered across the detector.

Following \citet[][]{sandin10}, we provide here a brief overview of some of the common tools for the reduction of data
from optical and near-IR IFUs \citep[see also][]{bershady09}, including both fiber-fed IFUs
with data formats similar to MaNGA and lenslet-
and slicer-based IFUs by way of comparison.
As shown in Table \ref{software.table}, the {\sc iraf} environment remains a common 
framework for the reduction of data from many facilities, especially Gemini, WIYN, and the WHT.  Similarly, the various IFUs
at the VLT can all be reduced with software from a common ISO C-based pipeline library, although some other packages 
\citep[e.g., GIRBLDRS,][]{blecha00} are also capable of reducing data from some VLT IFUs.  Substantial effort has been 
invested in the {\sc p3d} \citep{sandin10} and {\sc r3d} \citep{sanchez06} packages as well, 
which together are capable of reducing data from a wide variety of  fiber-fed instruments
(including PPAK/LARR, VIRUS-P, SPIRAL, GMOS, VIMOS, INTEGRAL, and SparsePak) for which 
similar extraction and calibration algorithms are generally possible.
For survey-style operations, the SAMI survey has adopted a two-stage approach, combining a general-purpose
spectroscopic pipeline {\sc 2dfdr} \citep{hopkins13} with a custom three-dimensional stage to assemble IFU data cubes
from individual fiber spectra \citep{sharp15}.

Similarly, the MaNGA Data Reduction Pipeline ({\mangadrp}; hereafter the DRP) is also divided into two components.  Like
the {\sc kungifu} package \citep{bolton07}, the 2d stage of the DRP is based largely
on the SDSS BOSS spectroscopic reduction pipeline {\sc idlspec2d} (Schlegel et al., in prep), and 
processes the raw CCD data to produce sky-subtracted, flux-calibrated spectra for each fiber.  The 3d stage of the DRP
is custom built for MaNGA, but adapts core algorithms from the CALIFA \citep{sanchez12} and VENGA \citep{blanc13} pipelines
in order to produce astrometrically registered composite data cubes.
In the present contribution, we describe  version v1\_5\_4 of
the MaNGA DRP corresponding to the first public release of science data products in SDSS Data-Release 13 (DR13)\footnote{DR13
is available at http://www.sdss.org/dr13/}.

We start by providing a brief overview of the MaNGA hardware and operational strategy in \S \ref{hardwareops.sec},
and give an overview of the DRP and related systems in \S \ref{drpoverview.sec}.  We then discuss the individual elements of the DRP in detail,
starting with the basic spectral extraction technique (including detector preprocessing, fiber tracing, flatfield and wavelength calibration) in \S \ref{drp2d.1}.
In \S \ref{skysub.sec} we discuss our method of subtracting the sky background (including the bright atmospheric OH features) from the science spectra, and demonstrate that we achieve 
nearly Poisson limited performance shortward of 8500 \AA.  In \S \ref{fluxcal.sec} we discuss the method for spectrophotometric calibration of the MaNGA spectra,
and in \S \ref{waverect.sec} our approach to resampling and 
combining all of the individual spectra onto a common wavelength solution.
We describe the astrometric calibration in \S \ref{astrometry.sec}, combining a basic approach that takes 
into account fiber bundle metrology, differential
atmospheric refraction, and other factors (\S \ref{basicast.sec}) and an `extended' astrometry module that registers the MaNGA spectra against SDSS-I broadband imaging (\S \ref{eam.sec}).
Using this astrometric information we combine together individual fiber spectra into composite 3d data cubes in \S \ref{cubes.sec}.
Finally, we assess the quality of the MaNGA DR13 data products in \S \ref{dq.sec}, focusing on the effective angular and spectral resolution, wavelength calibration accuracy, and typical
depth of the MaNGA spectra compared to other extant surveys.  We summarize our conclusions in \S \ref{summary.sec}.  Additionally, we provide an Appendix \ref{datamodel.sec} in which
we outline the structure of the MaNGA DR13 data products and quality-assessment bitmasks.


\section{MaNGA Hardware and Operations}
\label{hardwareops.sec}

\subsection{Hardware}
\label{hardware.sec}

The MaNGA hardware design is described in detail by \citet{drory15}; here we provide a brief summary  of the major elements
that most closely pertain to the DRP.
MaNGA uses the Baryonic Oscillation Spectroscopic Survey (BOSS) optical fiber spectrographs \citep{smee13} installed on the Sloan Digital Sky Survey 2.5m telescope
\citep{gunn06}
at Apache Point Observatory (APO) in New Mexico.  These two spectrographs interface with a removable cartridge and plugplate system; each of the
six MaNGA cartridges contains
a full complement of 1423 fibers that can be plugged into holes in pre-drilled plug plates $\sim 0.7$ meters ($3^{\circ}$) in diameter and which feed pseudo-slits
that align with the spectrograph entrance slits when a given cartridge is mounted on the telescope.

These 1423 fibers are bundled into IFUs ferrules with varying sizes; each cartridge has twelve 7-fiber IFUs that are used for spectrophotometic
calibration and 17 science IFUs of sizes varying from 19 to 127 fibers (see Table \ref{ifutable.table}).
As detailed by \citet{wake16}, this assortment of sizes is chosen to best correspond to the angular diameter distribution of the MaNGA 
target galaxy sample.
The orientation of each IFU on the sky is fixed by use of a locator pin and pinhole a short distance West of the IFU.
Additionally, each IFU ferrule has a complement of associated sky fibers (see Table \ref{ifutable.table}) amounting to a total of 92 individually pluggable sky fibers.

\begin{deluxetable}{lcccc}
\tablecolumns{5}
\tabletypesize{\scriptsize}
\tablecaption{MaNGA IFU Complement per Cartridge}
\tablehead{
\colhead{IFU size} & \colhead{Purpose} & \colhead{Number}  & \colhead{N$_{\rm sky}$\tablenotemark{a}} & \colhead{Diameter\tablenotemark{b}}\\
\colhead{(fibers)} & & of IFUs & & \colhead{(arcsec)}}
\startdata
7 & Calibration & 12 & 1 & 7.5\\
19 & Science & 2 & 2 & 12.5 \\
37 & Science & 4 & 2 & 17.5 \\
61 & Science & 4 & 4 &  22.5 \\
91 & Science & 2 & 6 & 27.5 \\
127 & Science & 5 & 8 & 32.5 \\
\enddata
\tablenotetext{a}{Number of associated sky fibers per IFU ferrule.}
\tablenotetext{b}{Total outer-diameter IFU footprint.}
\label{ifutable.table}
\end{deluxetable}

\begin{figure*}
\epsscale{0.9}
\plotone{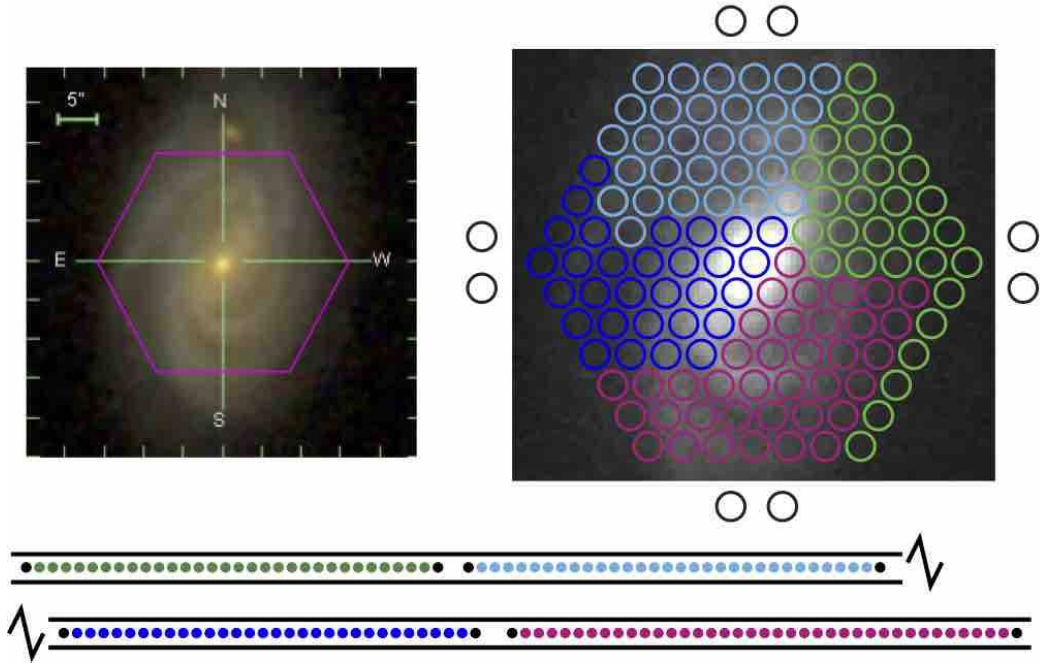}
\caption{Schematic diagram of a 127-fiber IFU on MaNGA galaxy 7495-12704.  The left-hand panel shows the SDSS 3-color RGB image of the galaxy overlaid with a hexagonal
bounding box showing the footprint of the MaNGA IFU.  The right-hand panel shows a 
zoomed-in greyscale $g$-band image of the galaxy overlaid with circles indicating the  locations
of each of the 127 optical science fibers (colored circles) and schematic locations of the 8 sky fibers (black circles).  
These fibers are grouped into 4 physical blocks on the spectrograph entrance slit (schematic diagram at bottom), with the sky fibers located at the ends of each block.
Note that the orientation of this figure is flipped in relation to Figure 9 of \citet{drory15} as the view presented here is on-sky (North up, East left).}
\label{ifudiagram.fig}
\end{figure*}

Each  fiber is 150 $\mu$m in diameter, consisting of a 120 $\mu$m glass core surrounded by a doped cladding and 
protective buffer.  The 120 $\mu$m core diameter subtends 1.98 arcsec
on the sky at the typical plate scale of $\sim 217.7$ mm degree$^{-1}$.
These fibers are terminated into 44 V-groove blocks with 21-39 fibers each that are mounted on the two pseudo-slits.  As illustrated in Figure \ref{ifudiagram.fig}, the sky fibers
associated with each IFU are located at the ends of each block to minimize crosstalk from adjacent science fibers.
In total, spectrograph 1 (2) is fed by 709 (714) individual fibers.

Within each spectrograph a dichroic beamsplitter reflects light blueward of 6000 \AA\ into a blue sensitive camera 
with a 520 l/mm grism and transmits red light into a camera with a 400 l/mm grism (both grisms consist of a VPH
transmission grating
between two prisms).  There are
therefore 4 `frames' worth of data taken for each MaNGA exposure, one each from the cameras b1/b2 (blue cameras on spectrograph 1/2 respectively) 
and r1/r2 (red cameras on spectrograph 1/2 respectively).  The blue cameras use 
blue-sensitive 4K $\times$ 4K e2V CCDs  while the red cameras use 4K $\times$ 4K fully-depleted LBNL CCDs, all with 15 micron pixels
\citep{smee13}.
The combined wavelength coverage of the blue and red cameras  is $\sim 3600 - 10,300$ \AA, with a 
$400$ \AA\ overlap in the dichroic region (see Table \ref{detector.table} for details).
The typical spectral resolution ranges from 1560 -- 2650, and is a function of 
the wavelength, telescope focus, and the location of an individual fiber
on each detector \citep[see, e.g., Fig. 37 of][]{smee13}; we discuss this further in \S \ref{wavecal.sec} and \ref{specres.sec}.

\begin{deluxetable}{lcccc}
\tablecolumns{5}
\tabletypesize{\scriptsize}
\tablecaption{BOSS Spectrograph Detectors}
\tablehead{
\colhead{} & \colhead{Blue Cameras} & \colhead{Red Cameras}}
\startdata
Type & e2V & LBNL fully depleted\\
Grism (l/mm) & 520 & 400\\
Wavel. Range (\AA)\tablenotemark{a} & 3600 -- 6300 & 5900 -- 10,300\\
Resolution\tablenotemark{a} & 1560 -- 2270 & 1850 -- 2650 \\
Detector Size & 4352 $\times$ 4224 & 4352 $\times$ 4224\\
Active Pixels\tablenotemark{b} & [128:4223, 56:4167] & [119:4232,48:4175]\\
Pixel Size ($\mu m$) & 15 & 15 \\
Read noise (e-/pixel)\tablenotemark{a} & $\sim 2.0$ & $\sim 2.5$\\
Gain (e-/ADU)\tablenotemark{a} & $\sim 1.0$ & $\sim 1.5-2.0$
\enddata
\tablenotetext{a}{Values are approximate; see \citet{smee13} for details.}
\tablenotetext{b}{0-indexed locations of active pixels between overscan regions.}
\label{detector.table}
\end{deluxetable}

While each of the IFUs is assigned a specific
plugging location on a given plate, the sky fibers 
are plugged non-deterministically  (although all are kept within 14 arcmin of the
galaxy that they are associated with).  Each cartridge is  mapped after plugging by scanning a laser along the pseudo-slitheads and recording the corresponding illumination pattern
on the plate.  In addition to providing a complete mapping of fiber number to on-sky location, this also serves to identify any broken or misplugged fibers.  This information is recorded in a central svn-based metadata repository called \mangacore\
(see \S \ref{metadata.sec}).

\subsection{Operations}
\label{ops.sec}

Each time a plate is observed, the cartridge on which it is installed is wheeled from a storage bay to the telescope and mounted at the Cassegrain focus.
Observers acquire a given field using a set of 16 coherent imaging fibers that feed a guide camera; these provide the necessary information to adjust focus,
tracking, plate scale, and field rotation using bright guide stars throughout a given set of observations.  In addition to simple tracking, constant corrections are required to compensate
for variations in temperature and altitude-dependent atmospheric refraction.

At the start of each set of observations, the spectrographs are first focused using a pair of hartmann exposures; the best focus is chosen to optimize the line spread function (LSF) across
the entire detector region (see \S \ref{extraction.sec} and \ref{wavecal.sec}).
25-second quartz calibration lamp flatfields and 4-second Neon-Mercury-Cadmium arclamp exposures are then obtained by closing the 8 flat-field petals covering the end of the telescope.
These provide information on the fiber-to-fiber relative throughput and wavelength calibration respectively; since both are mildly flexure dependent they are repeated every hour of observing at the relevant hour angle and declination.

After the calibration exposures are complete, science exposures are obtained in sets of three 15-minute dithered exposures.  As detailed by \citet{law15}, this integration time is a compromise
between the minimum time necessary to reach background limited performance in the blue while simultaneously minimizing astrometric drift due to differential atmospheric refraction (DAR)
between the individual exposures.
Since MaNGA is an imaging spectroscopic survey, image quality is important and the 56\% fill factor of circular fiber apertures within the hexagonal MaNGA IFU footprint \citep{law15}
naturally suffers from substantial gaps in coverage.  To that end, we obtain data in `sets' of 3 exposures dithered to the vertices of an equilateral triangle
with 1.44 arcsec to a side.  As detailed by \citet{law15}, this provides optimal coverage of the target field and permits complete reconstruction of the focal plane image.  Since atmospheric
refraction (which is wavelength dependent, time-dependent through the varying altitude and parallactic angle, and field dependent through 
uncorrected quadrupole scale changes over our 3$^{\circ}$ field)
degrades the uniformity of the effective dither pattern, each set of three exposures is
obtained in a contiguous hour of observing.\footnote{In practice, weather constraints sometimes make this impossible.  MaNGA scheduling software therefore takes into account 
observing conditions so that uniform-coverage sets can be assembled from exposures taken at similar hour angles on different nights.}
These sets of three exposures are repeated until each plate reaches a summed
signal-to-noise ratio (SNR) squared of 20 pixel$^{-1}$ fiber$^{-1}$ in $g$-band at $g = 22$ AB
and 36 pixel$^{-1}$ fiber$^{-1}$  in $i$-band at $i = 21$ AB
\citep[typically 2-3 hours of total integration, see][]{yan16b}.

All MaNGA galaxy survey observations are obtained in dark or grey-time for which the moon illumination is less than 35\% or below the horizon \citep[see][for details]{yan16b}.
Since MaNGA shares cartridges with the infrared SDSS-IV/APOGEE  spectrograph however \citep{wilson10}, both instruments are able to collect data simultaneously.
MaNGA and APOGEE therefore typically co-observe, meaning that data are also obtained with the MaNGA instrument during bright-time with up to 100\% moon illumination.
These bright-time data are not dithered, have substantially higher sky backgrounds, and are generally used for ancillary science observations
of bright stars with the aim of amassing a library of stellar reference spectra over the lifetime of SDSS-IV.  These bright-time data are processed with the same 
MaNGA software pipeline as the dark-time galaxy data, albeit with some modifications and unique challenges that we will address in a future contribution.


\section{Overview: MaNGA Data Reduction Pipeline (DRP)}
\label{drpoverview.sec}

In this section we give a broad overview of the MaNGA DRP and related systems in order to provide a framework for the detailed discussion of
individual elements presented in \S \ref{drp2d.1} - \ref{cubes.sec}.

\subsection{Data Reduction Pipeline (DRP)}

The MaNGA Data Reduction Pipeline (DRP) is tasked with producing fully flux calibrated data for each galaxy that has been spatially rectified and combined across
all individual dithered exposures in a multi-extension FITS format that may be used for scientific analysis.  This \mangadrp\ software is written primarily in IDL, with some C bindings for speed
optimization and a variety of python-based automation scripts.  Dependencies include the SDSS \idlutils\ and  NASA Goddard IDL astronomy users libraries; namespace collisions with these
and other common libraries have been minimized by ensuring that non-legacy DRP routines are prefixed by either `ml\_' or `mdrp\_'.
The DRP runs automatically on all data using the collaboration supercluster
at the University of Utah\footnote{Presently 27 nodes with 16 CPUs per node.}, is publicly accessible
in a subversion \svn\ repository at https://svn.sdss.org/public/repo/manga/mangadrp/tags/v1\_5\_4
with a BSD 3-clause license, and has been designed to run on individual users' home systems with relatively little overhead.\footnote{Installation
instructions are available at 
https://svn.sdss.org/public/repo/manga/mangadrp/tags/v1\_5\_4/pdf/userguide.pdf}
Version control of the \mangadrp\ code and dependencies is done via \svn\ repositories and traditional trunk/branch/tag methods; the version of \mangadrp\
described in the present contribution corresponds to tag v1\_5\_4 for public release DR13.  We note that v1\_5\_4 is nearly identical to v1\_5\_1 (which has been used for
SDSS-IV internal release MPL-4) save for minor improvements in cosmic ray rejection routines and data quality assessment statistics.

\begin{figure*}
\epsscale{1.0}
\plotone{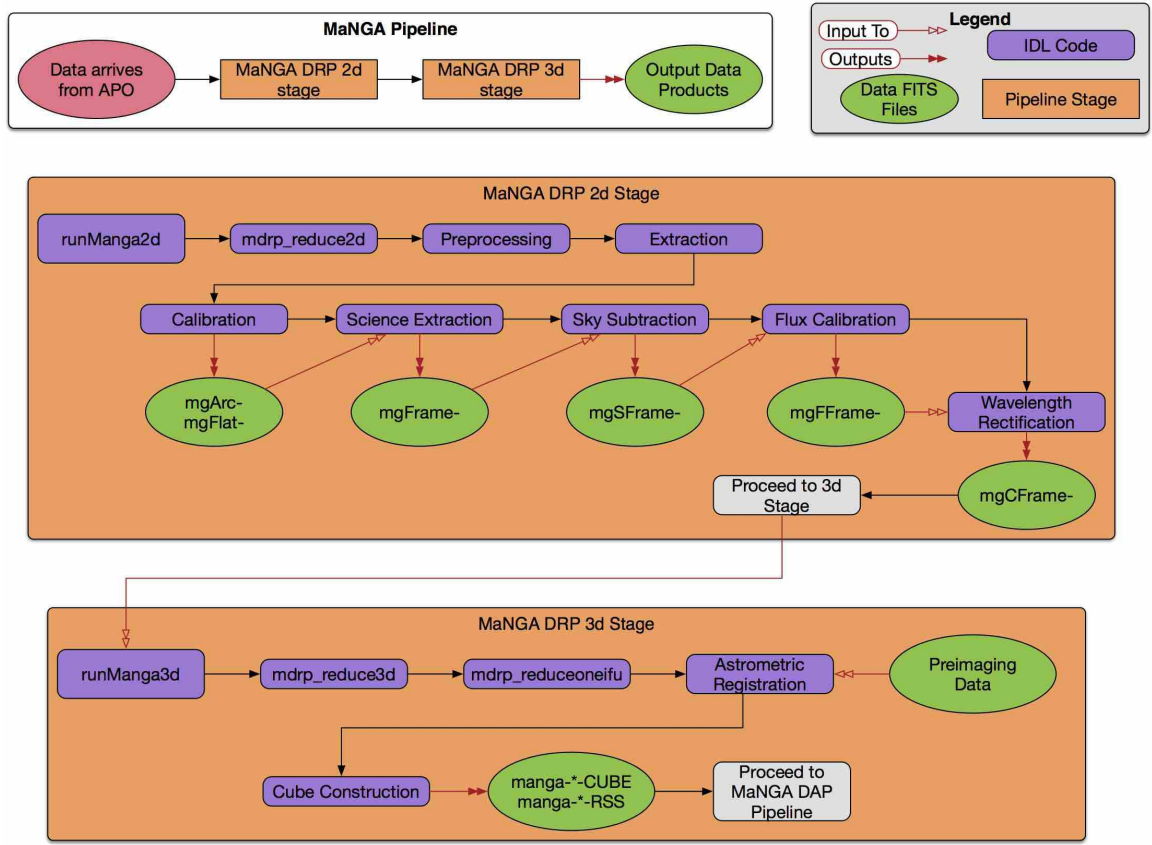}
\caption{Schematic overview of the MaNGA data reduction pipeline.  The DRP is broken into two stages: mdrp\_reduce2d and mdrp\_reduce3d.  The 2d pipeline data products are
flux calibrated individual exposures corresponding to an entire plate; the 3d pipeline products are summary data cubes and row-stacked-spectra for a given galaxy combining information
from many exposures.}
\label{flowchart.fig}
\end{figure*}

The DRP consists of two primary parts: the 2d stage that produces flux calibrated fiber spectra from individual exposures, and the 3d stage that combines individual
exposures with astrometric information to produce stacked data cubes.  The overall organization of the DRP is illustrated in Figure \ref{flowchart.fig}.
Each day when new data are automatically transferred from APO to the SDSS-IV central computing facility at the University of Utah
a cronjob triggers automated scripts that run the 2d DRP on all new exposures from the previous modified Julian date (MJD).  
These are processed
on a per-plate basis, and consist of a mix of science and calibration exposures (flatfields and arcs).  

The 2d stage of the MaNGA DRP is largely derived from the BOSS \idlspec\
pipeline \citep[see, e.g.,][Schlegel et al. in prep.]{dawson13}\footnote{The \idlspec\ software has also been used for the DEEP2 survey; see \citet{newman13}.} 
that has been modified
to address the different hardware
design and science requirements of the MaNGA survey (we summarize the numerous differences in Appendix \ref{pipeevol.sec}).
Each  frame undergoes basic
preprocessing  to remove overscan regions and variable-quadrant bias  before the 1-d fiber spectra are extracted from the CCD detector image.
The DRP first processes all of the calibration exposures
to determine the spatial trace of the fiber spectra on the detector and extract
fiber flatfield and wavelength calibration vectors, and applies these to the corresponding science frames.  
The science exposures are in turn extracted, 
flatfielded, and wavelength calibrated using the corresponding calibration files.  Using the sky fibers present in each exposure we create a super-sampled
model of the background sky spectrum, and subtract this off from the spectra of the individual science fibers.  Finally, the twelve minibundles targeting
standard stars in each exposure are used to determine the flux calibration vector for the exposure compared to stellar templates.  The final product of the 2d stage is a
single FITS file per exposure (mgCFrame) containing row-stacked spectra (RSS; i.e., a two-dimensional array in which each row corresponds
to an individual one-dimensional spectrum) of each of the 1423 fibers interpolated to a common wavelength grid and combined
across the four individual detectors.

Once a sufficient number of exposures have been obtained on a given plate, it is marked as complete at APO and 
a second automated script triggers the 3d stage DRP to combine each of the mgCFrame files
resulting from the 2d DRP.  For each IFU (including calibration minibundles) on the plate, the 3d pipeline identifies the relevant spectra in the mgCFrame files and assembles them into a
master row-stacked format consisting of all spectra for that target.  The astrometric solution as a function of wavelength for each of these spectra is computed on a per-exposure basis using
the known fiber bundle metrology and dither offset for each exposure, along with a variety of other factors including field and chromatic differential refraction \citep[see][]{law15}.
This astrometric solution is further refined using SDSS broadband imaging of each galaxy to adjust the position and rotation of the IFU fiber coordinates.
Using this astrometric information  the DRP combines the fiber spectra from individual exposures into a rectified data cube and associated inverse variance and mask cubes.
In post processing,  the DRP additionally computes mock broadband $griz$ images derived from the IFU data, estimates of the reconstructed PSF at $griz$, and a variety of quality control metrics
and reference information.

The final DRP data products in turn feed into the MaNGA Data Analysis Pipeline (DAP) which 
performs spectral modeling, kinematic fitting, and other analyses to produce
science data products such as H$\alpha$ velocity maps, kinemetry, spectral emission line ratio maps, etc. from the data cubes.
DAP data products will be made public in a future release and described in a forthcoming contribution by \citet{westfall16}.

\subsection{Quick-reduction pipeline (DOS)}

Rather than running the full DRP in realtime at the observatory, we instead use a pared-down version of the code that has been optimized
for speed that we refer to as DOS.\footnote{Daughter-of-Spectro.  This pipeline is a sibling to the Son-of-Spectro quick reduction pipeline
used by the BOSS and eBOSS surveys, both of which are descended from the original SDSS-I Spectro pipeline.}
The DOS pipeline shares much of its code with the DRP, performing reduction of the calibration and science exposures up 
through sky subtraction.  The primary difference is in the spectral extraction; while the DRP performs an optimized profile fitting 
technique to extract the spectra of each fiber (see \S \ref{extraction.sec}), DOS instead uses a simple boxcar extraction that sacrifices some accuracy
and robustness for substantial gains in speed.

\begin{figure}
\epsscale{1.2}
\plotone{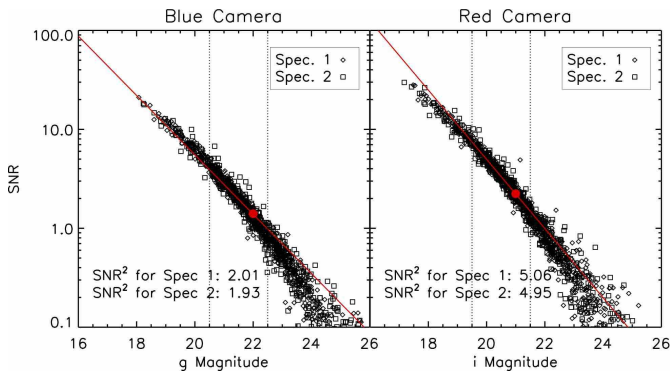}
\caption{SNR 
as a function of extinction-corrected fiber magnitude for blue (left panel) and red cameras (right panel), for spectrographs 1 and 2 (diamond vs square symbols respectively).  
The red line indicates the logarithmic
relation derived from fitting points in the magnitude range indicated by the vertical dotted lines.  The filled red circle indicates the derived
fit at the nominal magnitudes $g = 22$ and $i = 21$, with the SNR$^2$ values given for each spectrograph.
This example corresponds to
MaNGA plate 7443, MJD 56741, exposure 177378.}
\label{fiber2mag.fig}
\end{figure}

The primary purpose of DOS is to provide real-time feedback to APO observers on the quality and depth of each exposure.
Each exposure is characterized by an effective depth given by the mean SNR squared at a fixed fiber2mag \footnote{Fiber2mag is a magnitude measuring
the flux contained within a 2 arcsec diameter aperture; see http://www.sdss.org/dr13/algorithms/magnitudes/\#mag\_fiber}
of 22 ($g$-band) and 21 ($i$-band).  
The SNR of each fiber is calculated empirically by DOS from the sky-subtracted continuum 
fluxes and inverse variances, while nominal fiber2mags for each fiber in a galaxy IFU are calculated by applying aperture photometry to SDSS broadband imaging data at the known
locations of each of the IFU fibers (see \S \ref{basicast.sec}) and correcting for Galactic foreground extinction following \citet{schlegel98}.
As illustrated in Figure \ref{fiber2mag.fig},
the SNR as a function of fiber2mag for all fibers in a given exposure forms a logarithmic relation
that can be fitted and extrapolated to the effective achieved SNR at fixed nominal magnitudes $g = 22$ and $i = 21$.
This calculation is done independently for all four cameras using a $g$-band effective wavelength range $\lambda\lambda$ 4000-5000 \AA\
and an $i$-band effective wavelength range $\lambda\lambda 6910-8500$ \AA.
As described above in \S \ref{ops.sec}, we integrate on each plate until the cumulative SNR$^2$ in all complete sets of exposures reaches 
20 pixel$^{-1}$ fiber$^{-1}$ in $g$-band and 36 pixel$^{-1}$ fiber$^{-1}$  in $i$-band at the nominal magnitudes defined above.

\subsection{Metadata}
\label{metadata.sec}

MaNGA is a complex survey which requires tracking of multiple levels of metadata (e.g., fiber bundle metrology, cartridge layout, fiber plugging locations, etc.), any of which may
change on the timescale of a few days (in the case of fiber plugging locations) to a few years (if cartridges and/or fiber bundles are rebuilt).  At any point, it must be possible
to rerun any given version of the pipeline with the corresponding metadata appropriate for the date of observations.  This metadata must also be used throughout the different phases
of the survey from planning and target selection, to plate drilling, to APO operations, to eventual reduction and post-processing.

To this end, MaNGA maintains a central metadata repository \mangacore\ which is automatically synchronized between APO and the Utah data reduction hub using
daily crontabs.  Version control of files within \mangacore\ is maintained by a combination of modified Julian date (MJD) datestamps and periodic \svn\ tags corresponding to major data releases
(v1\_2\_3 for DR13).

\subsection{Quality Control}
\label{qc.sec}

Given the volume of data that must be processed by the MaNGA pipeline ($\sim 10$ million reduced galaxy spectra and $\sim 100$ million raw-frame 
spectra over the six-year lifetime of SDSS-IV\footnote{Assuming an average of 3 clear hours per night between the bright and dark time programs, 
5 exposures per hour (including calibrations), and $\sim$ 3000 spectra per exposure amongst 4 individual CCDs.}),
automated quality control is essential.  To that end, multiple
monitoring routines are in place.  The 2d and 3d stage DRP has bitmasks (MANGA\_DRP2PIXMASK and MANGA\_DRP3PIXMASK
respectively) associated with the primary flux extensions that can be used to 
indicate individual pixels (or spaxels\footnote{Spatial picture element.} in the case of the 3d data cubes) 
that are identified as problematic.  In the 2d case (spectra of all 1423 individual fibers within a single 
exposure), this pixel mask  indicates such things as cosmic ray events, bad flatfields, missing fibers, extraction problems, etc.  In the 3d stage
(a composite cube for a single galaxy that combines many individual exposures into a regularized grid), this pixel mask indicates things like
low/no fiber coverage, foreground star contamination, and other issues that mean a given spaxel should not be used for science.

Additionally, there are overall quality bits MANGA\_DRP2QUAL and MANGA\_DRP3QUAL that pertain to an entire exposure or data cube respectively and indicate
potential issues during processing.  In the 2d case, this can include effects like heavy cloud cover, missing IFUs, or abnormally
high scattered light.  In the 3d case, this can include warnings for bad astrometry, bad flux calibration, or (rarely) a critical problem suggesting that a
galaxy should not be used for science.  As of DR13, 22 of the 1390 galaxy data cubes are flagged as critically problematic for a variety of reasons ranging from the severe and unrecoverable (e.g., poor focus due to hardware failure, $\sim 5$ objects)
to the potentially recoverable in a future data release (e.g., failed astrometric registration due to a bright star at the edge of the IFU bundle) to the mundane (errant unflagged cosmic ray confusing the flux calibration QA routine).

All of these pixel-level and exposure-level data quality flags are used by the pipeline in deciding how and whether to 
continue to process data (e.g., flux calibration will not be attempted on an exposure flagged as completely cloudy).
We provide a reference table of the key MaNGA quality control bitmasks in Appendix \ref{bitmasks.sec}.



\section{Spectral Extraction}
\label{drp2d.1}

MaNGA exposures are differentiated from
BOSS/eBOSS exposures taken with the same spectrographs using FITS header keywords, and a planfile\footnote{A planfile is a plaintext ascii file that is both machine and human readable 
(see  http://www.sdss.org/dr13/software/par/) and contains a
list of the science and calibration exposures to be processed through a
given stage of the pipeline.} is created for each plate on a given MJD detailing each of the 
exposures obtained for which the quality was deemed by DOS at APO to be excellent.  The MaNGA DRP parses this planfile 
and performs pre-processing, spectral extraction, flatfielding, wavelength calibration, sky subtraction, and flux calibration on a per-exposure basis.

\subsection{Preprocessing}
\label{preproc.sec}

Raw data from each of the four CCDs (b1, r1, b2, r2) are in the format of 16 bit images with 4352 columns and 4224 rows (Table \ref{detector.table}),
with a 4096 x 4112 pixel active area (for the blue CCDs; 4114 $\times$ 4128 pixel active area for the red CCDs) and overscan regions along each edge of the detector.
As  described by \citet{dawson13}, 
the CCDs are read out with 4 amplifiers, one for each quadrant, resulting in variable bias levels.
Each exposure is preprocessed to remove the overscan regions of the detector, 
subtract off quadrant-dependent
biases, convert from bias corrected ADUs to electrons using quadrant-dependent gain factors derived from the overscan regions\footnote{Typical read noise 
and detector gains are given in Table \ref{detector.table}; these are slightly different for each quadrant of each detector, and can evolve
over the lifetime of the survey.  See \citet{smee13} for details.}, 
and divide by a flatfield containing the relative pixel-to-pixel response measured from a uniformly illuminated calibration image (see Figure \ref{rawdata.fig}).

A corresponding inverse variance image is created using the measured read noise and photon counts in each pixel; this inverse variance array is capped so that no
pixel has a reported SNR greater than 100.\footnote{This helps resolve problems arising when extracting extremely bright spectral emission lines.}
Finally, potential cosmic rays 
(which affect $\sim$ ten times as many pixels in the red cameras as in the blue) 
are identified and flagged using the same algorithm adopted previously by the SDSS imaging and spectroscopic surveys.
As discussed by R.~H. Lupton (see http://www.astro.princeton.edu/$\sim$rhl/photo-lite.pdf), this algorithm is a first-pass approach that
successfully detects {\it most} cosmic rays by looking for features sharper than the known detector PSF 
but sometimes incompletely flags pixels around the edge of cosmic ray tracks. 
A second-pass approach that addresses these residual features is applied later in the pipeline, as described in \S \ref{waverect.sec}.
The inverse variance image is combined
with this cosmic ray mask and a reference bad pixel mask so that affected pixels are assigned an inverse variance of zero (and hence have zero weight in the reductions).

\subsection{Calibration Frames}

All flatfield and arc calibration frames from a planfile are reduced prior to processing any science frames.  These provide estimates of the fiber-to-fiber flatfield and the wavelength solution,
and are also critical for determining the locations of individual fiber spectra on the detectors.  Since there are 4 cameras, each reduced flatfield (arc) exposure corresponds to four
mgFlat (mgArc) multi-extension FITS files as described in the data model in \S \ref{datamodel.sec}.

\subsubsection{Spatial Fiber Tracing}
\label{tracing.sec}

\begin{figure*}
\epsscale{1.2}
\plotone{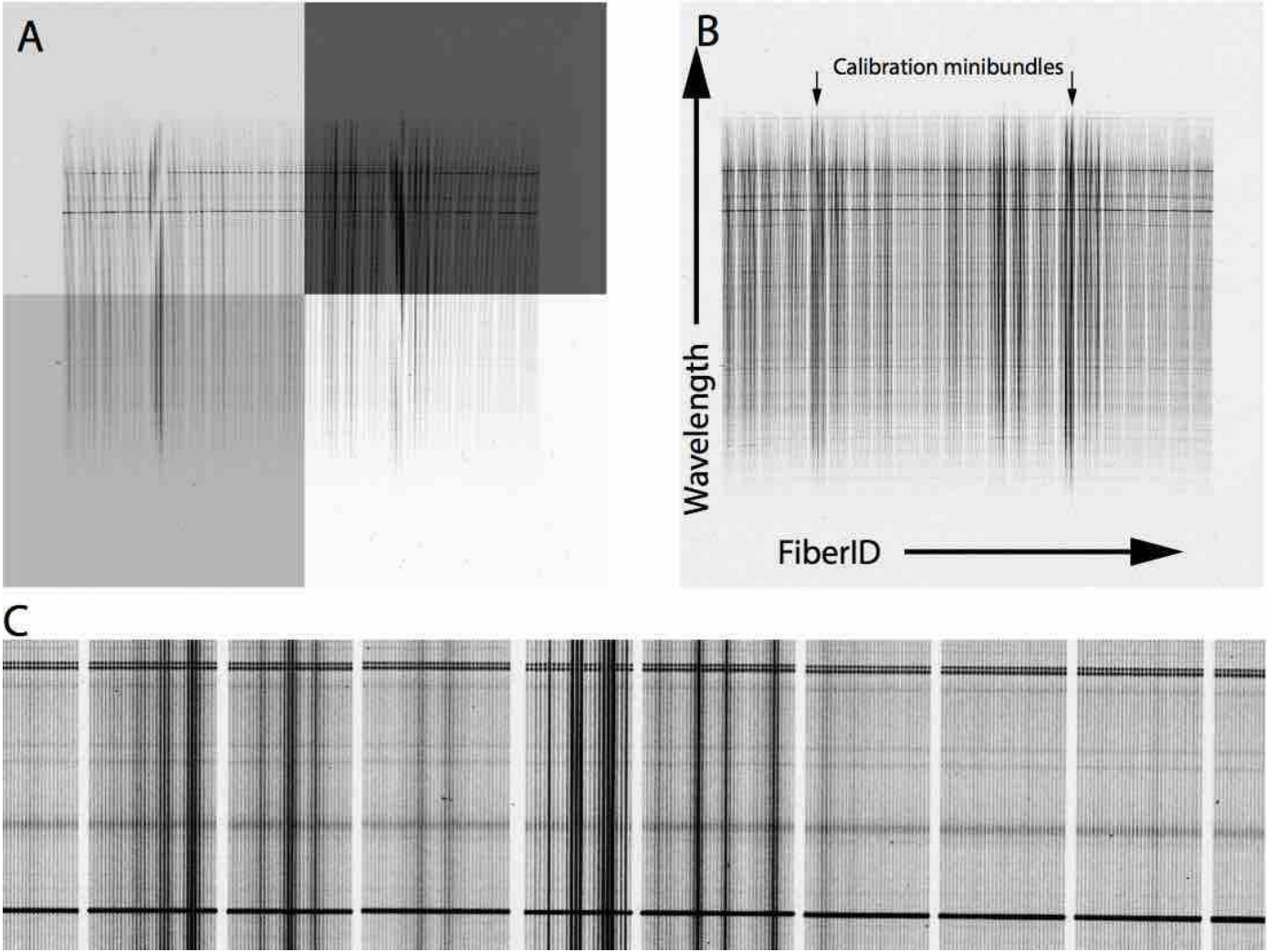}
\caption{Illustration of the MaNGA raw data format before (A) and after (B) preprocessing to remove the overscan and quadrant-dependent bias.  This image shows a color-inverted typical
15-minute science exposure for the b1 camera (exposure 177378 for plate 7443 on MJD 56741).  There are 709 individual fiber spectra on this detector, grouped into 22
blocks.  Bright spectra represent central regions of the target galaxies and/or spectrophotometric calibration stars; bright horizontal features are night-sky emission lines.
Panel C zooms in on 10 blocks in the wavelength regime of the bright [O I 5577] skyline. }
\label{rawdata.fig}
\end{figure*}

As illustrated in Figure \ref{rawdata.fig}, MaNGA fibers are arranged into blocks of 21-39 fibers with 22 blocks on each spectrograph, with individual spectra running vertically along each CCD.
The fiber spacing within blocks is 177 $\micron$ for science IFUs ($\sim$ 4 pixels), and 204 $\micron$ for spectrophotometric calibration IFUs,
with $\sim 624$ $\micron$ between each block.
Fibers are initially identified in a uniformly illuminated flat field image
using a cross-correlation technique to match the 1d profile along the middle row of the detector against a
reference file describing the nominal location of each fiber in relative pixel units.
The cross correlation technique matching against all fibers on a given slit allows for shifts due to flexure-based optical distortions while ensuring
robustness against missing or broken individual fibers and/or entire IFUs.  Fibers that are missing within the central row are flagged
as dead in \mangacore\ .

With the initial x-positions of each fiber in the central row thus determined, the centroids of each fiber in the other rows are then
determined using a flux-weighted mean with a radius of 2 pixels.  This algorithm sequentially steps up and down
the detector from the central row, using the previous row's position as the initial input to the flux-weighted mean.  Fibers with problematic centroids 
(e.g. due to cosmic rays) are masked out, and replaced with estimates based on neighboring traces.
These flux-weighted centroids are further refined using a per-fiber cross-correlation technique matching a gaussian model fiber profile (see \S \ref{extraction.sec})
against the measured profile in a given row.  This fine adjustment 
is required in order  to remove sinusoidal variations in the flux-weighted centroids at the $\sim$ 0.1 pixel level caused by 
discrete jumps in the pixels included in the 
previous flux-weighted centroiding.

Once the positions of all fibers across all rows of the detector have been computed, the discrete pixel locations are stored as a traceset\footnote{A traceset is
a set of coefficient vectors defining functions over a common independent-variable domain specified by `xmin' and `xmax' values. The functions in the set are defined in terms of a linear combination of basis functions (such as Legendre or Chebyshev polynomials) up to a specified maximum order, weighted by the values in the coefficient vectors, and evaluated using a suitable affine rescaling of the dependent-variable domain (such as [xmin, xmax] $\rightarrow$ [-1, 1] for Legendre polynomials). For evaluation purposes, the domain is assumed by default to be a zero-based integer baseline from xmin to xmax such as would correspond to a digital detector pixel grid.}
of 7th order Legendre polynomial coefficients.  An iterative rejection method
accounts for scatter and uncertainty in the centroid measurement of individual rows
and ensures realistically smooth variation of a given fiber trace as a function of wavelength along the detector.
The best-fit  traceset coefficients are stored as an extension in the per-camera mgFlat files (Table \ref{mgflat.tab}).

\subsubsection{Spectral Extraction}
\label{extraction.sec}

\begin{figure*}
\epsscale{1.1}
\plotone{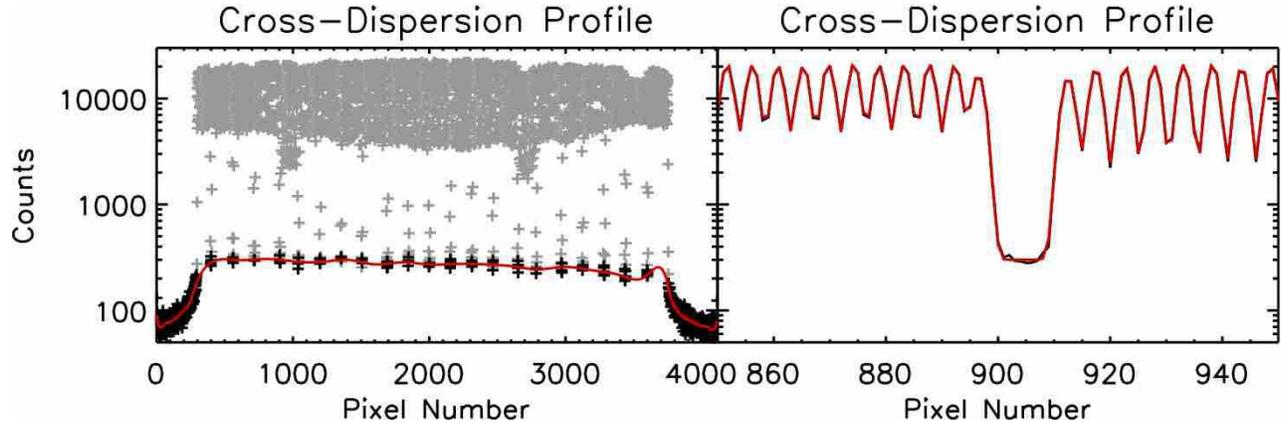}
\caption{Left panel: Cross-dispersion flatfield profile cut for the R1 camera.  Grey points lie
within 5 pixels of the measured fiber traces, black points are more than 5 pixels from the nearest fiber trace.  The solid
red line indicates the bspline fit to the inter-block values.
Right panel: Cross-dispersion profile zoomed in around CCD column 900.  The solid black line shows the individual pixel values, the solid red line
overplots the Gaussian profile fiber fit plus the bspline background term convolved with the pixel boundaries.
The trough around pixel 900-910 represents a gap between V-groove blocks.
Both panels show row 2000 from plate 8069 observed on MJD 57278.}
\label{fiberprofile.fig}
\end{figure*}

\begin{figure}
\epsscale{1.2}
\plotone{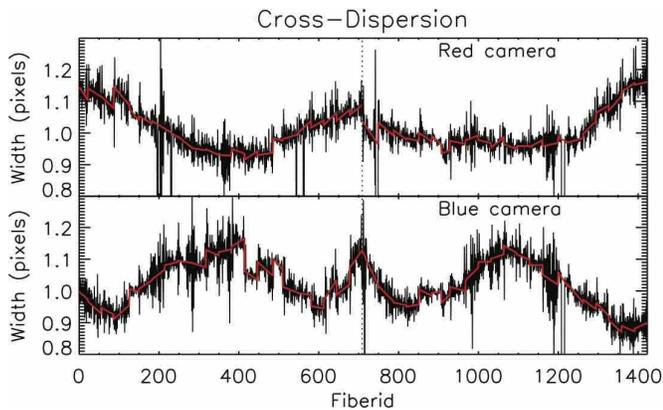}
\caption{Example spatial  width ($1\sigma$) for the cross-dispersion gaussian fiber profile as a function of fiberid for the middle row
of all 4 cameras.  This example is for plate 8618, observed on MJD 57199.
The solid black line represents individual measurements for each fiber in this row; the solid red line represents the
adopted fit that assumes smooth variation of the widths with wavelength and as a function of fiberid within each block. 
The vertical dotted
line represents the transition between the first and second spectrographs (fiberid 1-709 and 710-1423).
Similar plots are produced automatically by the DRP for each flatfield processed, and are used for quality control.}
\label{pwidths.fig}
\end{figure}

Similarly to the BOSS survey \citep{dawson13}, 
we extract individual fiber spectra from the 2d detector images using a row-by-row optimal extraction algorithm
that uses a least-squares profile fit to obtain an unbiased estimate of the total counts \citep{horne86}.
The counts in each row are modeled by a linear combination of $N_{\rm fiber}$\footnote{$N_{\rm fiber}$ is the number of fibers on a given detector ($N_{\rm fiber} = 709$ for spectrograph 1,
$N_{\rm fiber} = 714$ for spectrograph 2).}
 Gaussian profiles plus a low-order polynomial (or cubic basis-spline; see \S \ref{scattered.sec})
background term.  As we illustrate in Fig. \ref{fiberprofile.fig} (right panel), the resulting model is an extremely good fit to the observed profile.
MaNGA uses the extract\_row.c code (dating back to the original SDSS spectroscopic survey) which creates a pixelwise model of the gaussian profile integrated 
over fractional pixel positions (i.e. the profile is assumed to be gaussian prior to pixel convolution), describes deviations to the line centers and widths as linear 
basis modes (representing the first and second spatial derivatives, respectively), and solves for the banded matrix inversion
by Cholesky decomposition.
An initial fit to the flatfield calibration images allows both the amplitude and the width of the Gaussian profiles
in each row to vary freely, with the centroid set to the positions determined via
fiber tracing in \S \ref{tracing.sec}.  
These individual width measurements are noisy however, and for each block of fibers we therefore
fit the derived widths with a linear relation as a function of fiberid along the slit in order to reject errant values
and determine a fixed set of fiber widths that vary smoothly (within a given block) with both fiberid and wavelength.
As illustrated in Figure \ref{pwidths.fig}, low-frequency variation of the widths with fiberid reflects the telescope focus
(which we choose to ensure that the widths are as constant as possible across the entire slit), while discontinuities at the block boundaries are due to
slight differences in the slithead mounting.
These fixed widths are then used in a second fit to the detector images
in which only the polynomial background and the amplitude of the Gaussian terms are allowed to vary.

The final value adopted for the total flux in each row is the integral of the theoretical gaussian 
profile fits to the observed pixel values, while the inverse variance is taken to be the diagonal of the covariance matrix
from the Cholesky decomposition.
This approach allows us to be robust against cosmic rays
or other detector artifacts that cover some fraction of the spectrum, since unmasked pixels in the cross-dispersion profile can still be used to model the Gaussian profile (Figure \ref{fiberprofile.fig}).  
Additionally, this technique naturally
allows us to model and subtract crosstalk arising from the wings of a given profile overlapping any adjacent fibers, and to estimate the variance on the extracted spectra at each wavelength.
This step transforms our 4096 $\times$ 4112 CCD images  (4114 $\times$ 4128 for the red cameras) to row-stacked spectra with 
dimensionality 4112 $\times N_{\rm fiber}$ (4128 $\times N_{\rm fiber}$ for the
red cameras) 

\subsubsection{Scattered Light}
\label{scattered.sec}

The DRP automatically assesses the level of scattered light in the MaNGA data by taking 
advantage of the hardware design in which gaps of $\sim 16$ pixels
were left between each v-groove block (compared to $\sim 4$ pixels peak-to-peak 
between each fiber trace within a block) so that the interstitial regions contain negligible light from the gaussian fiber profile cores
\citep{drory15}.
By masking out everything within 5 pixels of the fiber traces we can identify those pixels on the edge of the detector and in empty regions between individual blocks whose counts are dominated
by diffuse light on the detector.  This light is a combination of 1) genuine scattered light that enters the detector via multiple reflections 
from unbaffled surfaces
and 2) highly-extended non-gaussian wings to the individual fiber profiles that can extend to hundreds of pixels and contain
$\sim 1-2$\% of the total light of a given fiber.

For MaNGA dark-time science exposures (which typically peak at about 30 counts pixel$^{-1}$ fiber$^{-1}$ for the sky continuum) both components are 
small and can be satisfactorily modeled by a low-order polynomial term in each extracted row.  For some bright-time exposures
used in the stellar library program however, the moon illumination can approach 100\% and produce larger scattered light counts $\sim$ a quarter
of the sky background seen by individual fibers.  Additionally, for our flat-field calibration exposures the summed contribution of the 
non-gaussian wings to the fiber profiles can reach $\sim 300$ counts pixel$^{-1}$ in the interstial regions between blocks
(compared to $\sim 20,000$ counts pixel$^{-1}$ in the fiber profile cores).  In both cases the simple polynomial background term can prove
unsatisfactory, and we instead fit the counts in the interstitial regions row-by-row with a fourth-order basis spline model
that allows for a greater degree of spatial variability in the background than is warranted for the dark-time science
exposures.
This bspline model is evaluated at the locations
of each intermediate pixel and smoothed along the detector columns by use of a 10-pixel moving boxcar to mitigate the impact of individual bad pixels.
The resulting bspline scattered light model is subtracted from the raw
counts before performing spectral extraction.


\subsubsection{Fiber Flatfield}
\label{flats.sec}

\begin{figure}
\plotone{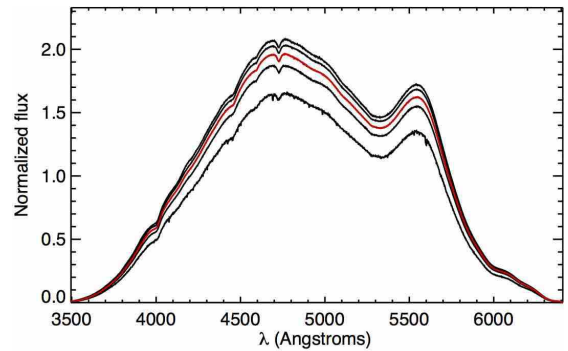}
\caption{Example of a typical superflat spectrum for the b1 camera normalized to a median of unity.  
Solid red line shows the superflat fit to the median fiber, solid black lines indicate the $1\sigma$ and $2\sigma$ deviations
about this median.}
\label{superflat.fig}
\end{figure}

Each flatfield calibration frame is extracted into individual fiber spectra using the above techniques and matched to the nearest (in time) arc-lamp calibration frame which has been processed
as described in \S \ref{wavecal.sec}.  Using the wavelength solutions derived from the arc frames, we combine the individual flatfield spectra (first normalized to a median of unity) into a single composite spectrum
with substantially greater spectral sampling than any individual fiber.\footnote{Since each fiber has a slightly different wavelength solution we effectively supersample the intrinsic input spectrum.}
We fit this composite spectrum with a cubic basis spline function to obtain the superflat vector 
describing the global flatfield response (i.e., the quartz lamp spectrum convolved with the detector response
and system throughput).  This global superflat is shown in
Figure \ref{superflat.fig}, and illustrates the falloff in system throughput toward the wavelength extremes of the detector \citep[see also Fig. 4 of][]{yan16}.

We evaluate the superflat spline function 
on the native wavelength grid of each individual fiber and divide it out from the individual fiber spectra 
in order to obtain the relative fiber-to-fiber flatfield spectra.  
So normalized, these fiber-to-fiber flatfield spectra have values near unity, vary only slowly (if at all) with wavelength,
and easily show any overall throughput differences between the individual fibers.
Each such spectrum
is in turn fitted with a bspline in order to minimize the contribution of photon noise to the resulting fiber flats and interpolate across bad pixels.
In the end, we are left with two flatfields to store in the mgFlat files (see Table \ref{mgflat.tab}); a 
single {\it superflat} spectrum describing the global average response as a function of wavelength, and a {\it fiberflat} of size 
4112 $\times N_{\rm fiber}$ (4128 $\times N_{\rm fiber}$ for the
red cameras) describing the relative
throughput of each individual fiber as a function of wavelength.

The individual MaNGA fibers typically have high throughput \citep[see discussion by][]{drory15} within 5-10\% of each other.  The relative distribution of throughputs is monitored daily
to trigger cleaning of the IFU surfaces when the DRP detects noticeable degradation in uniformity or overall throughput.  Individual fibers with throughput less than 50\% that of the best fiber
on a slit are flagged by the pipeline and ignored in the data analysis.  This may occur when a fiber and/or IFU falls out of the plate (a rare occurrence), or when a fiber breaks.  Such breakages
in the IFU bundles occur at the rate of about 1 fiber per month across the entire MaNGA complement of 8539 fibers.

\subsubsection{Wavelength and spectral resolution calibration}
\label{wavecal.sec}

The Neon-Mercury-Cadmium arc-lamp spectra are extracted in the same manner as the flat-fields, except that they use the fiber traces determined from the corresponding flatfield (with allowance for a continuous 2d polynomial
shift in the traces as a function of detector position to account for flexure differences) and allow only the gaussian profile amplitudes to vary.
These spectra are normalized by the fiber flatfield\footnote{In practice this is iterative; the flatfields prior to separation of the superflat and fiberflat are used to 
normalize the arc-lamp spectra, the wavelength
solution from which in turn allows construction of the superflat.}
and an initial wavelength solution is computed as follows.

A representative spectrum is constructed from the median of the 5 closest spectra closest to the central fiber on the CCD.  This spectrum is cross-correlated with a model spectrum 
generated using a reference table of known strong emission features in the Neon-Mercury-Cadmium arc lamps,\footnote{There are $\sim 50$ such features with counts in the range
10$^3$ -- 10$^5$ pixel$^{-1}$ in each of the blue and red cameras; see full list at https://svn.sdss.org/public/repo/manga/mangadrp/tags/v1\_5\_4/etc/lamphgcdne.dat}
 and iterated to determine the best-fit coefficients to map pixel locations to wavelengths.
These best-fit coefficients are used to contruct initial guesses for the wavelength solution of each fiber, which are then iterated on a fiber-to-fiber basis to obtain the final wavelength solutions.
Several rejection algorithms are run to ensure reliable arc-line centroids across all fibers.
A final 6th order Legendre polynomial fit converts the wavelength solutions into a 
series of polynomial traceset coefficients.  The higher order coefficients are 
forced to vary smoothly as a function of fiberid since they predominantly arise from optical distortions along the slit (whereas lower order terms
represent differences arising from the fiber alignment).
These coefficients are stored as an extension in the output mgArc file  (see Table \ref{mgarc.tab}), 
and used to reconstruct the wavelength solutions at all  fibers and positions on the CCD.      

\begin{figure}
\epsscale{1.2}
\plotone{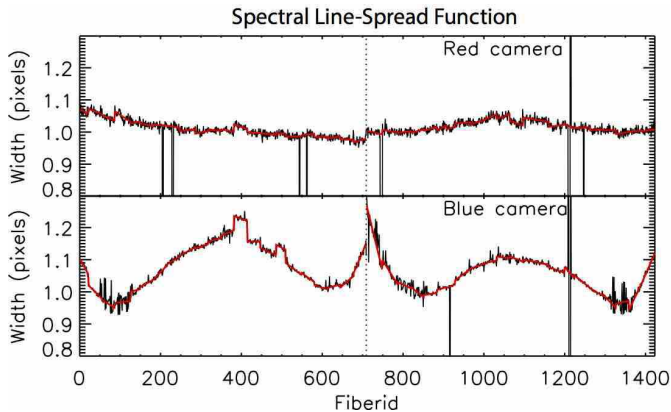}
\caption{As Figure \ref{pwidths.fig}, but showing the
spectral line spread function ($1\sigma$ LSF) for the gaussian arcline profile as a function of fiberid for an emission line near the middle row of all 4 detectors
(Cd I $5085.822$ \AA\ for the blue cameras, Ne I $8591.2583$ \AA\ for the red cameras).}
\label{lwidths.fig}
\end{figure}

The arc-lamp spectral resolution (hereafter the line spread function, or LSF) 
is computed by fitting the extracted spectra around the strong 
arc lamp emission lines in each fiber with a Gaussian profile integrated over each pixel 
(note that we {\it integrate} the fitted profile shape across each pixel rather than simply {\it evaluating} the profile
at the pixel midpoints; see discussion in \S \ref{specres.sec})
 and allowing both the width and amplitude
of the profile to vary.  
As illustrated in Figure \ref{lwidths.fig}, these widths are intrinsically noisy and the DRP therefore fits them with a linear relation as a function of fiberid along the slit in order to reject errant values
and determine a fixed set of line widths that vary smoothly (within a given block) with fiberid.
These arcline widths are then fit with a Legendre polynomial traceset that is stored in the mgArc files and evaluated at each pixel to compute
the LSF at wavelengths between the bright arc lines.

Both wavelength and LSF solutions derived from the arc frames are later adjusted for each individual science frame to account for instrumental flexure during and between
(see discussion in \S \ref{sciextract.sec}).


All calibrations are additionally complicated in the red cameras since the middle row of pixels on these detectors is 
oversized by a factor of 1/3, causing a discontinuity in both the wavelength solution and the LSF for each fiber as a function
of pixel number.  All of the algorithms described above therefore allow for such a discontinuity across the CCD quadrant boundary.
The primary impact of this discontinuity on the final data products is to produce a spike of low spectral resolution around $8100$ \AA, the exact
wavelength of which can vary from fiber to fiber based on the curvature of the wavelength solution along the detector.

\subsection{Science Frames}
\label{sciextract.sec}

Each science frame is associated with the arc and flat pair taken closest to it in time (generally within one hour since calibration frames are taken at the start of each plate
and periodically thereafter), and extracted row-by-row following the method outlined in \S \ref{extraction.sec}.  During this extraction only the profile amplitudes and background polynomial
term are allowed to vary freely; the trace centroids are tied to the flatfield traces with a global 2d polynomial shift to account
for instrument flexure, and the cross-dispersion widths are fixed to the values derived from the flatfield.
The extracted spectra are normalized by the superflat and fiber flat vectors derived from the flatfield.

The wavelength solutions derived from the arcs are adjusted for each science frame 
to match the known wavelengths of bright night-sky emission lines in the science spectra by fitting a low-order polynomial shift
as a function of detector position to allow for instrumental flexure (these shifts are typically less than a quarter pixel).
The final wavelength solution for each exposure is corrected to the vacuum heliocentric restframe using header keywords
recording atmospheric conditions and the time and date of a given pointing.  As we explore in \S \ref{waveacc.sec}, we achieve a $\sim 10$ \kms\
or better rms wavelength calibration accuracy with zero systematic offset to within 2 \kms.

Similarly, in order to account for flexure and varying spectrograph focus with time
the spectral LSF measurements derived from the arc-lamp exposures are also adjusted 
for each science frame to match the LSF of bright skylines that are known to be unblended in high-resolution
spectra \citep[e.g.,][]{osterbrock96}.
Starting from the original arcline
LSF model, we derive a quadrature correction term for the profile widths
$Q^2 = {\sigma_{\rm sky}}^2 - {\sigma_{\rm arc}}^2$.  $Q$ is taken to be constant as a function of wavelength for each camera, and is based
on the strong auroral O I 5577 line in the blue (since the Hg I lines are too weak
and broadened to obtain a reliable fit) and an average of many isolated bright lines in the red.\footnote{See
https://svn.sdss.org/public/repo/manga/mangadrp/tags/v1\_5\_4/etc/skylines.dat for a complete list.}
  The measured quadrature correction term
is fitted with a cubic basis spline to ensure that the correction applied varies smoothly with fiberid.  Across the $\sim$ 1100 individual exposures
in DR13 the average correction $Q^2 = 0.08 \pm 0.04$ pixel$^2$ in the blue cameras
and $Q^2 = 0.05 \pm 0.02$ pixel$^2$ in the red cameras (likely due to the flatter and more stable focus in the red cameras).

The final row-stacked spectra, inverse variances, pixel masks, wavelength solutions, and broadened LSF are all stored as extensions in the output mgFrame FITS file
(Table \ref{mgframe.tab}).


\section{Sky Subtraction}
\label{skysub.sec}

\begin{figure*}
\plotone{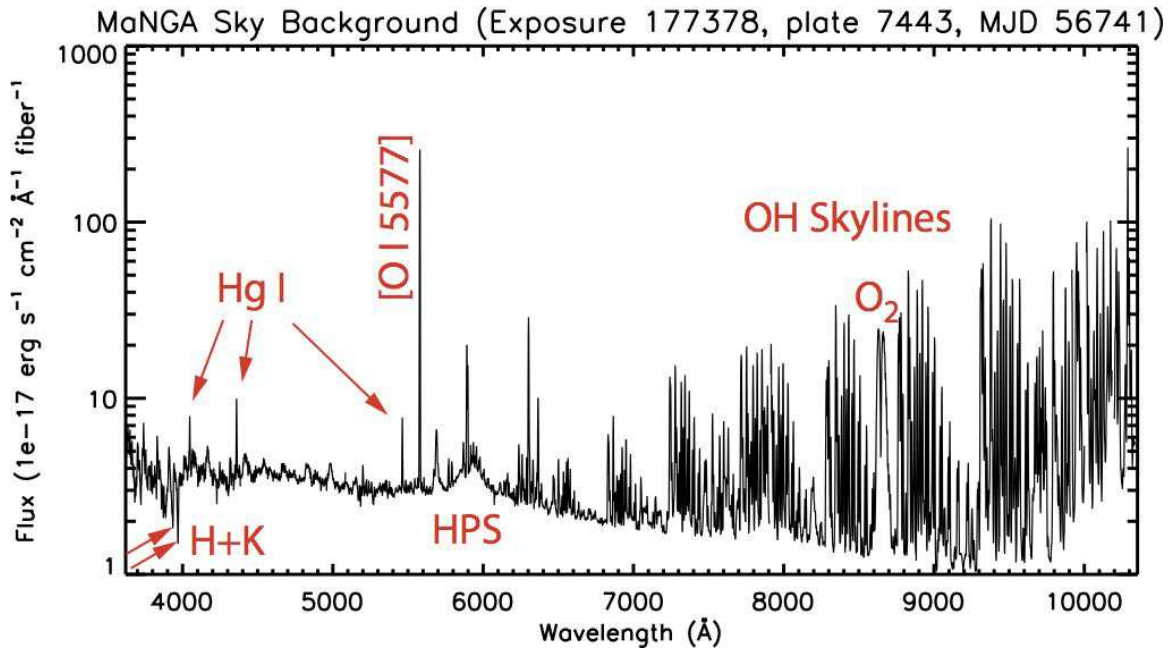}
\caption{Typical flux-calibrated MaNGA night-sky background spectrum seen by a single optical fiber (2 arcsec core diameter).  Bright features longward of 7000 \AA\ represent blended OH and $O_2$ skyline emission \citep[see, e.g.,][]{osterbrock96}.  The bright feature at 5577 \AA\ is atmospheric [O I], the broad feature around 6000 \AA\ is high-pressure sodium (HPS) from streetlamps; Hg I from mercury vapor lamps contributes most of the discrete features at short wavelengths \citep[see, e.g.,][]{massey00}.  Absorption features around 4000 \AA\ are zodiacal Fraunhofer H and K lines.}
\label{IRskyBG.fig}
\end{figure*}

Unlike previous SDSS spectroscopic surveys targeting bright central regions of galaxies, MaNGA will explore out to $\geq 2.5$ effective radii ($R_{\rm e}$) where galaxy
flux is decreasing rapidly relative to the sky background.  
As illustrated in Figure \ref{IRskyBG.fig}, 
this night sky background is especially bright at near-IR  wavelengths longwards of $\sim 8000$ \AA,
where bright emission lines from OH radicals \citep[e.g.,][]{rousselot00}
dominate the background flux.  These OH features vary in strength with both time and angular position depending on the coherence scale of the atmosphere, 
posing challenges for measuring 
faint stellar atmospheric features such as the Wing-Ford  \citep{wingford69} band of iron hydride absorption lines around $9900$ \AA.
In many cases such faint features will be detectable only in stacked bins of spectra, driving the need to reach the Poisson-limited noise regime so that stacked spectra are not limited by systematic sky subtraction residuals.

We therefore design our approach to sky subtraction with the aim of reaching Poisson-limited performance at all wavelengths from $\lambda\lambda 4000-10,000$ \AA\ 
(beyond which the increasing read noise of the BOSS cameras prohibits such performance).  
Our sky subtraction algorithm is closely based on the routines developed for the 
BOSS survey, and relies on using the dedicated 92 sky fibers (46 per spectrograph) 
on each plate to construct a highly-sampled model background sky that can be subtracted from each of the science fibers.
These sky fibers are plugged into regions identified during the plate design process
as blank sky `objects' within a 14 arcmin patrol radius of their associated IFU fiber bundle (see Fig. \ref{ifudiagram.fig}).

\subsection{Sky Subtraction Procedure}

Sky subtraction is performed independently for each of the 4 cameras using the flatfielded, wavelength-calibrated fiber spectra contained in the mgFrame files,
and is a multi-step iterative process.  Broadly speaking, we build a super-sampled sky model from all of the sky fibers,
scale it to the sky background level of a given block, and evaluate it on the native solution of each fiber within that block.
In detail:

\begin{enumerate}

\item The metadata associated with the exposure is used to identify the $N_{\rm sky}$ individual sky fibers in each frame based on their FIBERTYPE.

\item Pixel values for these $N_{\rm sky}$ sky fibers are resorted as a function of wavelength into a single one-dimensional array of length $N_{\rm sky} \times N_{\rm spec}$
(where $N_{\rm spec}$ is the length of a single spectrum).
Since each fiber has a unique wavelength solution, this super-sky vector 
has much higher effective sampling of the 
night sky background spectrum than any individual fiber 
and provides an accurate line spread function (LSF) for OH airglow features.
 An example of this procedure is shown in Figure \ref{skymodel.fig}.

\item Similarly, we also construct a super-sampled weight vector by comining individual sky fiber
{\it inverse variance spectra} that have first been smoothed by a boxcar
of width 100 pixels ($\sim 100-200$ \AA) in the continuum and 2 pixels ($\sim 2-3$ \AA) within 3 \AA\ of bright atmospheric emission features. 

\item The super-sky spectrum is then weighted by the smoothed inverse variance spectrum (convolved with the bad-pixel mask) and fitted with a cubic basis-spline 
as a function of wavelength, with the number of breakpoints set to $\sim
N_{\rm spec}$ so that high-frequency variations (due, e.g., to shot noise or bad pixels) are not picked up by the resulting model
(see, e.g., green line in Figure \ref{skymodel.fig}).\footnote{The number of breakpoints is reduced slightly in the blue cameras as there are few narrow spectral features that need to be fit.} 
The breakpoint spacing is set automatically to maintain
approximately constant S/N ratio between breakpoints.  The B-spline fit itself is iterative, with upper and lower rejection threshholds set to mask bad or deviant pixels.
{\it We note that the smoothing of the inverse variance in determining the weight function is
critical as otherwise the 
weights (which are themselves estimated from the data) would modulate with the Poisson scatter and bias the fit towards slightly lower values, resulting in systematic undersubtraction of the sky background, especially near the wavelength extrema
where the overall system throughput is low.}

\begin{figure}
\plotone{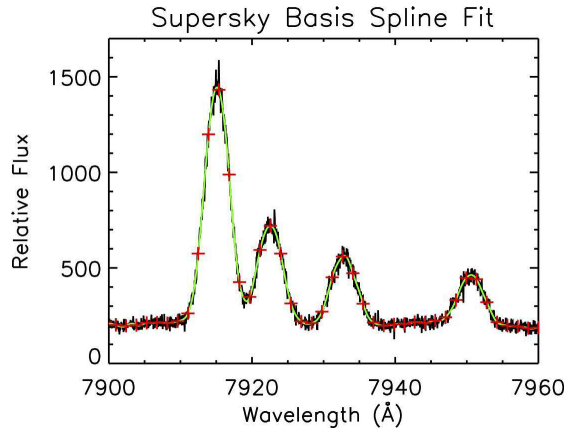}
\caption{Example MaNGA super-sky spectrum created by the wavelength-sorted combination of all sky fiber spectra (black line) in the OH-emission dominated wavelength region $\lambda\lambda 7900-7960$ \AA. 
Overlaid in green is the b-spline model fit to the super-sky spectrum; red points represent the b-spline model after evaluation on the native pixellized wavelength solution of a single fiber.}
\label{skymodel.fig}
\end{figure}

\item This B-spline function is evaluated on the native wavelength solution of each of the sky fibers.  Dividing the original sky fiber spectra by this functional model, and collapsing
over wavelengths using a simple mean we arrive at a series of scale factors describing the relative sky background seen by the fiber compared to all other fibers on the detector.  For each harness
(i.e., each IFU plus associated sky fibers) we compute the median of these scale factors to obtain a single averaged scale factor for each harness.  
 These scale factors 
help account for nearly-grey variations in the true sky continuum across our large field produced by a combination of intrinsic background variations and patchy cloudcover.
The variability
in sky background between harnesses is about 1.5\% rms, with some larger deviations $> 5$\% observed during the bright time stellar library program
when pointing near a full moon can produce strong background gradients.

\item Repeat steps 2-4 after first scaling each individual sky fiber spectrum by the value appropriate for its harness in order to obtain a super-sky spectrum in which per-harness scaling effects have been removed.

\item Evaluate the new B-spline function on the native pixellized wavelength solution of each fiber (sky plus science), and multiply it by the scaling factor for the harness to obtain the 
first-pass model sky spectrum for each fiber.  Subtract this from the spectra to obtain the first pass sky subtracted spectra. 

\item Identify deviant sky fibers in which the median sky-subtracted residual SNR$^2 > 2$ (this is extremely rare, and generally corresponds to a case where a sky fiber location was
chosen poorly, or a fiber was misplugged and not corrected before observing).  Eliminate these sky fibers from consideration, and repeat steps 2-7 to obtain
the second-pass model sky spectrum for each fiber.  We refer to this as the {\bf 1-d sky model}.

\item Repeat steps 2-4, this time allowing the bspline fit to accommodate a smoothly varying 3rd order polynomial of values at each breakpoint as a function of fiberid 
(i.e., rather than requiring
the model to be constant for all fibers, it is allowed to vary slowly as a function of slit position).  This polynomial term is introduced in order to model variations in the LSF 
along each slit; empirically, increasing polynomial orders up to 3 results
in an improvement of the skyline residuals, while no further gains are observed at greater than 3rd order.
Evaluate the new B-spline function on the native pixellized wavelength solution of each fiber (sky plus science) to obtain the {\bf 2-d sky model}.  
Notably, this 2-d model does not
use the explicit scaling used by the 1-d model.  This is partially because a similar degree of freedom is introduced by the 2d polynomial, and partially because OH features can vary in strength
independently from the underlying continuum background \citep[see, e.g.,][]{davies07}.

\item The {\bf final sky model} 
is a piecewise hybrid of the 1-d and 2-d models; in continuum regions
it is taken to be the 1-d model, and in the skyline regions (i.e., within 3 \AA\ of any wavelength for which the sky background is $> 5\sigma$
above a bspline fit to the interline continuum) it is taken to be the 2-d model.  
We opt for this hybrid model as it optimizes our various performance metrics:  In the continuum far from night sky lines, our performance is limited by the poisson-based RMS of the model
sky spectrum subtracted from each science fiber.  Therefore, we use the 1d model that is based on all 46 sky fibers on a given spectrograph.  In contrast, near bright skylines our performance is
instead limited by our ability to accurately model the shape of the skyline wings, which can vary along the slit (see, e.g., Fig \ref{lwidths.fig}).   Therefore, in skyline regions we use the 2d model
which improves the model LSF fidelity at the expense of some SNR.
There is no measurable discontinuity between the sky-subtracted spectra at the piecewise 1d/2d model boundaries.

\end{enumerate}

The final sky model is subtracted from the mgFrame spectra; these sky-subtracted spectra are stored in mgSFrame files (Table \ref{mgsframe.tab}), 
which contain the spectra, inverse variances (with appropriate error propagation), pixel masks, applied sky models, etc. in a row-stacked format identical to the input mgFrame files.

\subsection{Sky Subtraction Performance: All-Sky Plates}
\label{dq.skysub}


We estimate the accuracy of our calibration and sky subtraction up to this point by using specially designed ``all-sky'' plates
in which every science IFU is placed on a region of sky determined to be empty of visible sources according to the 
SDSS imaging data (calibration minibundles are still placed on standard stars so that these all-sky plates can be properly
flux calibrated).  The resulting sky-subtracted sky spectra can then be used to estimate the accuracy of our noise model,
extraction algorithms, and sky-subtraction technique.

\begin{figure*}
\epsscale{1.1}
\plotone{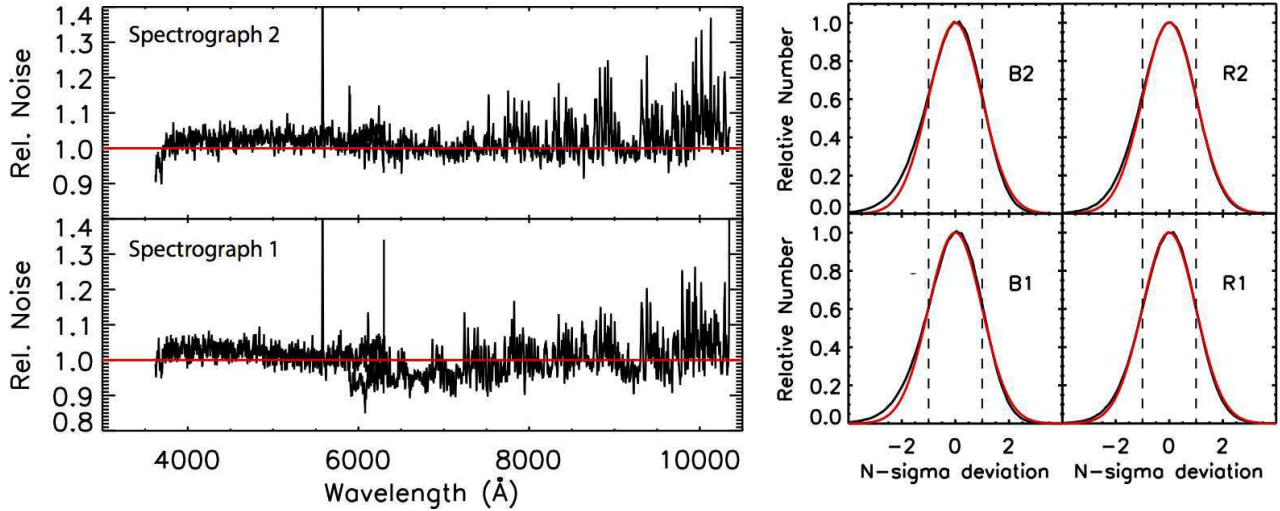}
\caption{Left-hand panel: Actual noise in sky-subtracted spectra (from all-sky plate 8069, observed on MJD 56901) divided by
that expected based on detector read noise and Poisson counting statistics 
as a function of wavelength for each spectrograph.  The overlapping region from $\lambda\lambda 5900-6300$ \AA\ is
the dichroic region over which blue and red cameras are combined.  The solid red line indicates unity; if sky subtraction was
done perfectly (and the noise properties of the spectra were estimated correctly) the black lines should nearly follow the red line
at all wavelengths.  Right-hand panel: Distribution of noise values divided by the expected for all 4 cameras (B1/B2/R1/R2).  Black curves represent the measured distribution of values (3621-6300 \AA\ for B1 and B2, 5900-10354 \AA\ for R1 and R2), red curves represent the Gaussian ideal distribution with width $N \sigma = 1$.  Vertical
dashed black lines represent the $1\sigma$ range.}
\label{skypanels.fig}
\end{figure*}

Working with the row-stacked mgSFrame spectra (i.e., prior to flux calibration and wavelength rectification) we construct `Poisson ratio' images for each camera
by multiplying the sky-subtracted residual counts by the square root of the inverse variance (which accounts for both shot noise and detector read noise).
If the sky-subtraction is perfect, and the noise model properly estimated, these poisson ratio images should be devoid of structure with a Gaussian distribution of values
with mean of 0 and $\sigma =$ 1.0.
In Figure \ref{skypanels.fig} (right-hand panels) we show the actual distribution of values for the sky-subtracted science fibers for exposure 183643 (cart 4, plate 8069, MJD 56901)
for each of the four cameras (solid black lines) compared to the ideal theoretical expectations (solid red line; note that this is {\it not} a fit to the data).  We find that the overall distribution
of values is broadly consistent with theoretical models in all four cameras
\citep[c.f. Fig 23 of][which shows similar plots for the DEEP2 survey]{newman13},
albeit with some evidence for slight oversubtraction on average and a non-gaussian wing in the blue cameras (pixels in this asymmetric wing do not correspond to particular wavelengths or fiberid).

We examine this behavior as a function of wavelength in Figure \ref{skypanels.fig} (left-hand panel) by plotting the $1\sigma$ width of the gaussian that best fits the distribution of unflagged pixel
values at a given wavelength across all science fibers.\footnote{Since each fiber has a different wavelength solution we can't simply use all pixels in a given column, and therefore instead
use the three pixels whose wavelengths are closest to a given wavelength in each fiber.}
As before, perfectly noise-limited sky subtraction with a perfect noise model would correspond to a flat distribution of $\sigma$ around 1.0 at all wavelengths; we note that the blue cameras and the continuum regions of the r2 camera are close to this level of performance with up to a 
3\% offset from nominal (suggesting that the read noise in some quadrants may be marginally underestimated).
In the r1 camera the read
noise may be overestimated by $\sim 10$\% in some quadrants
(as $\sigma < 1$ for r1 in the wavelength range $\lambda\lambda 5700-7600$ \AA), but is otherwise well-behaved in the continuum region.  In the skyline
regions of the red cameras, performance is within 10\% of Poisson expectations
out to $\sim 8500$ \AA.  Longward of
$\sim 8500$ \AA\ (where skylines are brighter, and the spectra have greater curvature on the detectors) sky subtraction performance in skyline regions is $\sim 10-20$ \% above theoretical
expectations.  This is likely due to systematic residuals in the subtraction caused by block-to-block variations in the spectral LSF that are difficult
to model completely.  Indeed, such
an analysis during commissioning revealed the OH skyline residuals were significantly worse in R1 than in the R2 camera.  
This led to the discovery of optical coma in R1 that was fixed
during Summer 2014 prior to the formal start of SDSS-IV, but which nonetheless affected the commissioning plates 7443 and 7495.

\begin{figure}
\epsscale{1.2}
\plotone{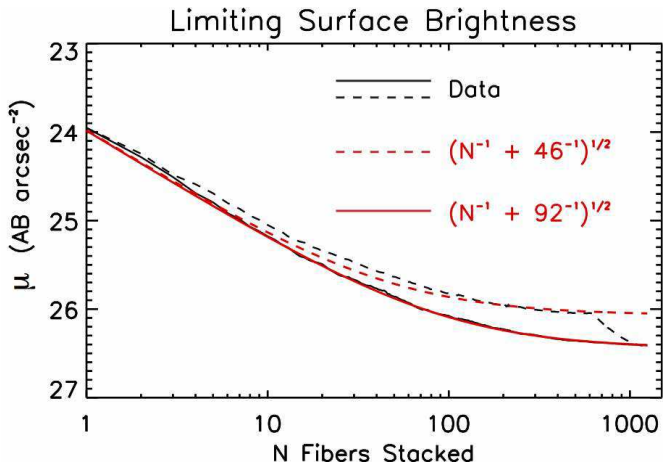}
\caption{$1\sigma$ limiting surface brightness reached in the wavelength range $\lambda\lambda 4000-5500$ \AA\ in a single 15-minute exposure
by a composite spectrum stacking $N$ sky-subtracted science fibers (based on all-sky plate 8069,
observed on MJD 56901).  The solid black line indicates results from stacking $N$ science fibers selected randomly from across both spectrographs; this is extremely
well reproduced by the theoretical curve (solid red line) representing expected performance based on $\sqrt{N^{-1} + 92^{-1}}$.
The dashed black line indicates results from stacking $N$ science fibers as a function of fiberid along the spectrograph slit; this improves more slowly at first
as $\sqrt{N^{-1} + 46^{-1}}$ (red dashed line).}
\label{skysub.fig}
\end{figure}

Overall, the results in Figure \ref{skypanels.fig} indicate excellent performance from the MaNGA DR13 data pipeline sky subtraction, 
albeit with some room for further improvement in
future data releases.  Finally, we assess whether any systematics exist within the data that would prohibit stacking of multiple fiber 
spectra in order to reach faint surface brightness levels (e.g.,
in the outer regions of the target galaxies).  Using the flux-calibrated, camera-combined mgCFrame data 
(again corresponding to exposure 183643 from MJD 56901) we compute the
limiting $1\sigma$ surface brightness reached in the largely skyline-free wavelength range $4000 - 5500$ \AA\ 
as a function of the number of individual fiber spectra stacked.
As shown in Figure \ref{skysub.fig},
when $N$ fibers are stacked randomly from across both spectrographs (solid black line) the limiting surface brightness 
decreases as $\sqrt{N^{-1} + 92^{-1}}$
(i.e., improving
as $\sqrt{N}$ for small $N$, and becoming limited by the statistics of the 92-fiber sky model as $N$ becomes large).
If fibers are stacked sequentially along the slit (dashed black line) the
limiting surface brightness decreases as  $\sqrt{N^{-1} + 46^{-1}}$ at first (since only the 46 sky fibers on a single slit are being used in the sky model)
but approaches nominal performance again once fibers from both spectrographs are included in the stack ($N > 621$).

\subsection{Sky Subtraction Performance: Skycorr}
\label{dq.skysub2}

Another way to check the sky subtraction quality of the DRP is to compare its performance for a typical
galaxy plate against the results obtained using the skycorr tool \citep{noll14}.
Skycorr was
designed as a data reduction tool to remove sky emission lines for
astronomical spectra using physically motivated scaling relations, and has been found to consistently perform better than
the popular algorithm of \citet{davies07}.
As input, skycorr needs the science spectrum and a
sky spectrum, preferably taken around the time as the science spectrum. 
After subtracting the continuum from both spectra, it then scales the sky
emission lines from the sky spectrum to fit these lines in the science
spectrum by comparing groups of sky lines which should vary in similar
ways.  

\begin{figure}
\epsscale{1.2}
\plotone{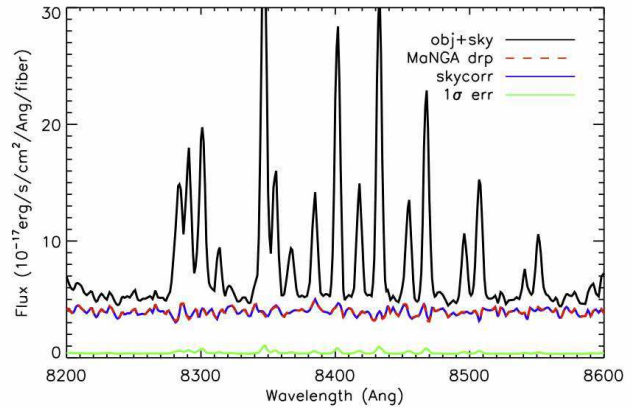}
\caption{Example spectrum of a randomly chosen MaNGA fiber from a typical galaxy plate (plate 7443, MJD 56745, exposure 177685) in the vicinity
of an OH airglow region.  The black line
represents the as-observed galaxy $+$ sky spectrum, the dashed red line the science spectrum after
sky subtraction using the MaNGA DRP, and the blue line the science spectrum using instead the skycorr \citep{noll14} algorithm.
The two methods of sky subtraction produce nearly identical results to within the DRP-estimate uncertainty (green line).}
\label{skycorr.fig}
\end{figure}

In Figure \ref{skycorr.fig} we compare a typical sky-subtracted MaNGA science spectrum obtained using the DRP algorithms described in \S \ref{skysub.sec} with the spectrum
obtained using skycorr instead.  The two sky-subtracted spectra are nearly indistinguishable, indicating comparable performance between the two techniques.

\section{Flux calibration}
\label{fluxcal.sec}

Flux calibration for MaNGA \citep{yan16} has a different goal than in previous generations of SDSS spectroscopic fiber surveys.
The goal for single fiber flux calibration is often to retrieve the total flux of a point-like source, accounting for both flux lost due to
atmospheric attenuation (or instrumental response) and the flux lost due to the fraction of the point-spread function that falls outside the fiber aperture.
In contrast, IFU observations provide a  {\it sampling} of the seeing-convolved flux profile for which we do not desire to make any aperture corrections and must
therefore seperate the aperture loss factor from the system response loss factor. 

To achieve this goal, we allocate a set of twelve 7-fiber mini-ifu bundles to standard stars on every plate (6 per spectrograph). 
Using the guider system to provide
a first-order estimate for the seeing profile in a given exposure we construct a model PSF 
as seen by each IFU minibundle  by including the effects of wavelength-dependent seeing and the shape mismatch between the focal plane and the plate. 
This allows us to estimate the relative fluxes amongst the seven IFU fibers in several wavelength windows and fit for the
spatial location of the star within the IFU, the scale of the PSF, and the scale and rotation of the expected differential atmosphere refraction (see \S \ref{basicast.sec}). 
With the best fit PSF model, 
we can compute the aperture loss factor of the fibers and estimate the total flux that would have been observed for each standard star if the IFU had captured 100\%
of its light.

Given this aperture correction, we can then derive the system response as a function of wavelength in a similar way as BOSS \citep{dawson13} by 
selecting the best fitting template from a grid of theoretical spectra normalized to the observed SDSS broadband magnitudes.
The correction vectors derived from the individual standard stars in a given exposure are then averaged to obtain the best system throughput correction to apply to all
of the science fibers.  This process is described in  detail by \citet{yan16}.

The flux calibration vectors are derived on a per-exposure, per-camera basis, and hence result in four FITS files in which the sky-subtracted row-stacked spectra
have been divided by the appropriate flux calibration vector.  These mgFFrame files (Table \ref{mgfframe.tab}).
are identical in format to the mgFrame and mgSFrame files, but have radiometric units of $10^{-17}$ erg s$^{-1}$ cm$^{-2}$ \AA$^{-1}$ fiber$^{-1}$ (see Appendix \ref{datamodel.sec}).
The accuracy of the MaNGA flux calibration has been described in detail by \citet{yan16}.  In brief, we
 find that MaNGA's relative calibration is accurate to 1.7\%  between the wavelengths of \Hb\ and \Ha\ 
and  4.7\% between
 \otwo\ $\lambda3727$ to \ntwo\ $\lambda6584$, and that the absolute RMS calibration
 (based on independent measurements of the calibration vector)
  is better than 5\% for more than 89\% of MaNGA's wavelength range. 
  \citet{yan16} assessed 
the systematic error by comparing the derived MaNGA photometry against PSF-matched
  SDSS broadband imaging, and found
a median flux scaling factor of 0.98 in $g$-band with a sigma of 0.04 between
individual galaxies.  Since publication of the \citet{yan16} study, additional improvements to the DR13 DRP that better model flux
in the outer wings of the SDSS 2.5m telescope PSF have improved the median flux scaling factor in $g$-band to 1.01
 \citep[see discussion by][]{yan16b}.


\section{Wavelength rectification}
\label{waverect.sec}

\begin{figure}
\plotone{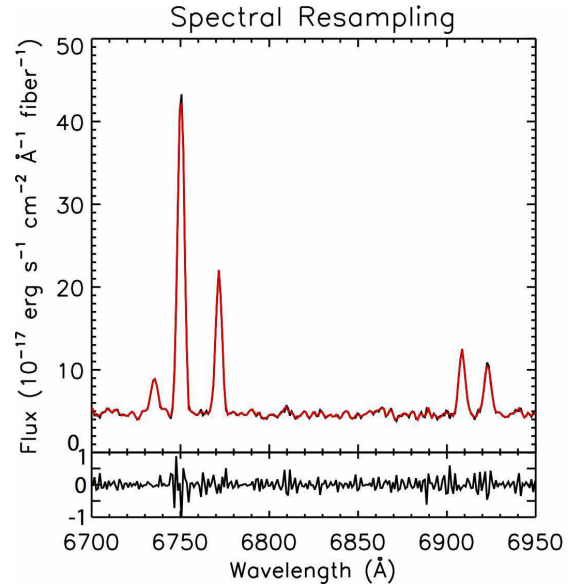}
\caption{Example MaNGA spectrum in the vicinity of \Ha, \ntwo, and \stwo\ emission on the native CCD pixel scale (solid black line) overlaid with the cubic bspline fit evaluated
on a constant logarithmic wavelength grid (solid red line).  The lower panel shows the difference between the native
spectrum and the wavelength rectified spline fit.}
\label{resampling.fig}
\end{figure}

The final step in the 2d section of the \mangadrp\ pipeline is to combine the four flux calibrated frames into a single frame that incorporates all 1423 fibers from both spectrographs
and combines together individual fiber spectra across the dichroic break at $\sim 6000$ \AA\ onto a common fixed wavelength grid.\footnote{The native CCD wavelength grid
varies from fiber to fiber, but is $\sim 1.0$ \AA\ pixel$^{-1}$ in the blue camera and $\sim 1.4$ \AA\ pixel$^{-1}$ in the red camera.}
Although this introduces slight covariance into the spectra (and degrades the effective spectral resolution by $\sim 6$\%; see \S \ref{specres.sec}),
it is required in order to ultimately coadd the individual spectra (each of which has its own unique wavelength solution) into a single composite 3-D data cube.
This rectification is achieved on a per-fiber basis by means of a cubic b-spline technique similar to that used previously in \S \ref{skysub.sec}, but with a fixed 
breakpoint spacing of 1.21 $\times 10^{-4}$ in units of logarithmic Angstroms (see Fig. \ref{resampling.fig}).
In order to mitigate the impact of biases in the data-derived variances on the mean of the resulting
spline fit (especially in the dichroic overlap region) we weight the data with a version of the inverse variance that has been smoothed with a 5 pixel boxcar; weights for the blue camera
are set to zero above 6300 \AA\ and weights for the red camera are set to zero below 5900 \AA.

We evaluate this bspline fit on two different fixed wavelength solutions, a decadal logarithmic and a linear.  The logarithmic wavelength grid 
runs from 3.5589 to 4.0151 (in units of logarithmic Angstroms) with a stepsize of $10^{-4}$ dex (i.e., 4563 spectral elements).  This corresponds to a wavelength range
of  3621.5960 to 10353.805 \AA\
with a dispersion ranging from 0.834 \AA\ channel$^{-1}$ to 2.384 \AA\ channel$^{-1}$ respectively.
The linear wavelength grid runs from 3622.0 to 10353.0 \AA\ with a stepsize of 1.0 \AA\ channel$^{-1}$ (i.e., 6732 spectral elements).
These endpoints are chosen such that the resulting spectra come from regions of the BOSS spectrographs where the throughput is sufficiently high for practical faint-galaxy science purposes.

Finally, we perform a second-pass cosmic ray identification on these camera-combined images by `growing' the previous cosmic-ray mask in both the fiberid and wavelength directions.
Pixels within a 1-pixel radius are included in the second-pass cosmic ray mask if their flux is more than $5\sigma$ away from the sigma-clipped mean for a given fiber within a 50-pixel
box in wavelength.  This additional step significantly reduces the occurrence of unflagged cosmic ray features in the final data products while only minimally ($\sim$ 2\%) increasing the total number of
flagged pixels.

The final flux calibrated, camera combined frames are saved as mgCFrame files (Table \ref{mgcframe.tab}).

\section{Astrometric Registration}
\label{astrometry.sec}

Once a sufficient number of exposures have been obtained on a given plate that the cumulative SNR$^2$ of
all complete sets exceeds the target threshhold (see \S \ref{ops.sec}) it is marked as complete in the observing database
and an `apocomplete' file is created in the \mangacore\ repository that contains a list of all corresponding exposure numbers.
This file serves as the trigger indicating that the  DRP at the University of Utah should enter the 3d stage of processing and combine together individual exposures
into final-form data cubes and RSS  for each IFU on the plate.

Using the metadata archived in \mangacore, spectra for each IFU target are pulled from the corresponding lines of the mgCFrame files
and collated into a single RSS file containing all of the spectra associated with a given object (manga-RSS; see Table \ref{rss.tab} and discussion in \S \ref{append.final.sec}).  
Typically, this corresponds to $3 \times N_{\rm set} \times N_{\rm ifu}$ spectra, where
$N_{\rm set}$ is the number of complete sets of exposures observed, and $N_{\rm ifu}$ is the number of fibers in the IFU.
After resorting the input spectra into a row-stacked format on a per-galaxy basis the DRP calculates the astrometric solution for each of the fibers.  This astrometric calibration 
has two stages: a basic module that computes fiber locations based on reference metadata and theoretical refraction models, and an advanced module that fine-tunes
the zero-point location and rotation of the basic solution by registering the spectra against SDSS broadband photometry.

\subsection{Basic Astrometry Module}
\label{basicast.sec}

The effective location of a particular IFU fiber in any given exposure is dictated by numerous optomechanical factors.  Many of these are possible to either measure or estimate
for an arbitrary source, and the MaNGA basic astrometry module combines these factors into a single wavelength-dependent position vector (in RA/DEC) for each fiber.
These factors include:

\begin{enumerate}

\item Relative and absolute fiber location within a given IFU ferrule based on the as-built fiber bundle metrology.  This is measured during the 
manufacturing process to a typical accuracy $\sim 0.3 \micron$ (relative)\footnote{$0.3 \micron$ corresponds to 5 milliarcsec at the SDSS focal plane ($\sim 60 \micron$ arcsec$^{-1}$)}
and $\sim 5 \micron$ (absolute) \citep[see][]{drory15} and recorded in \mangacore\ for each harness serial number.

\item Offset of an IFU from its base position due to the three-point dithering pattern.  We assume that the dithering exactly matches the commanded offsets; the accuracy
of this assumption is limited by the $\sim 0.1$ arcsec dithering accuracy of the telescope \citep[see][]{law15}.

\item Offset of drilled holes from the intended drilling location.  Although holes can be drilled to within an accuracy of $\sim 9 \micron$ rms, they are measured after the fact
to an accuracy $\sim 5 \micron$.  This information is recorded for each plate in \mangacore.

\item Chromatic differential atmospheric refraction (DAR) relative to the guide wavelength ($\sim 5500$ \AA).  
This shifts the effective location of each fiber as a function of wavelength (i.e., a given fiber
receives light from a different part of a target galaxy at blue versus red wavelengths).
The magnitude and direction of this effect is calculated 
using the SDSS plate design code model \citep[based in part on][]{filip82}
as discussed in detail by \citet{law15} and depends 
on the altitude, the parallactic angle, and the atmospheric temperature/pressure/humidity.  We calculate the expected effect 
at the midpoint of a given exposure for each of the 4563 wavelength channels (for the logarithmic case) given the known
location of each IFU on the sky and atmospheric conditions recorded in the headers of individual exposures.

\item Global shift of the IFU location at the guide wavelength due to field differential atmospheric refraction.  Over the 3$^{\circ}$ field of an SDSS plate there are changes in scale and
rotation of the sky image (in particular, altitude-dependent compression along the altitude axis).
The telescope guiding software corrects for these effects averaged over the plate, but cannot fully correct the quadrupole distortion in the effective location of a given IFU.
We estimate this effect as discussed in detail by \citet{law15}.

\item Wavelength-dependent distortions due to the SDSS telescope optics.  These are estimated based on upon optical models of the telescope in the SDSS plate design code \citep{gunn06}.

\end{enumerate}

The final product of the basic astrometry module is a pair of two-dimensional arrays 
(matched in size to the mgCFrame flux array) that
give the X and Y fiber positions (in units of arcsec in the tangent plane) relative to the nominal IFU center
IFURA, IFUDEC.  These arrays can thereafter be used to look up the effective on-sky location corresponding
to any wavelength, for any fiber.

\subsection{Extended Astrometry Module}
\label{eam.sec}

During operations within a single dark run a MaNGA cartridge will remain plugged with a given plate until observations for that plate are complete.  Since MaNGA
shares carts with APOGEE-2N however, at the end of a dark run it is typically necessary to unplug the MaNGA IFUs from unfinished plates and replug them again the following
dark run to continue observations.  This replugging introduces an uncertainty into the relative centering and rotation of each IFU in its hole at the level of $\sim 0.5$ arcsec (centering)
and $\sim 2-3^{\circ}$ (rotation)  from the required clearance of
locator pins within their holes.  The precise change cannot be measured directly and will change from plugging to plugging as it depends on
the torsional stresses arising from the routing of each IFU cable through a cartridge.
Such uncertainties\footnote{At the outer ring of a 127-fiber IFU such a 3 degree rotational uncertainty corresponds to $\sim 0.8$ arcsec.}
are significantly larger than any of the uncertainties derived from effects described in \S \ref{basicast.sec}.

We therefore follow the method employed by the VENGA survey \citep{blanc13} of registering the fiber spectra from each exposure against SDSS broadband imaging.
In this `extended astrometry module' (EAM)  we compute the synthetic broad-band flux of each fiber by integrating the flux-calibrated spectrum over the corresponding transmission
curve.  We then search a grid in right ascension, declination, and rotation of the fiber bundle relative to the base position determined in \S \ref{basicast.sec} (we keep the relative
positions of fibers within the bundle fixed).  At each position on the grid the fiber coordinates 
(collapsed from the basic astrometry solution over the appropriate wavelength range)
are shifted accordingly, and aperture photometry is performed on a PSF matched
SDSS broadband image using 2.0 arcsec diameter apertures for each fiber.  An additional overall flux normalization is permitted according to

\begin{equation}
f_{\rm SDSS} = A \, f_{\rm MaNGA} + B
\end{equation}

where $f_{\rm SDSS}$ and $ f_{\rm MaNGA}$ are the SDSS broadband and MaNGA fiber fluxes respectively, $A$ is a multiplicative scaling factor, and $B$ represent a 
zeropoint shift.\footnote{We note that these $A$ and $B$ terms are derived for informational purposes only; 
we do not apply corrections from them to the data but rather use them as an
independent check on the flux calibration of each frame and flag exposures as problematic where $A$ or $B$ deviate
substantially from 1.0 and 0.0 respectively.}
\citet{yan16} presented a discussion of these $A$ and $B$ coefficients, and determined that $A$ had a roughly gaussian distribution centered about 1.00 (in $i$-band)
with a sigma of 0.037 for $\sim 25,000$ IFU-exposures obtained during the first year of operations, indicating that the spectrophotometric accuracy of the MaNGA data is about 4\% with respect to the SDSS imaging data.
The best-fit values of position, rotation, and flux offsets are determined via $\chi^2$ minimization, with corresponding uncertainties drawn from the $\chi^2$ probability maps.
This exercise is repeated in each of the four $g$, $r$, $i$, and $z$ bands, with the final result a biweight mean of the four bands (this provides robustness against
occasional unmasked cosmic rays).

\begin{figure*}
\plotone{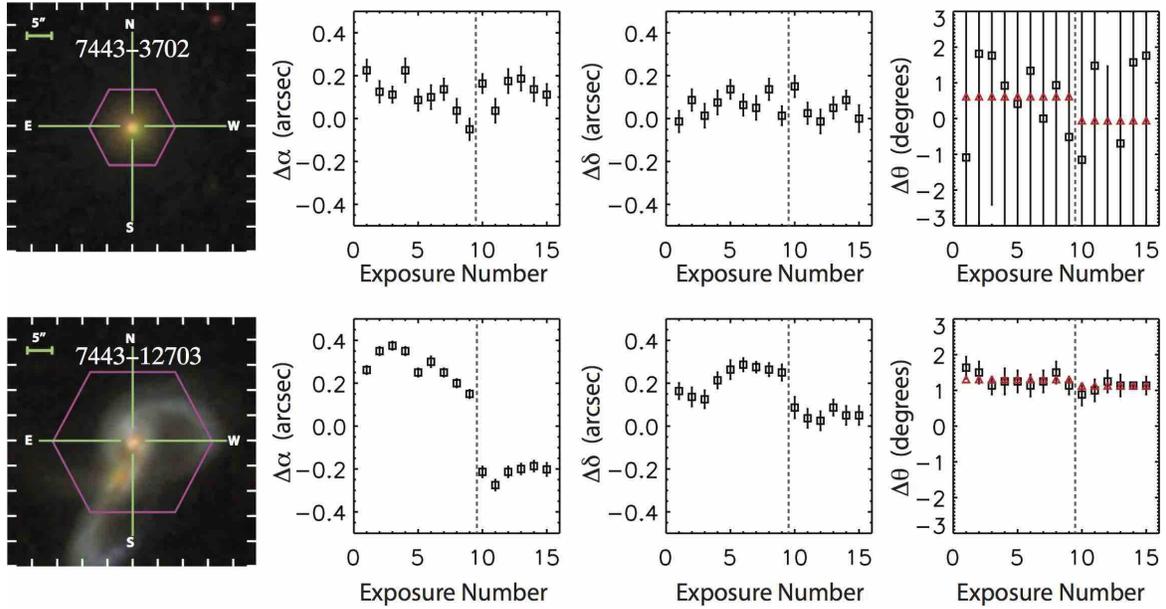}
\caption{MaNGA EAM performance for two commissioning galaxies 7443-12703 and 7443-3702
(mangaid 12-193481 and 12-84670 respectively).
The left-most panel shows a 3 color image of each galaxy based on SDSS imaging data, overlaid with a hexagonal bounding box indicating the footprint of the
MaNGA IFU.  The remaining boxes show the values calculated by the EAM for the relative shift in right ascension, declination, and bundle rotation between exposures (open black
boxes with associated $1\sigma$ uncertainties).  Red boxes in the right-hand panel show the average values in $\Delta \theta$ adopted for all exposures in a given plugging
in a second run of the EAM.  Values shown for the shifts $\Delta \alpha$ and $\Delta \delta$ are after this second-pass with fixed $\Delta \theta$.
The vertical dotted line represents a replugging of the plate between exposures 9 and 10.}
\label{extastrom.fig}
\end{figure*}

Unsurprisingly, the EAM can achieve better results for larger fiber bundles on galaxies with significant azimuthal structure than for smaller bundles on smooth
and circular galaxies.  In Figure \ref{extastrom.fig} we show EAM results for two commissioning galaxies 7443-12703 and 7443-3702.
For the large IFU  on a source with significant structure (7443-12703) the measurement uncertainties on both positional shift and global rotation are small, and reveal (in this case)
a $\sim 0.5$ arcsec shift in the IFU center across a cartridge replugging.  In contrast, for the small IFU on a rotationally symmetric source (7443-3702) the positional shift is
still well constrained but the global rotation is almost completely unconstrained in the range of values explored by the EAM ($\pm 5^{\circ}$).

In order to avoid introducing errors into our astrometry due to such noisy measurements, we therefore run the EAM iteratively.  In the first pass, each exposure
is fitted independently.  The derived values of $\Delta \theta$ are then averaged across all exposures in a given plugging, and the EAM run again holding
$\Delta \theta$ fixed at these average values in order to better determine positional shifts between exposures.  Since rotation is expected to change only between repluggings
(consistent with observed behavior based on galaxies with sufficient azimuthal structure to measure $\Delta \theta$ reliably) this allows us to mitigate the uncertainty
in any individual measurement of bundle rotation.  In contrast, such averaging is not justified for the positional shifts.  Although such shifts are dominated
by repluggings (e.g., Fig. \ref{extastrom.fig}), smaller shifts at the $\sim 0.1$ arcsec level are possible  due to uncertainties in the applied dither offsets that we wish to correct
through the EAM.  On average, we find that the median astrometric uncertainty of the  exposures making up the
1390 galaxies in DR13 
relative to the SDSS preimaging data is
$\sim 0.1$ arcsec ($1\sigma$) based on the reduced $\chi^2$ surface.

Since each set of three exposures is known to have uniform coverage \citep[see][]{law15,yan16b} sets from different pluggings of a given plate
(and indeed, even between different cartridges) can therefore be combined together onto a common astrometric solution using the EAM.  Since all MaNGA target galaxies are
drawn from the SDSS imaging footprint, this correction is automatically applied to all MaNGA galaxies.\footnote{The method will fail on point-like sources
and some ancillary targets outside the SDSS imaging footprint; for these objects the EAM is disabled and the basic astrometry module is used alone.}


\section{Data Cube Construction}
\label{cubes.sec}


\subsection{Basic Cube Building}
\label{cubebasic.sec}

Using the RSS  files and associated astrometric solutions derived in
\S \ref{astrometry.sec} we combine the individual fiber spectra into rectilinearly gridded cubes 
(with orientation RA, DEC, $\lambda$) for each IFU on both logarithmic and linearly-sampled wavelength solutions.  
Since these input spectra have already been resampled onto a common wavelength grid, this simplifies to
the 2-dimensional reconstruction of a regularly-gridded image from an irregularly sampled cloud of measurements of the intensity profile at a given wavelength channel.

Multiple methods exist for performing such image reconstruction (see \S \ref{cubealg.sec}); we choose to build
our data cubes one image slice at a time using a flux-conserving variant of Shepard's 
method similar to that used by the CALIFA pipeline \citep{sanchez12}. 
At each of the 4563 wavelength channels (for the logarithmically sampled data; 6732 for the linear), we 
describe our input data as one-dimensional vectors of intensity $f[i]$ and variance $g[i]$ with length 
$N = N_{\rm fiber} \times N_{\rm exp}$ where $N_{\rm fiber}$ is the number of fibers in the IFU (e.g., 127) and $N_{\rm exp}$ is the total number of exposures to combine together.  
Similarly, we can construct vectors $x$ and $y$ which describe the effective position of the center of each fiber based on the astrometric solution derived in \S \ref{astrometry.sec},
and converting to fractional pixel coordinates relative to some chosen origin and pixel scale.
We adopt a spatial pixel scale of 0.5 arcsec pixel$^{-1}$ and an output grid of size $X_{\rm max}$ by $Y_{\rm max}$ taken to be slightly larger than the dithered footprint of
the MaNGA IFU.

Each of the $M = X_{\rm max} \times Y_{\rm max}$ pixels in the output image can likewise be resorted into a one-dimensional array of values, with the pixel locations given by $X[j]$ and $Y[j]$
respectively for $j = 1$ to $M$.  The mapping between the $f[i]$ intensity measurements in the irregularly-sampled input and the $F[j]$ intensities in the regularly-sampled output
image are then determined by the weights $w[i,j]$ describing the relative contribution of each input point to each output pixel.  We take this weight function to be a circular Gaussian:
\begin{equation}
w[i,j] =  b[i] \, {\textrm {exp}} \left( -0.5 \frac{r[i,j]^2}{\sigma^2}\right)
\end{equation}
where $\sigma = 0.7$ arcsec is an exponential scalelength, $r[i,j] = \sqrt{(x[i]-X[j])^2+(y[i]-Y[j])^2}$ is the distance between the $i$'th fiber location and the $j$'th output grid square,
and $b[i]$ is a binary integer equal to zero where the inverse variance $g[i]^{-1}=0$ and one elsewhere.  Essentially, 
$b[i]$ functions as a mask that allows us to exclude known bad values in individual spectra from the final combined image.
Additionally, we set $w[i,j] = 0$ for all $r[i,j] > r_{\rm lim} = 1.6$ arcsec as an upper limit on the radius of influence of any given measurement.  These limiting radii and scale lengths
are chosen empirically based on observed performance; the present values are found to provide the smallest reconstructed FWHM for stellar targets observed as part of commissioning
(see \S \ref{angularres.sec}) whilst not introducing spurious structures by shrinking the impact-radius of individual fibers too severely.

In order to conserve flux we must normalize the weights such that the sum of the weights contributing to any given output pixel is unity.  The normalized weight function is therefore:
\begin{equation}
W[i,j] = \frac{w[i,j]}{\sum_{i=1}^{N} w[i,j]}
\end{equation}
where in order to avoid divide by zero errors we set $W[i,j] = 0$ where $w[i,j] = 0$ for all $i$ in the range $1$ to $N$ (e.g., outside the hexagonal footprint of the IFU).

The intensity distribution of the pixels in the output image may therefore be written as the matrix product of the normalized weights and the input intensity vector:
\begin{equation}
F = \alpha \, W \times f = \alpha
\begin{bmatrix}
    W_{11}  & \dots  & W_{N1} \\
    \vdots & \ddots & \vdots \\
    W_{1M} & \dots  & W_{NM}
\end{bmatrix}
\times
\begin{bmatrix}
    f_{1} \\
    \vdots  \\
    f_{N} 
\end{bmatrix}
\end{equation}
or alternatively as
\begin{equation}
F[j] = \alpha \sum_{i=1}^{N} f[i] \, W[i,j]
\label{eq:regrid}
\end{equation}
where $\alpha = 1/(4\pi)$ is a constant factor to account for the
conversion from flux per unit fiber area ($\pi$ arcsec$^2$) to flux per
unit spaxel area (0.25 arcsec$^2$).   The resulting $F[j]$ may then trivially be rearranged to form the output image at this wavelength slice given the known mapping
of the pixel coordinates $X[j]$ and $Y[j]$.

Similarly, the variance $G$ of the rectified output image may be written as
\begin{equation}
G[j] = \alpha^2 \sum_{i=1}^{N} g[i] \, (W[i,j])^2
\end{equation}
This calculation therefore
propagates the uncertainties in individual spectra through to the final
data cube, but does not use these uncertainties in constructing the
combined flux values (except for the simple masking of bad values where
inverse variance is equal to zero).

These rectified images of the intensity profile and the corresponding inverse variance maps at each wavelength channel are reassembled by the DRP
into three-dimensional cubes along with a 3d quality mask describing the effective coverage and data quality of each spaxel.  The final manga-CUBE 
files are discussed further in \S \ref{append.final.sec} (see also Table \ref{cubes.tab}).

\subsection{Algorithm Choice}
\label{cubealg.sec}

As stated in \S \ref{cubebasic.sec}, there are multiple algorithms that we could have adopted for building our data cubes, ranging from surface fitting techniques
(e.g., thin plate spline fits)
to drizzling and our adopted modified Shepard approach.  
Based on idealized numerical simulations performed prior to the start of the survey we found that the surface-fitting approach provided reasonable quality reconstructed images,
but was nonetheless undesirable because there is no simple means by which to propagate uncertainties in the resulting surface.  In contrast,
the modified Shepard approach allows for easy calculation of both the variance and covariance of the reconstructed data cubes, as described in \S \ref{covar.sec}.

The drizzle approach \citep{fruchter02} has been tested by ourselves and by the CALIFA
\citep{sanchez12}
and SAMI \citep{sharp15} surveys, all of whom have found that (1) it broadened the
final PSF, and (2) since fiber bundle IFUs have $< 100$\% fill factor
in a given exposure it can create artificial structures in the intensity distribution
following the footprint of the circular fibers.
To mitigate this problem the SAMI survey \citep[see discussion by][]{sharp15} adopted a weighting system
based on the ratio between the original fiber area and the area covered
by a final spaxel of a particular fiber (if the fiber is reduced by an arbitrary
amount smaller than the original size). This in essence
redistributes the flux following a weighting that depends on the distance to the
centroid of the fiber and is truncated at a maximum distance
controlled by the arbitrary reduction of the covered area of
the fibers.  This weighting function results in sharper images, but in order to smooth
out the artificial structure in the intensity distribution \citet[][see their Figs. 7 and 9]{sharp15}
found that a large number of dither positions ($\geq 7$) was required to sufficiently sample
the galaxy.

Such an approach is not viable for MaNGA (or CALIFA) for a variety of reasons.
First, the effective filling factor of the MaNGA IFUs is lower than that of SAMI \citep[56\% vs 75\%; see][]{law15}, meaning the gaps in coverage
for a given exposure are larger (although much more regular).  Secondly, the inner diameter of the MaNGA fibers (2 arcsec)
and the fiber-to-fiber spacing in the IFUs (2.5 arcsec) is large compared to the typical FWHM of the observational seeing ($\sim 1.5$ arcsec), meaning
that the spatial resolution incident upon the IFU bundles is drastically undersampled in a single exposure.  Most importantly, however,
the MaNGA survey strategy of reaching constant depth on each target field requires a different total number of exposures depending
on observational conditions and the Galactic foreground extinction.  The number of exposures on a given target can therefore range from 6 -- 21, obtained in
sets of 3 dithered exposures that must achieve uniform coverage and good reconstructed image quality.  Similarly, the SAMI approach also does not work for CALIFA since
CALIFA often has only a single visit to a given field.

In contrast, the modified Shepard approach adopted in \S \ref{cubebasic.sec} allows for
 high-quality image reconstruction from just 3 dithered exposures that can be repeated
as necessary to achieve the desired depth in a given field \citep[see discussion in][]{law15}.  This algorithm was found to perform well based on prior experience with the CALIFA
survey, and in numerical simulations designed to optimize the choice of the scale length and truncation radius for the exponential weighting function.
We note that although the MaNGA and SAMI approaches to cube building are {\it conceptually} different they are {\it mathematically} quite similar, albeit that the SAMI
weighting function does not follow a Gaussian
distribution and the kernel is in essence sharper (i.e., with smaller size and truncation radius).

\subsection{Covariance}
\label{covar.sec}

The redistribution of intensity measurements from individual fibers into a rectilinearly-sampled
datacube via the equations in \S \ref{cubebasic.sec}  leads to significant covariance among
spatially adjacent pixels at each wavelength slice. 
The formal covariance matrix of each slice of the data cube can be written via matrix multiplication as:
\begin{equation}
C = \alpha^2 \, W \times (g' \times W^{\intercal})
\end{equation}
where $\alpha$ is again a constant scale factor, and $g'$ is the diagonal variance matrix
\begin{equation}
g' = \begin{bmatrix}
    g_1  & 0 & \dots  & 0 \\
     0  & g_2 & \dots  & 0 \\
    \vdots & \vdots & \ddots & \vdots \\
   0 & 0 & \dots  & g_N
\end{bmatrix}
\end{equation}

The diagonal elements of $C$ represent the $M$ elements of the variance array $G[j]$ for the output image while the off-diagonal elements of $C$
represent the covariance introduced between different pixels in the output image by the chosen weighting method.  These may in turn be recast as the correlation matrix $\rho$, where
$\rho_{jk} = C_{jk} / \sqrt{C_{jj} C_{kk}}$ for all $j$ and $k$ from 1 to $M$.  $\rho$ is thus unity along the diagonal elements (since each pixel has unity correlation with itself).
Following this exercise, we
find that, generally, pixels separated by $0\farcs5$ (1 pixel) have
correlation coefficients of $\rho\approx 0.85$, decreasing to $\rho <
0.1$ (i.e., nearly uncorrelated) at separations of $\geq$ 2 arcsec.  Spatial covariance therefore becomes
important when, for example, one calculates the inverse variance in a
spectrum generated by coadding many adjacent spaxels.

Although $\rho$ is nominally a large matrix, in practice it is both symmetric and sparse,  containing
mostly zero-valued elements since we have truncated the weight function to be zero outside a radius of 1.6 arcsec.
Since the MaNGA reconstructed PSF is only a weak function of wavelength
$\rho$ also changes only slowly with wavelength, meaning that values of $\rho$ at a given wavelength may generally
be interpolated from adjacent wavelengths.
In a future data release, the DRP will therefore include the correlation matrix 
at the central wavelengths of the $griz$ bands
in the final data products of the cube building algorithm.
At the present time in DR13 however these correlation matrices are not yet available, and we therefore provide a rough calibration
of the typical covariance in the MaNGA data cubes following the conventions established by the CALIFA
survey \citep{2013A&A...549A..87H}.  Specifically, we provide a
calibration of the nominal calculation of the noise vector of a coadded
spectrum under the incorrect assumption of no covariance to one
determined from a rigorous calculation that includes covariance.

\begin{figure}
\epsscale{1.2}
\plotone{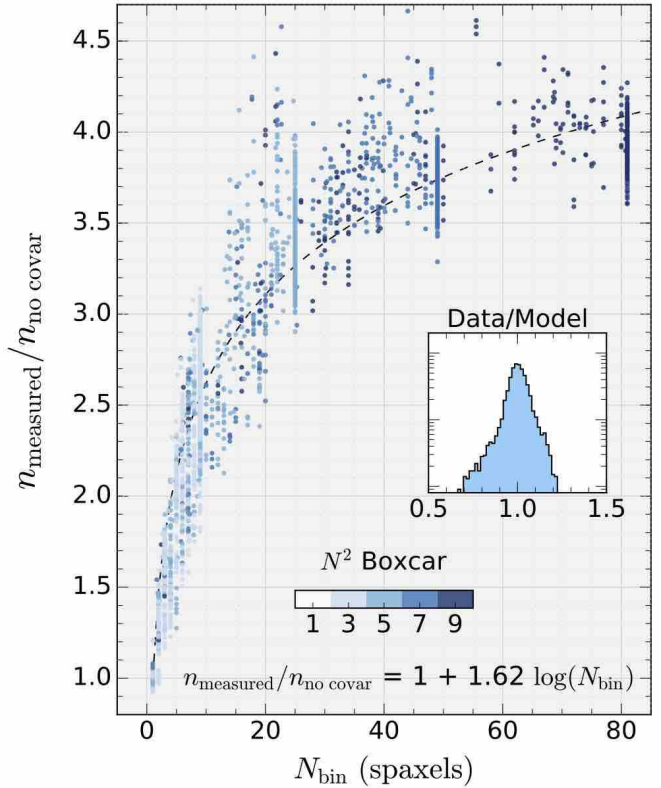}
\caption{
Ratio of the measured noise in a synthetic datacube, $n_{\rm measured}$,
(see text) to a nominal calculation of the noise in a binned spectrum
that does not include covariance, $n_{\rm no\ covar}$, as a function of
the number of spaxels included in the combined spectrum, $N_{\rm bin}$.
The point color provides the size of the boxcar used to create the bin.
Nominally, $N_{\rm bin} = N^2$, however some boxcar windows fell outside
of the IFU field-of-view in the synthetic datacube.  The equation at the
bottom right gives the best-fitting calibration of $n_{\rm no\ covar}$
to $n_{\rm measured}$ for values of $N_{\rm bin} \leq 100$.  The inset histogram shows the ratio of the model
to the data, demonstrating the calibration is good to about 30\%.
}
\label{fig:covariance}
\end{figure}

We have done so using an idealized experiment.  Using five datacubes
from plate 7495, one of each of the fiber-bundle sizes, we synthetically
replace each RSS spectrum with unity flux and Gaussian error.  We then
construct the datacube identically as done for our galaxy observations.
We bin the resulting spaxels using a simple boxcar of size $N^2$ where
$N=1,3,5,7,9$ and calculate the mean and standard deviation in the
resulting spectrum.  This noise estimate is our {\em measured} error,
$n_{\rm measured}$.  Alternatively, we can use the inverse-variance
vectors for each spaxel in the synthetic datacube that results from the
nominal calculation above to create a separate noise estimate, which
instead assumes each spaxel is independent.  This calculation follows
nominal error propagation, but does not account for the covariance
between spaxels; we refer to this as $n_{\rm no\ covar}$.  The ratio of
these two estimates is shown in Figure \ref{fig:covariance}.

Figure \ref{fig:covariance} demonstrates that the true error in a
combined spectrum is substantially larger than an error calculated by
ignoring spatial covariance.  The relationship of the errors with and without
covariance depends upon the number $N_{\rm bin}$ of spaxels combined.
For small $N_{\rm bin}$ the values in nearby spaxels are highly correlated
and the SNR is nearly constant with $N_{\rm bin}$ (i.e., both the signal and the true
error increase proportionally to $N_{\rm bin}$).  At large $N_{\rm bin}$ the
values in combined spaxels are nearly uncorrelated, and the SNR
increases proportionally to $\sqrt{N_{\rm bin}}$.

We have thus fit a functional form identical to
that used by \citet{2013A&A...549A..87H} to our measurements in Figure \ref{fig:covariance} and find that
\begin{equation}
n_{\rm measured}/n_{\rm no\ covar} \approx 1 + 1.62\log(N_{\rm bin}),
\label{eq:covariance_calibration}
\end{equation}
for $N_{\rm bin} \lesssim 100$, and
\begin{equation}
n_{\rm measured}/n_{\rm no\ covar} \approx 4.2
\end{equation}
for $N_{\rm bin} > 100$ (i.e., beyond $\sim 2$ times the FWHM where spaxels are uncorrelated).

It is important to
note that the binned spaxels must be adjacent for this calibration to
hold; i.e., a random selection of spaxels across the face of the IFU
will not show as significant an effect because they will not be as
strongly covariant.  The inset histogram shows the ratio of the data to
the fitted model in equation \ref{eq:covariance_calibration},
demonstrating the calibration is good to about 30\%.  We have confirmed
this result empirically by comparing the standard deviation of the
residuals of the best-fitting continuum model for a large set of galaxy
spectra, following an approach similar to \citet{2013A&A...549A..87H}.
However, we emphasize that the test we have performed to produce Figure
\ref{fig:covariance} is more idealized and controlled.  We also confirm
that a rigorous calculation of the covariance, following the matrix
multiplication discussed at the beginning of this section, and a
subsequent calculation of the noise vector in the binned spectra used in
Figure \ref{fig:covariance} are fully consistent with our meausurements
$n_{\rm measured}$.

\section{Data Quality}
\label{dq.sec}

\subsection{Data Cubes: Angular Resolution}
\label{angularres.sec}

An estimate of the spatial light profile of an unresolved point source (i.e., the `reconstructed PSF')
is automatically provided for each data cube using a numerical simulation tied to the specific observing conditions of each exposure.
Using the known fiber locations for a given exposure, the DRP computes the flux expected to be recorded by each fiber from an unresolved point source located at the center of the IFU.
This model flux is based on integration of the nominal PSF incident on the face of the IFU in the focal plane of the SDSS 2.5m telescope.  The focal-plane PSF is taken to be a 
double-gaussian that accounts for chromatic distortions due to the telescope optics and observational seeing recorded by the guide camera.  As detailed by \citet{yan16},
since the guide-camera reports image FWHM systematically larger than measured by the MaNGA IFU fiber bundles, the guider seeing measurements are also `shrunk' by a scale factor
determined by the flux calibration module to give an incident PSF that best matches differential fiber fluxes recorded by the 12 photometric standard star minibundles.
These simulated fiber fluxes are reconstructed into a data cube using the same algorithm as the science data, and slices of this cube corresponding to $g,r,i,z$ bands 
are attached to each data cube.

These $griz$ images (GPSF, RPSF, IPSF, ZPSF; see \S \ref{append.final.sec}) provide a reasonable estimate of the reconstructed PSF
in each data cube and are reported in each of the FITS headers.  
We confirmed the fidelity of these reconstructed PSF models by observing a plate during survey commissioning in which every MaNGA IFU targeted bright stars
with two sets of dithered observations (i.e., following the methodology of typical galaxy observations).  
This plate (7444) was processed by the DRP in an identical manner to standard galaxy plates, with the exception that only the
basic astrometry module was used to register the fiber locations since there is no extended structure against which to use extended astrometry module.

\begin{figure*}
\epsscale{1.2}
\plotone{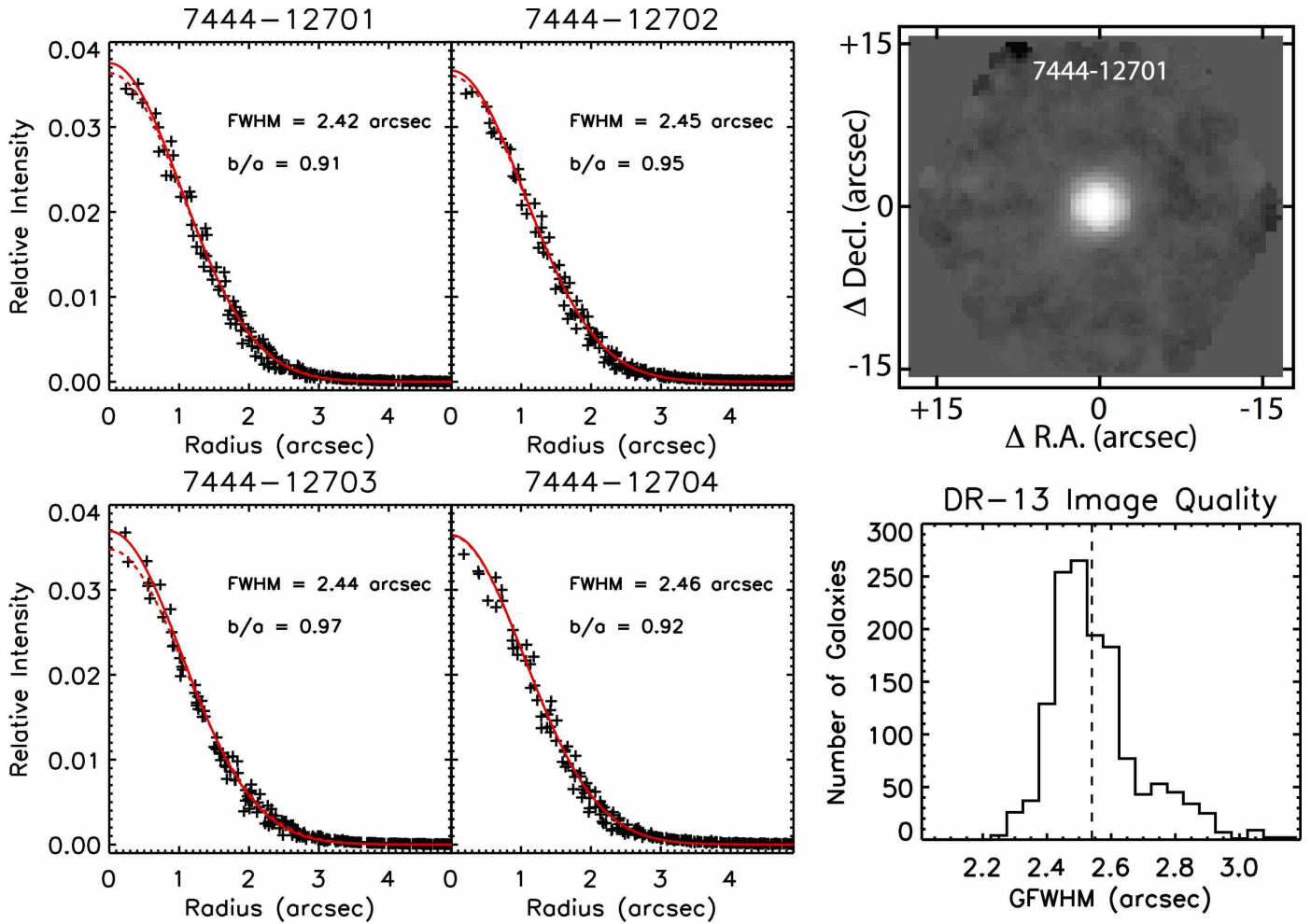}
\caption{Top right panel: Reconstructed image of a bright star observed in standard dithered observations (7444-12701); the data cube has been collapsed
over wavelength channels 300-700 ($\lambda\lambda 3881-4255$ \AA).  The greyscale stretch is logarithmic to illustrate the symmetrical nature of the
extended profile wings.  Left-hand panels: radial profiles of bright stars targeted by four of the largest IFUs on plate 7444.  Black points show the radial
profile of the reconstructed image (based on collapsing the corresponding data cube over the range $\lambda\lambda 3881-4255$ \AA).  The solid red lines show
the best 2d gaussian fitted to the black points, with characteristic FWHM and minor/major axis ratio ($b/a$) indicated.  The dashed red lines show the corresponding 2d
gaussian fitted to the PSF model provided by the pipeline based on known fiber locations and observing conditions for each exposure.  Lower right panel: Distribution
of $g$-band FWHM measured for all 1390 galaxy data cubes in DR13; vertical dashed line indicates the median of 2.54 arcsec.}
\label{psf.fig}
\end{figure*}

In Figure \ref{psf.fig} we show the  profiles of stars in four of the reconstructed data cubes compared to the simulated estimates.
We find that the actual reconstructed PSF of these data cubes is well described by a single 2d gaussian function with normalized intensity
\begin{equation}
I(r) = \frac{1}{2 \pi \sigma^2} \, {\rm exp}\left( -r^2/ 2 \sigma^2 \right)
\end{equation}
where $2.35 \sigma$ is the standard gaussian FWHM.  This profile is well matched to the model PSF estimated based on mock integrations of an artificial point source
at the known fiber positions; the model FWHM estimates agree with the measured values to within 1-2\%.
The measured FWHM of the reconstructed PSF for the other 13 IFUs on plate 7444 similarly lie in the range 
$2.4-2.5$ arcsec.\footnote{Except for one 19-fiber IFU, for which the reconstructed image is clearly out of focus, indicating that it partially fell out of the plate.  Such cases are rare,
and detected during quality-control checks by the extended astrometry module.}
Based on the simulations presented by \citet{law15} and the range of $\Omega$ uniformity values for DR13 reported by \citet{yan16b}
we expect that the reconstructed PSF FWHM should vary by less than 10\% across a given IFU.

As discussed in greater detail by \citet{yan16b}, the range of $g$-band reconstructed PSF FWHM in the 1390 DR13 galaxy data cubes is generally
distributed in the range $2.2 - 2.7$ arcsec, with a tail to about 3 arcsec (Figure \ref{psf.fig}).

\subsection{Data Cubes: Spectral Resolution}
\label{specres.sec}

\begin{figure*}
\epsscale{1.0}
\plotone{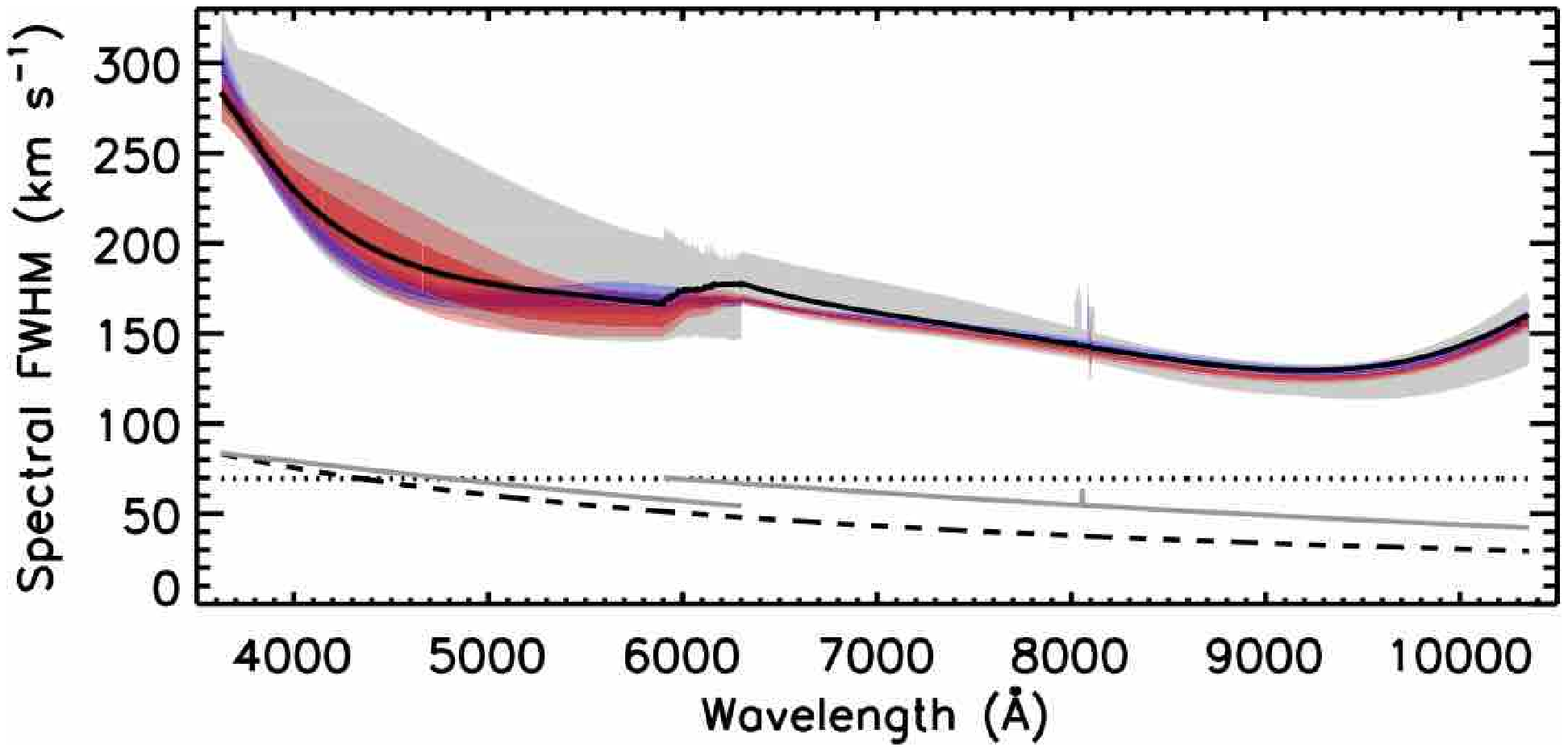}
\caption{MaNGA spectral resolution (FWHM) as a function of wavelength for the final wavelength-rectified data products.  
The solid black line represents the average FWHM across all 1390 galaxy data cubes in DR13, while the 
grey shaded region indicates the minimum and
maximum FWHM of all 11,916 fiber spectra obtained for example plate 8588.  Blue dark/light
shaded regions and red dark/light shaded regions show the $1\sigma$/$2\sigma$ variations about the the least-variable and
most-variable IFUs on this plate respectively (8588-12704 and 8588-12705).  The dotted/dashed black lines indicate the final pixel sampling
scale of the MaNGA LOG-format and LINEAR-format data respectively.  The solid grey lines represent the native pixel sampling
of the blue and red cameras.  The feature
around 8100 \AA\ indicates the two-phase detector discontinuity.
Note that the values shown here have been broadened by 10\% relative to the values reported by the DR13 data pipeline
to account for post-pixellization modeling and wavelength rectification
(see discussion in \S \ref{specres.sec}).
}
\label{specresifu.fig}
\end{figure*}

As indicated in \S \ref{wavecal.sec}, the line spread function varies along the spectrograph slit, and hence varies spatially within a
given IFU.  Similarly, the LSF can also vary between exposures with ambient temperature drifts and changes in the focus of the spectrograph.
The typical spectral resolution for DR13 galaxies is shown in Figure \ref{specresifu.fig}; typical IFUs show rms variability at the level of
$1-2$\% (blue shaded region), while the worst-case large IFUs on the ends of the spectrograph slit can show variability as high as $8-10$\%
at blue wavelengths (red shaded region).  This variability within the worst-case IFUs is dominated by the along-slit variability, but compounded by variations between exposures.
The focus in the red cameras is significantly flatter than in the blue cameras, meaning that variation in spectral 
resolution longward of $6000$ \AA\ is  1\% or less even for the worst-case IFUs.\footnote{Except around 8100 \AA\ where the red detectors
have a two-phase discontinuity (see \S \ref{wavecal.sec}).}

Each MaNGA data cube therefore has an associated extension (see \S \ref{append.final.sec}) describing 
both the mean and $1\sigma$ deviation about the mean spectral resolution for all fiber spectra contributing to the cube.  Detailed information on spectral resolution of the individual
fiber spectra used to create a given data cube are contained in the final RSS files.

After finalization of the DR13 data pipeline it was realized that the instrumental LSF estimates reported by the pipeline are systematically
underestimated.  There are two factors that contribute to this underestimation; first, the LSFs reported in DR13
correspond to native gaussian widths prior to convolution with the boxcar detector pixel boundaries (i.e., the gaussian function is integrated over the pixel boundaries), 
while many third-party analysis routines simply evaluate gaussian models at the pixel midpoints.  
Although neither approach is necessarily more `correct' than the other, this nonetheless represents a systematic difference between the values quoted and the values that would be measured
with most third-party routines.
Second, the wavelength rectification performed in \S \ref{waverect.sec}
effectively resamples the spectra and introduces a broadening into the LOG and LINEAR format spectra that is not accounted for by the DR13 data pipeline.
These issues are not unique to the MaNGA data and pipeline, but rather affect all previous generations of SDSS optical fiber spectra as well.

Efforts to address this discrepancy are ongoing \citep[see, e.g.,][]{westfall16} and will be detailed in a future version of the MaNGA data pipeline.
In the present contribution, we note that re-analysis of $\sim$ 2500 individual exposures suggests that multiplying the DR13 LSF  by a factor
of 1.10 gives a reasonable first-order correction (i.e., the spectral resolution of the DR13 data products is overestimated by $\sim$ 10\%).
This correction factor accounts for both the pre- vs post-pixellization
gaussian difference ($\sim 4\%$) and the wavelength rectification broadening ($\sim 6$\%).

\subsection{Wavelength calibration}
\label{waveacc.sec}

Based on previous calculations for the BOSS redshift survey \citep[e.g.,][their Figure 14]{bolton12}, the MaNGA spectra (which share the same instrument and much of the same
reduction pipeline software) should also have absolute wavelength calibration good to $\sim 5$ km s$^{-1}$.
We verify this estimate by comparing bright emission line features in the MaNGA data cubes against publicly-available SDSS-I single fiber spectra of each of the galaxies
in DR13.  For each galaxy, we obtain the corresponding SDSS-I spectrum from SkyServer\footnote{SkyServer is a web-based public interface to the SDSS archive;
see http://skyserver.sdss.org/dr12/en/home.aspx}
, and determine the effective location of the spectrum
from the PLUG\_RA and PLUG\_DEC header keywords.  We then perform aperture photometry in a 2 arcsec circular radius about this location
at every wavelength slice of the MaNGA data cube in order to construct a 1d MaNGA spectrum of the central pointing.  Both the SDSS-I
and MaNGA spectra are then fitted with single-Gaussian emission line components at the expected wavelengths of the H$\beta$, \othree\ $\lambda 5007$,
H$\alpha$, and \ntwo\ $\lambda 6583$ nebular emission lines given the known galaxy redshift from the NASA-Sloan Atlas \citep[NSA;][]{blanton11}.\footnote{http://www.nsatlas.org}

\begin{figure}
\epsscale{1.0}
\plotone{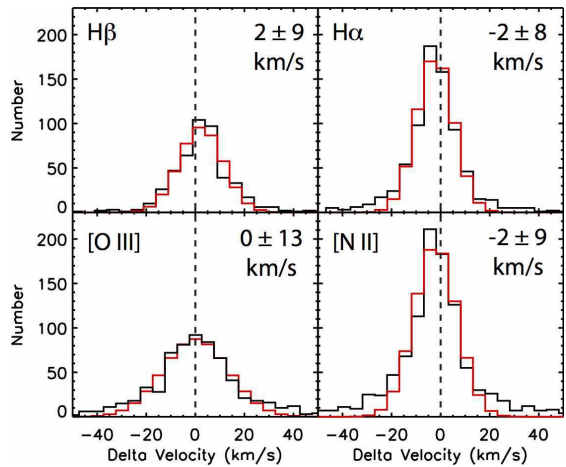}
\caption{Histograms of velocity difference between SDSS-I spectra and MaNGA IFU spectra extracted from a 2 arcsec radius circular aperture
centered on the location of the SDSS-I spectra.  The 4 panels show the results for H$\beta$, \othree\ $\lambda5007$, H$\alpha$, and \ntwo\ $\lambda6583$ for the 
1351 unique galaxies in DR13.  Note that the many galaxies with nebular emission lines too weak for reliable measurement have been omitted from the distribution.
Black histograms in each panel show the observed distribution, while red histograms illustrate the best-fit Gaussian model.  The values in each panel give the center
and $1\sigma$ width of the Gaussian model; this width may be driven largely by internal velocity gradients paired with uncertainties in the SDSS-I fiber
locations.}
\label{wavecal.fig}
\end{figure}

Although many of the MaNGA galaxies do not have strong emission line features in their central spectra, sufficiently many do in order to allow us to statistically compare
the MaNGA and SDSS-I spectra.  Considering only galaxies for which both MaNGA and SDSS fits are
within 5 \AA\ of the nominal wavelength, have $\sigma$ width of 0.5 to 5 \AA, and line fluxes $> 10^{-16}$ erg s$^{-1}$ cm$^{-2}$,
we find that 470/670/760/1063 galaxies fulfill the criteria for H$\beta$,\othree,H$\alpha$, and \ntwo\ respectively.
In Figure \ref{wavecal.fig} we plot the distribution of relative peak velocity offsets for each of these 4 emission lines.  We conclude that there is no systematic offset between the
MaNGA and SDSS-I spectra to within $\sim 2$ \kms, and that individual galaxies are distributed nearly according to a Gaussian with $1\sigma$ width $\sim 10$ \kms.

This width may in part however reflect intrinsic velocity gradients within the galaxies combined with
uncertainties at the few tenths of an arcsecond level in the effective location of the SDSS-I fibers due to hardware tolerances and differential atmospheric refraction.\footnote{Indeed, the 
SDSS-I spectra also have effective locations that {\it change} as a function of wavelenth due to chromatic atmospheric refraction.}
Using the MaNGA IFU spectra, we find that changes in location at the level of just 0.25 arcsec 
(compared to the typical MaNGA astrometric uncertainty of 0.1 arcsec; see \S \ref{eam.sec})
can easily result in 
$\sim 20$ \kms\ velocity shifts in the resulting spectra for galaxies with strong central velocity gradients (e.g., 8453-12703).
The actual wavelength accuracy of the MaNGA spectra may therefore more accurately be given by the rms agreement between repeat MaNGA observations
of a small sample of galaxies in DR13; indeed, although there are only
$\sim$ 10 repeat-observations with strong emission lines in DR13 we find a typical rms agreement of 5 \kms\ between the four emission line wavelengths
above.

The {\it relative} wavelength calibration accuracy of the individual fibers within a given IFU is more difficult to assess
in the absence of a calibration reference.  However, we can obtain a rough estimate by considering the rms scatter between
the measured centroids of bright skylines and the fitted value adopted by the pipeline as described in \S \ref{sciextract.sec}.
As a conservative estimate,\footnote{The rms of any individual line is closely related to the strength of the line (stronger lines have smaller rms), and the wavelength
solution is based upon a fit to many such lines (both skylines and arc lamp lines).} we assume that the smallest rms amongst the individual skyline measurements
is indicative of the relative wavelength calibration accuracy.  At 0.024 pixels at 8885 \AA, this suggests a relative fiber-to-fiber
wavelength calibration accuracy of better than 1.2 \kms\ rms.

\subsection{Typical depth}



\begin{figure*}
\epsscale{0.9}
\plotone{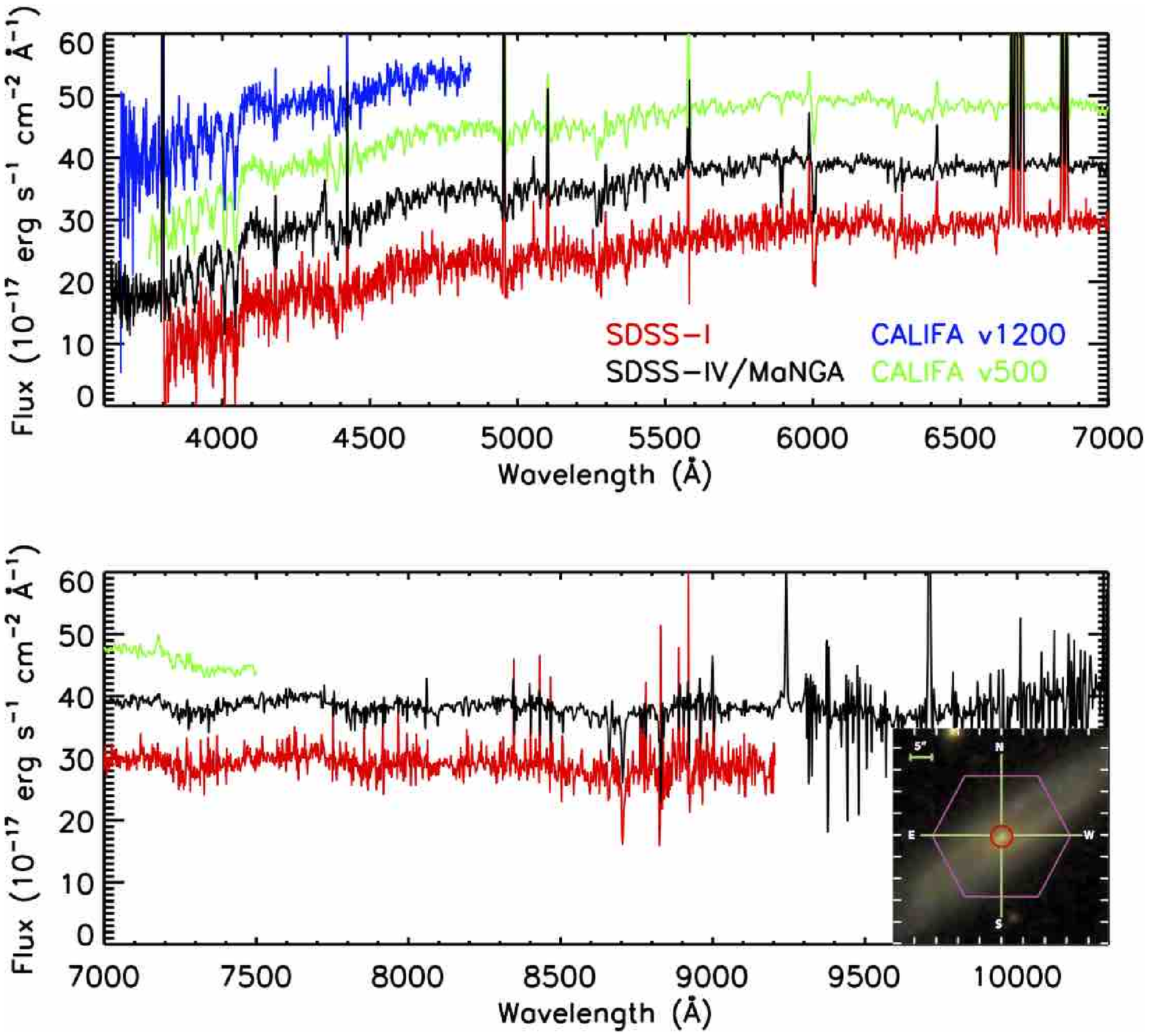}
\caption{Central spectrum of galaxy 7443-12704 (UGC 09873) extracted from the SDSS-IV MaNGA IFU data cube (black line) compared to 
the co-located SDSS-I single fiber spectrum (red line).  For comparison we also include spectra extracted from the CALIFA high-resolution
(v1200; blue line) and low-resolution (v500; green line) IFU data cubes.  SDSS-I and CALIFA spectra have been offset 
 vertically from the MaNGA spectrum to aid visual inspection.  
 The inset at lower right shows the SDSS 3-color image of UGC 09873 along with an indication of the MaNGA IFU
footprint (pink hexagon) and circular spectral extraction region (red circle).}
\label{cubecalib.fig}
\end{figure*}

Finally, we illustrate the overall quality of the MaNGA spectral data by comparing the spectrum of the central region of galaxy
7443-12704 (aka UGC 09873) from the MaNGA commissioning plate against previous SDSS-I single-fiber and CALIFA\footnote{Based 
on observations collected at the Centro Astron{\'o}mico Hispano Alem{\'a}n (CAHA) at Calar Alto, operated jointly by the 
Max-Planck-Institut f{\H u}r Astronomie and the Instituto de Astrof{\'i}sica de Andaluc'a (CSIC).  See http://califa.caha.es/}
DR-2 \citep{califadr2, walcher14, sanchez12} IFU observations of the same galaxy.
Such a direct comparison is intrinsically difficult as the total flux in a given circular aperture is strongly affected by both the observational seeing and chromatic differential refraction (for SDSS-I)
and by the effective spatial resolution of the reconstruction data cubes (MaNGA and CALIFA), especially in regions of the galaxy where there is a strong gradient in the intrinsic surface brightness
(i.e., near the center).  
This method is therefore good for comparing the {\it relative} shapes of spectra from different surveys, but not the overall normalization of the flux calibration \citep[which should instead be assessed
through PSF-matched broadband imaging, e.g.,][]{yan16}.

In this case, the SDSS-I spectrum (observed in May 2004, and obtained from the DR-12 Science Archive Server) 
corresponds to a circular fiber with core diameter 3 arcsec observed in $\sim 1.6$ arcsec seeing.
In contrast, the MaNGA and CALIFA cubes have an effective FWHM of $\sim 2.5$ arcsec, meaning that for a centrally concentrated source there will be systematically less flux 
within a 3 arcsec diameter aperture
within these cubes than in the original SDSS-I single-fiber spectrum.  We therefore extract the corresponding MaNGA and CALIFA spectra in a 5-arcsec diameter circular aperture
about the nominal location of the SDSS-I spectrum, and additionally allow for a constant multiplicative scaling factor between all of the spectra (derived from the average ratio of the spectra
interpolated to a common wavelength solution).

In Figure \ref{cubecalib.fig} we plot the resulting spectra for the SDSS-I (red line), SDSS-IV/MaNGA (black line), and CALIFA $R \sim 850$ (green line) and $R \sim 1650$ (blue line) data.
Although we cannot assess the absolute flux calibration from this plot, we note that the {\it relative} flux calibration between the four spectra is in extremely good agreement.
In the regions of common wavelength coverage, all four spectra show similar structure in the continuum and the emission/absorption lines, with the exception of a known downturn due to vignetting in the CALIFA low-resolution spectrum
 longwards of 7100 \AA.
Figure \ref{cubecalib.fig} also clearly demonstrates the longer wavelength baseline and higher SNR (especially in the far blue) 
of the MaNGA data compared to both SDSS-I and CALIFA.

\begin{figure}
\epsscale{1.2}
\plotone{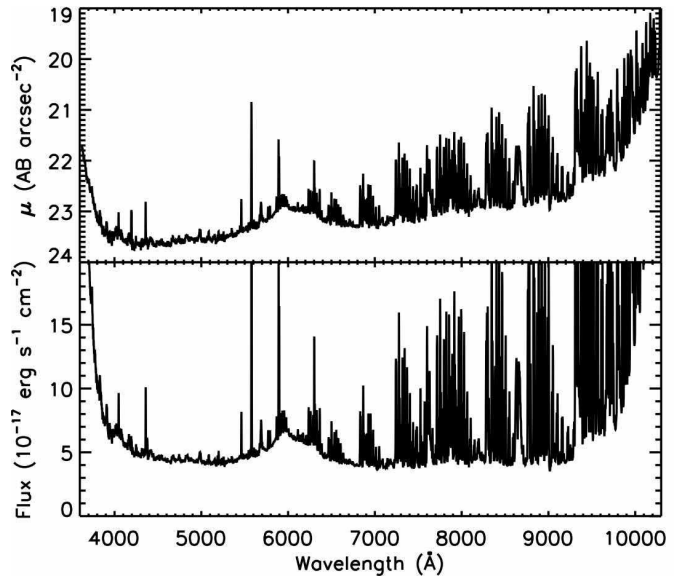}
\caption{Top panel: MaNGA $10\sigma$ limiting continuum surface brightness sensitivity within a 5 arcsec diameter aperture.  Bottom panel:
MaNGA $5\sigma$ limiting line sensitivity for a spectrally unresolved emission line in a 5 arcsec diameter aperture.
Both panels are based on the off-axis region far from the edge-on galaxy 8465-12704.}
\label{limflux.fig}
\end{figure}

Additionally, we estimate the typical sensitivity of the MaNGA data cubes based on the inverse variance reported by the pipeline for regions far along the minor axis away from edge-on disk galaxy 8465-12704.  We estimate the typical continuum surface brightness sensitivity by taking the square root of the sum of the variances of cube spaxels within 5 arcsec diameter region, 
multiplying by a covariance correction
factor based on the number of spatial elements summer (see Eqn. \ref{eq:covariance_calibration}), and converting the resulting
$1\sigma$ flux sensitivity to a $10\sigma$ sensitivity in terms of AB surface brightness.
Similarly, to determine the typical $5\sigma$ point source emission line sensitivity we 
sum the variance over twice the FWHM of the line spread function, sum over a 5 arcsec diameter aperture, and multiply the square
root of this by a covariance correction factor.  We note that both sensitivity estimates include only noise from the detector and background
sky, and do not account for any additional noise that may be introduced by astrophysical sources.
As illustrated in Figure \ref{limflux.fig}, the derived sensitivities within a 5 arcsec diameter aperture 
are strong functions of wavelength, varying from about 23.5 AB arcsec$^{-2}$ and $5 \times 10^{-17}$ erg s$^{-1}$ cm$^{-2}$
at blue wavelengths to about 20 AB arcsec$^{-2}$ and $2 \times 10^{-16}$ erg s$^{-1}$ cm$^{-2}$ in the vicinity of the strongest OH skylines.





\section{Summary}
\label{summary.sec}

The 13th data release of the Sloan Digital Sky Survey includes the raw MaNGA spectroscopic data, the fully reduced spectrophotometrically calibrated
data, and the pipeline software and metadata required for individual users to re-reduce the data themselves.
In this work, we have described the framework and algorithms of the MaNGA Data Reduction Pipeline software \mangadrp\ version v1\_5\_4
and the format and quality of the ensuing reduced data products.
The DRP operates in two stages; the first stage performs optimal extraction, sky subtraction, and flux calibration of
individual frames, while the second combines multiple frames together with astrometric information to create
calibrated individual fiber spectra (in a row-stacked format) and rectified coadded 
data cubes for each target galaxy.  The row-stacked spectra and coadded data cubes are provided for both a linear and a
logarithmically sampled wavelength grid, both covering the wavelength range 3622 - 10,354 \AA.

For the 1390 galaxy data cubes released in DR13 we demonstrate that the MaNGA data 
have nearly Poisson-limited sky subtraction shortward of $\sim$ 8500 \AA, with a residual pixel value distribution
in all-sky test plates nearly consistent with a Gaussian distribution whose width is determined by the expected contributions from detector
and Poisson noise.

Each MaNGA exposure is flux calibrated independently of all other exposures using mini-bundles placed on spectrophotometric
standard stars;  based on comparison to broadband imaging
the composite data cubes have a typical relative calibration of 1.7\% (between \Hb\ and \Ha) with an
absolute calibration of better than 5\% for more than 89\% of the MaNGA wavelength range.
These data cubes reach a typical $10\sigma$ limiting continuum surface brightness $\mu = 23.5$ AB arcsec$^{-2}$ in a five arcsec
diameter aperture in the $g$ band.
Additionally, we have demonstrated that:

\begin{itemize}

\item The wavelength calibration of the MaNGA data has an absolute accuracy of 5 \kms\ rms with a relative fiber-to-fiber accuracy of better than 1 \kms\ rms.

\item The astrometric accuracy of the reconstructed MaNGA data cubes is typically 0.1 arcsec rms, based on comparison to previous SDSS broadband imaging.

\item The spatial resolution of the MaNGA data is a function of the observational seeing, with a median of 2.54 arcsec FWHM.  We have shown that the effective reconstructed point source profile is well described by a single gaussian whose parameters are given in the header of each data cube.

\item The spectral resolution of the MaNGA data is a function of both both fiber number and wavelength, but has a median 
 $\sigma = 72$ \kms.
 
\end{itemize}

Despite these overall successes of the MaNGA DRP, 
we conclude by noting that there is still ample room for future improvements to be made in some key areas.
First, sky subtraction (while adequate for most purposes) shows some non-gaussianities in the residual distribution,
a slight overstimate in the read noise of one camera, and a possible systematic oversubtraction at the $\sim 0.1\sigma$ level
in the blue.  Work is ongoing to test whether better treatment of amplifier crosstalk or the scattered light model
can improve limiting performance in this area for the purposes of extremely deep spectral stacking.  Secondly, 
the spectral line spread functions given in the DR13 data products (and in previous SDSS optical fiber spectra) are effectively under-reported by about 10\%.  Work
is currently underway to use high spectral resolution observations of MaNGA target galaxies to constrain this effect
more precisely and fix it in future data releases.  Third, spatial covariance in the reconstructed data cubes (treated here
by a simple functional approximation) can also be treated more completely.  Finally, with additional data it will be possible
to fine tune the MaNGA quality control algorithms (which currently can be overly aggressive in flagging potentially
problematic cases) and likely recover some of the objects whose reduced data have been identified as unreliable for use
in DR13.

\acknowledgements 

This work was supported by the World Premier International Research Center Initiative (WPI Initiative), MEXT, Japan.
AW acknowledges support of a Leverhulme Trust Early Career Fellowship.
MAB acknowledges support from NSF AST-1517006.  G.B. is supported by CONICYT/FONDECYT, Programa de Iniciacion, Folio 11150220.
Funding for the Sloan Digital Sky Survey IV has been provided by the
Alfred P. Sloan Foundation and the Participating Institutions. SDSS-IV
acknowledges support and resources from the Center for
High-Performance Computing at the University of Utah. The SDSS web
site is www.sdss.org.

SDSS-IV is managed by the Astrophysical Research Consortium for the
Participating Institutions of the SDSS Collaboration including the
Carnegie Institution for Science, Carnegie Mellon University, the
Chilean Participation Group, Harvard-Smithsonian Center for
Astrophysics, Instituto de Astrof\'isica de Canarias, The Johns Hopkins
University, Kavli Institute for the Physics and Mathematics of the
Universe (IPMU) / University of Tokyo, Lawrence Berkeley National
Laboratory, Leibniz Institut f\"ur Astrophysik Potsdam (AIP),
Max-Planck-Institut f\"ur Astrophysik (MPA Garching),
Max-Planck-Institut f\"ur Extraterrestrische Physik (MPE),
Max-Planck-Institut f\"ur Astronomie (MPIA Heidelberg), National
Astronomical Observatory of China, New Mexico State University, New
York University, The Ohio State University, Pennsylvania State
University, Shanghai Astronomical Observatory, United Kingdom
Participation Group, Universidad Nacional Aut\'onoma de M\'exico,
University of Arizona, University of Colorado Boulder, University of
Portsmouth, University of Utah, University of Washington, University
of Wisconsin, Vanderbilt University, and Yale University.

\appendix{}


\section{Key Differences between \mangadrp\ and \idlspec}
\label{pipeevol.sec}

As discussed in previous section, the 2d stage of the MaNGA DRP (i.e., raw data through flux calibrated individual exposures)
is derived in large part from the \idlspec\ software that has been widely used in one form or another 
from the original SDSS spectroscopic survey
\citep{dr1.ref}, to the BOSS and eBOSS surveys \citep{dawson13,dawson16}, to the DEEP2 survey \citep{newman13}.  
Given this legacy, we summarize here for ease of reference
the key differences between our implementation of this code and its implementation during the BOSS survey for DR12.

\begin{itemize}

\item Spectral Preprocessing (\S \ref{preproc.sec}): \mangadrp\ and \idlspec\ use nearly identical algorithms, except that for MaNGA the cosmic-ray identification routine is run twice to flag additional features missed the first time.

\item Spatial Fiber Tracing (\S \ref{tracing.sec}): The \mangadrp\ fiber tracing code is substantially different from that used
by \idlspec.  For BOSS, the initial fiber locations in the starting row were determined by locating peaks and determining which
block of fibers a given peak must belong to (and which fibers were missing) 
based on the known (and constant) number of fibers in each v-groove block.
This method proved unreliable for MaNGA given the variable number of fibers per block and different potential
failure modes (in particular, if a large IFU falls out of the plate during observations there can be large regions of the
detector with only the block-edge sky fibers plugged).  After implementing a cross-correlation technique based on
the known nominal locations of each fiber, the MaNGA tracing routine has proven robust against all hardware failure modes.

The fine-adjustment of the flux-weighted fiber centroids in each row using cross-correlation of a gaussian model is also
new to the \mangadrp\ code.

\item Scattered Light (\S \ref{scattered.sec}): The bspline scattered light routine implemented in \mangadrp\ for bright-time data and flatfields
is entirely new compared to \idlspec.

\item Spectral Extraction (\S \ref{extraction.sec}): The spectral extraction technique used by \mangadrp\ is similar 
to that of \idlspec.  However, MaNGA uses the C-based implementation of the extraction used by the original SDSS-I
survey (which extracts an entire detector row at a time) while BOSS and eBOSS use an 
IDL-based implementation which operates on a given v-groove block of fibers at a time.  We found the latter to be undesirable
for MaNGA since discrete processing of individual blocks can produce discontinuities in the background term that can be
seen in the reduced all-sky data when a given IFU covers more than one block.

Additionally, MaNGA fits the derived fiber widths in a given v-groove block by a
linear relation as a function of fiberid where BOSS uses a constant value for each block.

\item Fiber Flatfield (\S \ref{flats.sec}): The fiber flatfield technique is nearly identical between \mangadrp\ and \idlspec.

\item Wavelength and LSF calibration (\S \ref{wavecal.sec}): The initial wavelength solution and LSF estimate
based on the arclamp calibration frames is nearly identical between \mangadrp\ and \idlspec, with the exception that MaNGA
fits the derived LSF in a given v-groove block by a
linear relation as a function of fiberid where BOSS uses a constant value for each block.

\item Science Frame extraction (\S \ref{sciextract.sec}): The science frame extraction process is largely similar between
\mangadrp\ and \idlspec, with the exception that BOSS makes no correction to the derived arcline LSF based on the skylines.

\item Sky subtraction (\S \ref{skysub.sec}): Although the general approach to sky subtraction is similar between \mangadrp\ and \idlspec, in the sense that both use basis splines to build a super-sampled sky model, the practical implementation differs
substantially.  This difference is largely due to the fundamental hardware differences between the two surveys; where BOSS has 1000 fibers (science plus sky and standard) distributed nearly randomly across the entire 3 degree field, MaNGA effectively has
large groups of fibers clustered at the same few locations on-sky with outrigger sky fibers surrounding them.  This means that
MaNGA samples a more discrete and discontinuous assortment of background sky locations, but can similarly use the locality
of sky and IFU fibers to contrain the background local to a given IFU.  In contrast to the assortment of scaling factors, smoothed inverse variance weighting,  local sky
adjustments, and 1d and 2d sky models used by MaNGA, BOSS simply uses a 2d basis spline model of the sky background
evaluated at the wavelengths of each fiber (although we note that eBOSS has also recently adopted a smoothed inverse variance 
weighting scheme similar to ours in order to avoid systematic undersubtraction of the sky background present in the
previous BOSS reductions).

\item Flux calibration (\S \ref{fluxcal.sec}): As discussed by \citet{yan16}, flux calibration techniques differ substantially between
\mangadrp\ and \idlspec\ since MaNGA and BOSS are attempting to solve different problems.  While BOSS must correct
for both system throughput losses and geometric fiber aperture losses, MaNGA must disentangle the two and correct only
for system losses.  Although the core of the stellar spectral library comparison is thus shared between the two codes, the 
implementation differs dramatically.

\item Wavelength rectification (\S \ref{waverect.sec}): The spline-based approach to the wavelength rectification is common
between both \mangadrp\ and \idlspec, but MaNGA uses a smoothed inverse-variance weighting approach where BOSS used 
simple inverse variance weighting (this has since been updated to smoothed inverse variance for eBOSS).  MaNGA also uses
a slightly different breakpoint spacing, and evaluates the bspline fit on both a logarithmic and a linear wavelength solution.  The second-pass cosmic ray identification by growing the previous cosmic ray mask is also unique to MaNGA.

\item Quality control (\S \ref{qc.sec}): The DRP2QUAL infrastructure to evaluate frame quality and stop reduction at various points
if necessary is entirely new to \mangadrp.

\end{itemize}


\section{MaNGA Data Model}
\label{datamodel.sec}

We provide here for convenient reference an overview of the primary data products delivered by the MaNGA DRP.
These are in the format of gzipped multi-extension FITS files, with a mixture of image data and binary table extensions.
For a detailed description including definitions of keyword headers see the online DR13 documentation at 
http://www.sdss.org/dr13/manga/manga-data/data-model/ .  This Appendix is split into 4 sections: \S \ref{append.int.sec}
describes the intermediate (2d DRP) products, \S \ref{append.final.sec} describes the final (3d DRP) products, \S \ref{drpall.sec} describes the 'drpall' summary table product, and \S \ref{bitmasks.sec} describes the key 3d pipeline quality bitmasks.

\subsection{Intermediate DRP data products}
\label{append.int.sec}

\begin{figure*}
\plotone{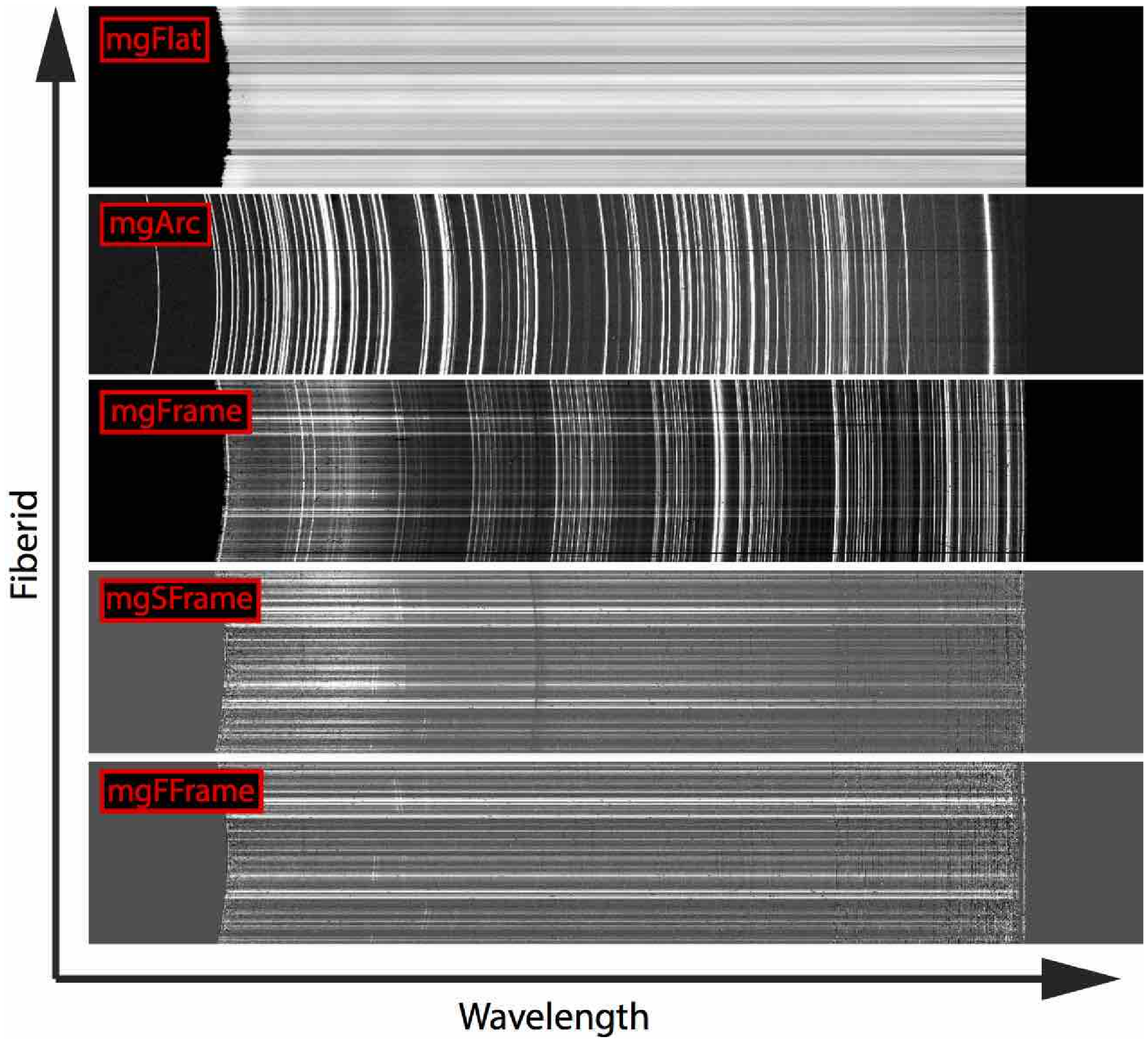}
\caption{MaNGA intermediate data products from individual exposures.  Shown here are extracted fiber flats (mgFlat), arc lamp spectra
(mgArc), extracted science frame spectra (mgFrame), sky-subtracted science frame spectra (mgSFrame), and flux-calibrated science
frame spectra (mgFFrame).  Note the curvature of the wavelength solution along the spectroscopic slit.  The examples shown here are for the r2 camera.  The greyscale stretch on the fiberflat image runs from 0.6 to 1.1.}
\label{filetypes.fig}
\end{figure*}

The intermediate data products are produced by the 2d stage of the MaNGA DRP.  These products are output during the calibration, flux extraction, sky subtraction, and flux calibration stages of the pipeline.  In Figure \ref{filetypes.fig} we show examples of the primary data extension of these types of files.  In Tables~\ref{mgarc.tab} - \ref{mgcframe.tab} we give the structure of the intermediate and calibration FITS files.  For the intermediate data products, the naming convention includes the camera name (except for the camera-combined mgCFrame file), and the zero-padded exposure number.  Since the MaNGA instrument has two spectrographs each with a red and blue camera, there are four camera designations: b1, r1, b2, r2.  

\subsubsection{mgArc}

These are the extracted arc frames, produced during wavelength calibration.  The format is similar to the BOSS spArc file, with the exception of a blank extension 0 and extension names instead of numbers.

{\bf Written by:}  https://svn.sdss.org/public/repo/manga/mangadrp/tags/v1\_5\_4/pro/spec2d/mdrp\_calib.pro

{\bf Data Model:} https://data.sdss.org/datamodel/files/MANGA\_SPECTRO\_REDUX/DRPVER/PLATE4/MJD5/mgArc.html

 \begin{deluxetable}{cccl}[ht]
\tabletypesize{\footnotesize}
\tablecolumns{4} 
\tablewidth{0pt}
 \tablecaption{mgArc-[camera]-[exposure] Data Structure
 \label{mgarc.tab}}
 \tablehead{
 \colhead{HDU} & \colhead{Extension Name} & \colhead{Format} & \colhead{Description}}
 \startdata 
 0 & ... & ... & Empty except for global header \\
 1 & FLUX & [CCDROW $\times$ NFIBER] & Extracted arclamp spectra \\
 2 & LXPEAK & [NFIBER+1 $\times$ NLAMP] & Wavelengths and x positions of arc lamp lines. \\
 3 & WSET & [BINARY FITS TABLE] & Wavelength solution as Legendre polynomials for all fibers \\
 4 & MASK & [NFIBER] & Fiber bit mask (MANGA\_DRP2PIXMASK) \\
 5 & DISPSET & [BINARY FITS TABLE] & Spectral LSF ($1\sigma$) in pixels as Legendre polynomials for each fiber \\
 \enddata
\tablecomments{NFIBER is the number of fibers in the camera, CCDROW the number of rows on the detector, and NLAMP the number of bright arc lines.}
\end{deluxetable}

\subsubsection{mgFlat}

These are the extracted flatfield frames, produced after the fiber tracing, wavelength calibration, and global quartz
lamp spectrum have been removed.  The format is similar to the BOSS spFlat files, with the exception of a blank extension 0 and extension names instead of numbers.
   
{\bf Written by:}  https://svn.sdss.org/public/repo/manga/mangadrp/tags/v1\_5\_4/pro/spec2d/mdrp\_calib.pro

{\bf Data Model:} https://data.sdss.org/datamodel/files/MANGA\_SPECTRO\_REDUX/DRPVER/PLATE4/MJD5/mgFlat.html
   
 \begin{deluxetable}{cccl}[ht]
\tabletypesize{\footnotesize}
\tablecolumns{4} 
\tablewidth{0pt}
 \tablecaption{mgFlat-[camera]-[exposure] Data Structure
 \label{mgflat.tab}}
 \tablehead{
 \colhead{HDU} & \colhead{Extension Name} & \colhead{Format} & \colhead{Description}}
 \startdata 
 0 & ... & ... & Empty except for global header \\
 1 & FLUX & [CCDROW $\times$ NFIBER] & Extracted flatfield lamp spectra \\
 2 & TSET & [BINARY FITS TABLE] & Legendre polynomial traceset containing the x,y centers of the fiber traces\\
3 & MASK & [NFIBER] &  Fiber bit mask (MANGA\_DRP2PIXMASK) \\
4 & WIDTH & [CCDROW $\times$ NFIBER] &  Profile cross-dispersion width ($1\sigma$) of each fiber \\
5 & SUPERFLATSET & [BINARY FITS TABLE] & Legendre polynomial traceset describing the quartz lamp response function \\
 \enddata
\end{deluxetable}

\subsubsection{mgFrame}

These are the extracted fiber spectra for each camera for the science exposures.

{\bf Written by:}  https://svn.sdss.org/public/repo/manga/mangadrp/tags/v1\_5\_4/pro/spec2d/mdrp\_extract\_object.pro

{\bf Data Model:} https://data.sdss.org/datamodel/files/MANGA\_SPECTRO\_REDUX/DRPVER/PLATE4/MJD5/mgFrame.html

 \begin{deluxetable}{cccl}[ht]
\tabletypesize{\footnotesize}
\tablecolumns{4} 
\tablewidth{0pt}
 \tablecaption{mgFrame-[camera]-[exposure] Data Structure
 \label{mgframe.tab}}
 \tablehead{
 \colhead{HDU} & \colhead{Extension Name} & \colhead{Format} & \colhead{Description}}
 \startdata 
 0 & ... & ... & Empty except for global header \\
1 & FLUX & [CCDROW $\times$ NFIBER] & Extracted spectra in units of flatfielded electrons \\
2 & IVAR & [CCDROW $\times$ NFIBER] & Inverse variance of the extracted spectra \\
3 & MASK & [CCDROW $\times$ NFIBER] & Pixel mask (MANGA\_DRP2PIXMASK) \\
4 & WSET & [BINARY FITS TABLE] & Legendre polynomial coefficients describing wavelength solution \\
& & & in log$_{10}$ \AA\ (vacuum heliocentric) \\
5 & DISPSET & [BINARY FITS TABLE] & Legendre polynomial coefficients describing spectral LSF \\
& & & ($1\sigma$) in pixels \\
6 & SLITMAP & [BINARY FITS TABLE] & Slitmap structure describing plugged plate configuration \\
7 & XPOS & [CCDROW $\times$ NFIBER] & X position of fiber traces on detector \\
8 & SUPERFLAT &  [CCDROW $\times$ NFIBER] & Superflat vector from the quartz lamps \\
 \enddata
\end{deluxetable}

\subsubsection{mgSFrame}

These are the science fiber spectra for each camera after the sky subtraction routine 
has been applied to the mgFrame files (the `S' in mgSFrame stands for Sky Subtracted).

{\bf Written by:}  https://svn.sdss.org/public/repo/manga/mangadrp/tags/v1\_5\_4/pro/spec2d/mdrp\_skysubtract.pro

{\bf Data Model:} https://data.sdss.org/datamodel/files/MANGA\_SPECTRO\_REDUX/DRPVER/PLATE4/MJD5/mgSFrame.html

 \begin{deluxetable}{cccl}
\tabletypesize{\footnotesize}
\tablecolumns{4} 
\tablewidth{0pt}
 \tablecaption{mgSFrame-[camera]-[exposure] Data Structure
 \label{mgsframe.tab}}
 \tablehead{
 \colhead{HDU} & \colhead{Extension Name} & \colhead{Format} & \colhead{Description}}
 \startdata 
 0 & ... & ... & Empty except for global header \\
1 & FLUX & [CCDROW $\times$ NFIBER] & Sky-subtracted spectra in units of flatfielded electrons \\
2 & IVAR & [CCDROW $\times$ NFIBER] & Inverse variance of the sky-subtracted spectra \\
3 & MASK & [CCDROW $\times$ NFIBER] & Pixel mask (MANGA\_DRP2PIXMASK) \\
4 & WSET & [BINARY FITS TABLE] & Legendre polynomial coefficients describing wavelength solution \\
& & & in log$_{10}$ \AA\ (vacuum heliocentric) \\
5 & DISPSET & [BINARY FITS TABLE] & Legendre polynomial coefficients describing spectral LSF \\
& & & ($1\sigma$) in pixels \\
6 & SLITMAP & [BINARY FITS TABLE] & Slitmap structure describing plugged plate configuration \\
7 & XPOS & [CCDROW $\times$ NFIBER] & X position of fiber traces on detector \\
8 & SUPERFLAT &  [CCDROW $\times$ NFIBER] & Superflat vector from the quartz lamps \\
9 & SKY & [CCDROW $\times$ NFIBER] & Subtracted model sky spectra in units of flatfielded electrons \\
 \enddata
\end{deluxetable}

\subsubsection{mgFFrame}

These are the science fiber spectra for each camera after the flux calibration routine has been applied to the mgSFrame
files (the `F' in mgFFrame stands for Flux calibrated).

{\bf Written by:}  https://svn.sdss.org/public/repo/manga/mangadrp/tags/v1\_5\_4/pro/spec2d/mdrp\_fluxcal.pro

{\bf Data Model:} https://data.sdss.org/datamodel/files/MANGA\_SPECTRO\_REDUX/DRPVER/PLATE4/MJD5/mgFFrame.html

 \begin{deluxetable}{cccl}
\tabletypesize{\footnotesize}
\tablecolumns{4} 
\tablewidth{0pt}
 \tablecaption{mgFFrame-[camera]-[exposure] Data Structure
 \label{mgfframe.tab}}
 \tablehead{
 \colhead{HDU} & \colhead{Extension Name} & \colhead{Format} & \colhead{Description}}
 \startdata 
 0 & ... & ... & Empty except for global header \\
1 & FLUX & [CCDROW $\times$ NFIBER] & Flux calibrated spectra in units of $10^{-17}$ erg s$^{-1}$ cm$^{-2}$ \AA$^{-1}$ fiber$^{-1}$ \\
2 & IVAR & [CCDROW $\times$ NFIBER] & Inverse variance of the flux-calibrated spectra \\
3 & MASK & [CCDROW $\times$ NFIBER] & Pixel mask (MANGA\_DRP2PIXMASK) \\
4 & WSET & [BINARY FITS TABLE] & Legendre polynomial coefficients describing wavelength solution \\
& & & in log$_{10}$ \AA\ (vacuum heliocentric) \\
5 & DISPSET & [BINARY FITS TABLE] & Legendre polynomial coefficients describing spectral LSF \\
& & & ($1\sigma$) in pixels \\
6 & SLITMAP & [BINARY FITS TABLE] & Slitmap structure describing plugged plate configuration \\
7 & XPOS & [CCDROW $\times$ NFIBER] & X position of fiber traces on detector \\
8 & SUPERFLAT &  [CCDROW $\times$ NFIBER] & Superflat vector from the quartz lamps \\
9 & SKY & [CCDROW $\times$ NFIBER] & Subtracted model sky spectra in units of $10^{-17}$ erg s$^{-1}$ cm$^{-2}$ \AA$^{-1}$ fiber$^{-1}$ \\
 \enddata
\end{deluxetable}

\subsubsection{mgCFrame}

These are the science fiber spectra after the individual-camera flux calibrated mgFFrame files have been combined together across the dichroic break and fibers from spectrograph 2 have been appended atop those from spectrograph 1 (i.e., in order of increasing fiberid).  All spectra in this file have been resampled to a common wavelength grid across the entire MaNGA survey using a basis spline technique described in \S \ref{waverect.sec} (the `C' in mgCFrame stands for Calibrated and Camera Combined on a Common wavelength grid).
There are two versions of this file; the first uses a logarithmic wavelength sampling from log10($\lambda$/\AA)=3.5589 to 4.0151 (NWAVE=4563 spectral elements).  The second uses a linear wavelength sampling running from 3622.0 to 10353.0 \AA\ (NWAVE=6732 spectral elements). 

{\bf Written by:}  https://svn.sdss.org/public/repo/manga/mangadrp/tags/v1\_5\_4/pro/spec2d/mdrp\_combinecameras.pro

{\bf Data Model:} https://data.sdss.org/datamodel/files/MANGA\_SPECTRO\_REDUX/DRPVER/PLATE4/MJD5/mgCFrame.html

 \begin{deluxetable}{cccl}
\tabletypesize{\footnotesize}
\tablecolumns{4} 
\tablewidth{0pt}
 \tablecaption{mgCFrame-[exposure] Data Structure
 \label{mgcframe.tab}}
 \tablehead{
 \colhead{HDU} & \colhead{Extension Name} & \colhead{Format} & \colhead{Description}}
 \startdata 
 0 & ... & ... & Empty except for global header \\
1 & FLUX & [NWAVE $\times$ NFIBER] & Camera combined, resampled spectra in units of $10^{-17}$ erg s$^{-1}$ cm$^{-2}$ Ang$^{-1}$ fiber$^{-1}$ \\
2 & IVAR & [NWAVE $\times$ NFIBER] & Inverse variance of the camera-combined spectra \\
3 & MASK & [NWAVE $\times$ NFIBER] & Pixel mask (DRP2PIXMASK) \\
4 & WAVE &  [NWAVE] & Wavelength vector in units of \AA (vacuum heliocentric) \\
5 & DISP & [NWAVE $\times$ NFIBER]  & Spectral resolution  ($1\sigma$ LSF) in units of \AA \\
6 & SLITMAP & [BINARY FITS TABLE] & Slitmap structure describing plugged plate configuration \\
9 & SKY & [NWAVE $\times$ NFIBER] & Resampled model sky spectra in units of  $10^{-17}$ erg s$^{-1}$ cm$^{-2}$ Ang$^{-1}$ fiber$^{-1}$ \\
 \enddata
 \tablecomments{Both LINEAR and LOG format versions of this file are produced, with either logarithmic or linear wavelength sampling respectively.  
 NWAVE is the total number of wavelength channels (6732 for LINEAR, 4563 for LOG).  NFIBER $=$ 1423 total fibers.}
\end{deluxetable}

\subsection{Final DRP data products}
\label{append.final.sec}

Depending on the science case, different final summary products are desirable.  The MaNGA DRP provides both RSS files and regularly-gridded combined data cubes, with both logarithmic and linear wavelength solutions.

These have the naming convention of {\bf manga-[PLATEID]-[IFUDESIGN]-[BIN][MODE].fits.gz}.  
PLATEID refers to the four or five digit plate identifer.  IFUDESIGN refers to the design id of the IFU bundle.  BIN refers to the wavelength sampling of the output data product, LOG for logarithmic sampling, or LIN for linear sampling.  MODE refers to the output structure, whether an RSS file or a CUBE file.  The combination of plateID-ifuDesign provides a unique identifier to a MaNGA target, and output final-DRP products.  While the identifier of manga-id maps to a unique galaxy, it does not map to a unique set of output data products.  If a given galaxy is observed on more than one plate, it will have different final-DRP outputs associated with it by default. 

The RSS files (Table \ref{rss.tab}) are a two-dimensional array in row-stacked-spectra format with horizontal size $N_{\rm spec}$ and vertical size $N = \sum N_{\rm fiber} (i)$ where $N_{\rm fiber} (i)$ is the number of fibers in the IFU targeting this galaxy for the $i$'th exposure and the sum runs over all exposures.  In contrast, the cubes (Table \ref{cubes.tab}) are three-dimensional arrays in which the first and second dimensions are spatial (with regular 0.5 arcsec square spaxels) and the third dimension represents wavelength.

In each case, there are associated image extensions describing the inverse variance, pixel mask, and a binary table `OBSINFO' that describes full information about each exposure that was combined to produce the final file (exposure number, integration time, hour angle, seeing, etc.). This structure is appended to each file with one line per exposure (Table \ref{obsinfo.tab}) both for quality control purposes (so that delivered data can be tracked back to individual exposures easily), and so that future forward modeling efforts
can read from this extension everything necessary to know about the instrument and observing configuration of each exposure.

Additionally, each RSS-format file has an extension listing the effective X \& Y position (calculated by the astrometry module) corresponding to each element in the flux array.  Because of chromatic DAR, each wavelength for a given fiber has a slightly different position, and therefore the positional arrays have the same dimensionality as the corresponding flux array.  Each data cube also has four extensions corresponding to reconstructed broadband images 
obtained by convolving the data cube with the SDSS $griz$ filter response
functions, and four extensions illustrating the reconstructed PSF in the $griz$ bands (see discussion in \S \ref{angularres.sec}).

As detailed by http://www.sdss.org/dr13/manga/manga-data/data-model/ 
there are an assortment of FITS header keycards specifying information such as World Coordinate Systems (WCS),
average reconstructed PSF FWHM in $griz$ bandpasses, total exposure time, Milky Way dust extinction, etc.
The WCS adopted for the logarithmic wavelength solution follows the CTYPE=WAVE-LOG 
convention \citep{greisen06} convention in which

\begin{equation}
\lambda = {\rm CRVALi} \times {\rm exp}({\rm CDi\_i} \times {\rm (p-CRPIXi)/CRVALi})
\end{equation}

{\bf Written by:}  https://svn.sdss.org/public/repo/manga/mangadrp/tags/v1\_5\_4/pro/spec3d/mdrp\_reduceoneifu.pro

{\bf RSS Data Model:} https://data.sdss.org/datamodel/files/MANGA\_SPECTRO\_REDUX/DRPVER/PLATE4/stack/manga-RSS.html

{\bf CubeData Model:} https://data.sdss.org/datamodel/files/MANGA\_SPECTRO\_REDUX/DRPVER/PLATE4/stack/manga-CUBE.html

 \begin{deluxetable}{cccc}
\tabletypesize{\footnotesize}
\tablecolumns{4} 
\tablewidth{0pt}
 \tablecaption{manga-[plate]-[ifudesign]-LOGRSS Data Structure
 \label{rss.tab}}
 \tablehead{
 \colhead{HDU} & \colhead{Extension Name} & \colhead{Format} & \colhead{Description}}
 \startdata 
  0 & ... & ... & Empty except for global header \\
1 & FLUX & [NWAVE $\times$ (NFIBER $\times$ NEXP)] & Row-stacked spectra in units of $10^{-17}$ erg s$^{-1}$ cm$^{-2}$ \AA$^{-1}$ fiber$^{-1}$ \\
2 & IVAR &  [NWAVE $\times$ (NFIBER $\times$ NEXP)]  & Inverse variance of row-stacked spectra \\
3 & MASK & [NWAVE $\times$ (NFIBER $\times$ NEXP)]  & Pixel mask (MANGA\_DRP2PIXMASK) \\
4 & DISP & [NWAVE $\times$ (NFIBER $\times$ NEXP)]  & Spectral LSF  ($1\sigma$) in units of \AA \\
5 & WAVE & [NWAVE] & Wavelength vector in units of \AA (vacuum heliocentric) \\
6  & SPECRES & [NWAVE] & Median spectral resolution vs wavelength \\
7 & SPECRESD & [NWAVE] & Standard deviation ($1\sigma$) of spectral resolution vs wavelength  \\
8 & OBSINFO  & [BINARY FITS TABLE] & Table detailing exposures combined to create this file. \\
9 & XPOS &  [NWAVE $\times$ (NFIBER $\times$ NEXP)]  & Array of fiber X-positions (units of arcsec) relative to the IFU center \\
10 & YPOS &  [NWAVE $\times$ (NFIBER $\times$ NEXP)]  & Array of fiber Y-positions (units of arcsec) relative to the IFU center \\
 \enddata
 \tablecomments{Both LINEAR and LOG format versions of this file are produced, with either logarithmic or linear wavelength sampling respectively.  
 NWAVE is the total number of wavelength channels (6732 for LINEAR, 4563 for LOG).  NFIBER is the number of fibers in the IFU, NEXP is the number of exposures.}
\end{deluxetable}

 \begin{deluxetable}{cccc}
\tabletypesize{\footnotesize}
\tablecolumns{4} 
\tablewidth{0pt}
 \tablecaption{manga-[plate]-[ifudesign]-LOGCUBE Data Structure
 \label{cubes.tab}}
 \tablehead{
 \colhead{HDU} & \colhead{Extension Name} & \colhead{Format} & \colhead{Description}}
 \startdata 
 0 & ... & ... & Empty except for global header \\
1 & FLUX & [NX $\times$ NY $\times$ NWAVE] & 3d rectified cube in units of $10^{-17}$ erg s$^{-1}$ cm$^{-2}$ \AA$^{-1}$ spaxel$^{-1}$ \\
2 & IVAR &  [NX $\times$ NY $\times$ NWAVE]  & Inverse variance cube \\
3 & MASK &  [NX $\times$ NY $\times$ NWAVE] & Pixel mask cube (MANGA\_DRP3PIXMASK) \\
4 & WAVE & [NWAVE] & Wavelength vector in units of \AA (vacuum heliocentric) \\
5  & SPECRES & [NWAVE] & Median spectral resolution vs wavelength \\
6 & SPECRESD & [NWAVE] & Standard deviation ($1\sigma$) of spectral resolution vs wavelength \\
7 & OBSINFO  & [BINARY FITS TABLE] & Table detailing exposures combined to create this file. \\
8 & GIMG & [NX $\times$ NY] &  Broadband SDSS $g$ image created from the data cube \\
9 & RIMG & [NX $\times$ NY] &  Broadband SDSS $r$ image created from the data cube \\
10 & IIMG & [NX $\times$ NY] &  Broadband SDSS $i$ image created from the data cube \\
11 & ZIMG & [NX $\times$ NY] &  Broadband SDSS $z$ image created from the data cube \\
12 & GPSF & [NX $\times$ NY] &  Reconstructed SDSS $g$ point source response profile \\
13 & RPSF & [NX $\times$ NY] &  Reconstructed SDSS $r$ point source response profile \\
14 & IPSF & [NX $\times$ NY] &  Reconstructed SDSS $i$ point source response profile \\
15 & ZPSF & [NX $\times$ NY] &  Reconstructed SDSS $z$ point source response profile \\
 \enddata
 \tablecomments{Both LINEAR and LOG format versions of this file are produced, with either logarithmic or linear wavelength sampling respectively.  
 NWAVE is the total number of wavelength channels (6732 for LINEAR, 4563 for LOG).}
\end{deluxetable}

 \begin{deluxetable}{cccl}
\tabletypesize{\footnotesize}
\tablecolumns{4} 
\tablewidth{0pt}
 \tablecaption{ObsInfo Binary Table Extension
 \label{obsinfo.tab}}
 \tablehead{
 \colhead{ColumnNo} & \colhead{ColumnName} & \colhead{Format} & \colhead{Description}}
 \startdata 
1 & SLITFILE & str & Name of the slitmap \\
2 & METFILE & str & Name of the metrology file \\
3 & HARNAME & str & Harness name \\
4 & IFUDESIGN & int32 & ifudesign (e.g., 12701) \\
5 & FRLPLUG & int16 & The physical ferrule matching this part of the slit \\
6 & MANGAID & str & MaNGA identification number \\
7 & AIRTEMP & float32 & Temperature in Celsius \\
8 & HUMIDITY & float32 & Relative humidity in percent \\
9 & PRESSURE & float32 & Pressure in inHg \\
10 & SEEING & float32 & Best guider seeing in Arcsec \\
11 & PSFFAC & float32 & Best-fit PSF size relative to guider measurement \\
12 & TRANSPAR & float32 & Guider transparency \\ 
13 & PLATEID & int32 & Plate id number \\
14 & DESIGNID & int32 & Design id number \\
15 & CARTID & int16 & Cart id number \\
16 & MJD & int32 & MJD of observation \\
17 & EXPTIME & float32 & Exposure time (seconds) \\
18 & EXPNUM & str & Exposure number \\
19 & SET & int32 & Which set this exposure belongs to \\
20 & MGDPOS & str & MaNGA dither position (NSEC) \\
21 & MGDRA & float32 & MaNGA dither offset in RA (arcsec) \\
22 & MGDDEC & float32 & MaNGA dither offset in DEC (arcsec) \\
  23-27 & OMEGASET\_[UGRIZ] & float32 & Omega value of this set in ugriz bands \\
  & & & at [3622, 4703, 6177, 7496, 10354]~\AA, respectively \\
  28-39 & EAMFIT\_[PARAM] & float32 & Parameters from the Extended Astrometry Module$^a$ \\
40 & TAIBEG & str & TAI at the start of the exposure \\
41 & HADRILL & float32 & Hour angle plate was drilled for \\
42 & LSTMID & float32 & Local sidereal time at midpoint of exposure \\
43 & HAMID & float32 & Hour angle at midpoint of exposure for this IFU \\
44 & AIRMASS & float32 & Airmass at midpoint of exposure for this IFU \\
45 & IFURA & float64 & IFU right ascension (J2000) \\
46 & IFUDEC & float64 & IFU declination (J2000) \\
47 & CENRA & float64 & Plate center right ascension (J2000) \\
48 & CENDEC & float64 & Plate center declination (J2000) \\
49 & XFOCAL & float32 & Hole location in xfocal coordinates (mm) \\
50 & YFOCAL & float32 & Hole location in yfocal coordinates (mm) \\
51 & MNGTARG1 & int32 & manga\_target1 maskbit for galaxy target catalog \\ 
52 & MNGTARG2 & int32 & manga\_target2 maskbit for non-galaxy target catalog \\ 
53 & MNGTARG3 & int32 & manga\_target3 maskbit for ancillary target catalog  \\
54 & BLUESN2 & float32 & SN2 in blue for this exposure \\
55 & REDSN2 & float32 & SN2 in red for this exposure \\
56 & BLUEPSTAT & float32 & Poisson statistic in blue for this exposure \\
57 & REDPSTAT & float32 & Poisson statistic in red for this exposure \\
58 & DRP2QUAL & int32 & DRP-2d quality bitmask \\
59 & THISBADIFU & int32 & 0 if good, 1 if this IFU was bad in this frame \\
  60-63 & PF\_FWHM\_[GRIZ] & float32 & FWHM (arcsec) of a single-gaussian fitted to the point source \\
  & & & response function Prior to Fiber convolution in bands [$griz$] \\
 \enddata
 \tablecomments{\\
 a - EAM Parameters: RA, Dec, Theta, Theta0, A, B, RAerr, DECerr, ThetaErr, Theta0Err, Aerr, Berr.  \\
 See https://data.sdss.org/datamodel/files/MANGA\_SPECTRO\_REDUX/DRPVER/PLATE4/stack/manga-CUBE.html\#hdu7
 for a full description of the obsinfo data model \\
 }
\end{deluxetable}

\subsection{DRPall summary table}
\label{drpall.sec}

The 3d stage reductions of the MaNGA DRP (including calibration minibundles) are summarized in the DRPall FITS file, {\bf drpall-[version].fits}.  
This file aggregates metadata pulled from all individual reduced data cube files (plus spectrophotometric standard stars), as well as the NSA targeting catalog.  Each row in this table corresponds to an individual observation.  The DRPall summary file is a convenient place to quickly look for information regarding, for example, unique cube identifiers, achieved S/N, data quality, observing conditions, targeting bitmasks and basic NSA catalog parameters.  The complete data model for the DRPall summary file can be found at
https://data.sdss.org/datamodel/files/MANGA\_SPECTRO\_REDUX/DRPVER/drpall.html

\subsection{DRP Data Quality Bitmasks}
\label{bitmasks.sec}

The MaNGA DRP 2d pixel bitmasks applicable to individual reduced frames and composite RSS files are given in Table \ref{drp2pixmask.tab}.  These indicate the quality of entire
fibers or individual pixels within these frames, accounting for cases of broken and/or unplugged fibers, cosmic rays, sky-subtraction failures, etc.  A catch-all summary bit 3DREJECT
is set when a given pixel should be excluded from use in building a 3d composite data cube.

The MaNGA DRP 3d spaxel masks applicable to these composite data cubes are given in Table \ref{drp3pixmask.tab}.  Since these cubes combine across many individual exposures,
the 3d spaxel masks are necessarily less detailed than the 2d pixel masks, and indicate simply the overall quality of individual spaxels
within a given data cube.  This includes whether there is no coverage (i.e., outside the footprint of the IFU bundle), low coverage (near the edges of the IFU bundle), a dead
fiber (which will in turn cause low and/or no coverage within the bundle), or a foreground star that should be masked for many science analyses.  {\bf These foreground
stars are identified manually using a combination of SDSS imaging and the MaNGA data cubes, and stored in a reference list read by the DRP.}  A catch-all DONOTUSE
flag indicates a superset of all  pixels that should not be used for science.

 \begin{deluxetable}{rrll}
\tabletypesize{\footnotesize}
\tablecolumns{4} 
\tablewidth{0pt}
 \tablecaption{MANGA\_DRP2PIXMASK Data Quality Bits
 \label{drp2pixmask.tab}}
 \tablehead{
 \colhead{Bit} & \colhead{Value} & \colhead{Label} & \colhead{Description}}
 \startdata  
 \multicolumn{3}{c}{Mask bits per fiber}\\
0 &	1 &	NOPLUG &	Fiber not listed in plugmap file  \\
1 &	2 &	BADTRACE &	Bad trace \\
2 &	4 &	BADFLAT  &	Low counts in fiberflat \\
3 &	8 &	BADARC 	& Bad arc solution \\
4 &	16 &	MANYBADCOLUMNS  &	More than 10\% of pixels are bad columns \\
5 &	32 &	MANYREJECTED &	More than 10\% of pixels are rejected in extraction \\
6 &	64 &	LARGESHIFT 	& Large spatial shift between flat and object position \\
7 &	128 &	BADSKYFIBER &	Sky fiber shows extreme residuals \\
8 &	256 &	NEARWHOPPER &	Within 2 fibers of a whopping fiber (exclusive) \\
9 &	512 &	WHOPPER &	Whopping fiber, with a very bright source. \\
10 &	1024 &	SMEARIMAGE 	& Smear available for red and blue cameras \\
11 &	2048 &	SMEARHIGHSN &	S/N sufficient for full smear fit \\
12 &	4096 &	SMEARMEDSN &	S/N only sufficient for scaled median fit \\
13 &	8192 &	DEADFIBER &	Broken fiber according to metrology files \\
 \multicolumn{3}{c}{Mask bits per pixel}\\
15 &	32768 &	BADPIX &	Pixel flagged in badpix reference file. \\
16 &	65536 &	COSMIC &	Pixel flagged as cosmic ray. \\
17 &	131072 &	NEARBADPIXEL &	Bad pixel within 3 pixels of trace. \\
18 &	262144 &	LOWFLAT &	Flat field less than 0.5 \\
19 &	524288 &	FULLREJECT 	& Pixel fully rejected in extraction model fit (INVVAR=0) \\
20 &	1048576 & 	PARTIALREJECT &	Some pixels rejected in extraction model fit \\
21 &	2097152 &	SCATTEREDLIGHT 	& Scattered light significant \\
22 &	4194304 &	CROSSTALK &	Cross-talk significant \\
23 &	8388608 &	NOSKY &	Sky level unknown at this wavelength (INVVAR=0) \\
24 &	16777216 &	BRIGHTSKY &	Sky level $>$ flux + 10$\ast$(flux\_err) AND sky $>$ 1.25 $\ast$ median(sky,99 pixels) \\
25 &	33554432 &	NODATA &	No data available in combine B-spline (INVVAR=0) \\
26 &	671108864 &	COMBINEREJ 	& Rejected in combine B-spline \\
27 &	134217728 &	BADFLUXFACTOR 	& Low flux-calibration or flux-correction factor \\
28 &	268435456 &	BADSKYCHI &	Relative chi2 $>$ 3 in sky residuals at this wavelength \\
29 &	536870912 &	REDMONSTER &	Contiguous region of bad chi2 in sky residuals (with threshold of relative chi2 $>$ 3). \\
30 &	1073741824 &	3DREJECT &	Used in RSS file, indicates should be rejected when making 3D cube 
 \enddata
 \tablecomments{}
 \end{deluxetable}

 \begin{deluxetable}{rrll}
\tabletypesize{\footnotesize}
\tablecolumns{4} 
\tablewidth{0pt}
 \tablecaption{MANGA\_DRP3PIXMASK Data Quality Bits
 \label{drp3pixmask.tab}}
 \tablehead{
 \colhead{Bit} & \colhead{Value} & \colhead{Label} & \colhead{Description}}
 \startdata 
0 &	1 &	NOCOV &	No coverage in cube  \\
1 &	2 &	LOWCOV &	Low coverage depth in cube \\
2 &	4 &	DEADFIBER &	Major contributing fiber is dead \\
3 &	8 &	FORESTAR &	Foreground star \\
10 &	1024 &	DONOTUSE &	Do not use this spaxel for science 
 \enddata
 \tablecomments{}
 \end{deluxetable}

The progress of a given exposure through the DRP is controlled by use of the MANGA\_DRP2QUAL maskbit (Table \ref{drp2qual.tab}, which indicates any potential problems that affect the
reduction of the exposure.  These range from the informative for operations (RAMPAGINGBUNNY indicates dust accumulation on the IFU surfaces that must be cleaned) to
the fatal (FULLCLOUD indicates that the transparency is too low to successfully flux calibrate the data).

 \begin{deluxetable}{rrll}
\tabletypesize{\footnotesize}
\tablecolumns{4} 
\tablewidth{0pt}
 \tablecaption{MANGA\_DRP2QUAL Data Quality Bits
 \label{drp2qual.tab}}
 \tablehead{
 \colhead{Bit} & \colhead{Value} & \colhead{Label} & \colhead{Description}}
 \startdata 
0 &	1 &	VALIDFILE &	File is valid \\
1 &	2 &	EXTRACTBAD 	& Many bad values in extracted frame \\
2 &	4 &	EXTRACTBRIGHT 	& Extracted spectra abnormally bright \\
3 &	8 &	LOWEXPTIME 	& Exposure time less than 10 minutes \\
4 &	16 &	BADIFU 	& One or more IFUs missing/bad in this frame \\
5 & 	32 &	HIGHSCAT &	High scattered light levels \\
6 &	64 &	SCATFAIL &	Failure to correct high scattered light levels \\
7 &	128 &	BADDITHER &	Bad dither location information \\
8 &	256 &	ARCFOCUS &	Bad focus on arc frames \\
9 &	512 &	RAMPAGINGBUNNY &	Rampaging dust bunnies in IFU flats \\
10 &	1024 &	SKYSUBBAD &	Bad sky subtraction \\
11 &	2048 &	SKYSUBFAIL &	Failed sky subtraction \\
12 &	4096 &	FULLCLOUD &	Completely cloudy exposure \\
13 &	8192 &	BADFLEXURE & 	Abnormally high flexure LSF correction 
 \enddata
 \tablecomments{}
 \end{deluxetable}

The final quality of a given object processed by the 3d stage of the DRP is indicated by the reduction quality bit
MANGA\_DRP3QUAL (Table \ref{drp3qual.tab}).  This single integer refers to the quality of an entire
galaxy data cube, and can indicate a variety of possible problems sorted roughly in increasing order of importance from low average depth (BADDEPTH) to a CRITICAL
failure that means that the data should be treated with great caution or (conservatively) omitted from science analyses.  We note that many of even the CRITICAL failure cases
may represent an overly-vigorous QA algorithm rather than any intrinsic problem in the data though; these routines will continue to be refined throughout SDSS-IV.

 \begin{deluxetable}{rrll}
\tabletypesize{\footnotesize}
\tablecolumns{4} 
\tablewidth{0pt}
 \tablecaption{MANGA\_DRP3QUAL Data Quality Bits
 \label{drp3qual.tab}}
 \tablehead{
 \colhead{Bit} & \colhead{Value} & \colhead{Label} & \colhead{Description}}
 \startdata 
0 &	1 &	VALIDFILE &	File is valid  \\
1 &	2 &	BADDEPTH &	IFU does not reach target depth  \\
2 &	4 &	SKYSUBBAD &	Bad sky subtraction in one or more frames  \\
3 &	8 &	HIGHSCAT &	High scattered light in one or more frames  \\
4 &	16 &	BADASTROM &	Bad astrometry in one or more frames  \\
5 &	32 &	VARIABLELSF &	LSF varies significantly between component spectra  \\
6 &	64 &	BADOMEGA &	Omega greater than threshhold in one or more sets  \\
7 &	128 &	BADSET &	One or more sets are bad  \\
8 &	256 &	BADFLUX &	Bad flux calibration  \\
9 &	512 &	BADPSF &	PSF estimate may be bad  \\
30 &	1073741824 &	CRITICAL &	Critical failure in one or more frames   
 \enddata
 \tablecomments{}
 \end{deluxetable}

We note that additional bits may be added to each of these quality control bitmasks over the lifetime of the survey.  An online version can be found at http://www.sdss.org/dr13/algorithms/bitmasks/
for DR13, and at similar locations for future data releases.

\end{document}